\documentclass[aps,psfig,prb,showpacs,superscriptaddress]{revtex4}
\usepackage{dcolumn}
\usepackage{amsmath}
\usepackage{graphicx}
\usepackage{latexsym}
\usepackage{amsfonts}
\usepackage{amssymb}
\usepackage{color}

\begin{document}

\title{Layer Response Theory: Energetics of layered materials from
semi-analytic high-level \ theory}
%\author{John F Dobson, Tim Gould \ and Sebastien Lebegue}
\author{John F Dobson}
\affiliation{Queensland Micro and Nano Technology Centre, Griffith
  University, Nathan, Queensland 4111, Australia}
\affiliation{Universit\'e de Lorraine,  Vandoeuvre-l\`es-Nancy, F-54506, France}
\author{Tim Gould}
\affiliation{Queensland Micro and Nano Technology Centre, Griffith
  University, Nathan, Queensland 4111, Australia}
\author{S\'ebastien~Leb\`egue}
\affiliation{Laboratoire de Cristallographie, R\'esonance Magn\'etique et Mod\'elisations (CRM2, UMR CNRS 7036)
 Institut Jean Barriol, Universit\'e de Lorraine  BP 239, Boulevard des Aiguillettes  54506 Vandoeuvre-l\`es-Nancy,France}
%\maketitle    must be after anbstract in Revtex 4


\begin{abstract}
We present a readily computable semi-analytic "Layer response theory" (LRT)
for analysis of cohesive energetics involving two-dimensional layers such as
BN or graphene. The theory approximates the Random Phase Approximation (RPA)
correlation energy. Its RPA character ensures that the energy has the
correct van der Waals asymptotics for well-separated layers, in contrast to
simple pairwise atom-atom theories, which fail qualitatively for layers with
zero electronic energy gap. At the same time our theory is much less
computationally intensive than the full RPA energy. It also gives accurate
correlation energies near to the binding minimum, in contrast to
Lifshitz-type theory.  We apply our LRT theory successfully to graphite and to BN, and to a graphene-BN heterostructure. 
\end{abstract}

\maketitle

\section{Introduction}

The interest in two dimensional compounds such as graphene\cite{novoselov_electric_2004}, hexagonal boron nitride\cite{novoselov_two-dimensional_2005,coleman_two-dimensional_2011},
 transition metal dichalcogenides\cite{novoselov_two-dimensional_2005,coleman_two-dimensional_2011,mak_atomically_2010}, and other promising compounds has grown exponentially
 over the last few years\cite{geim_rise_2007,castro_neto_electronic_2009,xu_graphene-like_2013,liao_chemistry_2014}. In particular, there has been much recent interest
 in their cohesive properties 
 and also in layered, heterostructured solids composed of loosely-bound stacks of such 2D layers \cite{geim_van_2013}.
Interesting properties include for example the binding energetics of atoms or
molecules on such systems\cite{amft,mebphthalo,liu_benzene_2012}, and the energetics of binding 
\cite{RPAGrapheneOnMedtalsDespojaPRB2014} 
%and of mechanical distortion 
of the layer systems themselves.
%(linear and nonlinear elastic constants) {\blue citations from Tim?}. 
%Applications include gas separation technologies and dynamics of
Possible applications include gas separation and storage technologies,
\cite{HAbsGrapheneOrigamiZhuACSNano2014} and 
%dynamics of
nano-electro-mechanical devices \cite{GraphitizedSiCMicrobeamsCunningNanotech2014,SiliceneTransistorsThTao,CatalGrapheneOnSiC_Si_Iacopi2015}.

The binding energetics of these layer systems typically show a significant
contribution from van der Waals (vdW) forces, which renders
their description by semilocal density functionals inaccurate. Much recent
effort has therefore been put into simple vdW corrections, most typically
involving a pairwise atom-based sum of vdW-type $R^{-6}$ atom-atom
interactions \cite{Grimme+FT+DWithBondCountingJCP2010,AchievmentsShortcomingsDFT_D3_chargedSupramolecular,Grimme:06,Tkatchenko:09,Grimme:10,kim_universal_2012}. Some more
sophisticated theories also can be expressed via pairwise summation 
\cite{vdWFnalGenGeomDionPRL04}. While such pairwise and triples approaches are
computationally efficient and often have good success with the inner,
close-contact parts of the binding energy curves, it is known that they give
an incorrect description of the far-distant energetics, especially for layers
that are highly polarizable, metallic or semi-metallic \cite{AsyDispIntPRL2006JFDWhiteRubio,TkatchenkoMB_vdW_PRL2012,GobreTkatvdWscalingNatComm2013,CasimirgrapheneSernelius2011,vdWFullerenesDefyConvWisdomRuzsEtalPRL2012}. 

%{\blue Seb: I SUGGEST TO REMOVE THIS PARAGRAPH SINCE IT IS IN OVERLAP WITH WHAT %IS WRITTEN 2 PARAGRAPHS BELOW. In the present work we
%propose a readily computable dispersion energy aproximation for layered
%systems.  This semi-analytic correction is based on the Random Phase %approximation (RPA) approach
%outlined below, and it has the correct RPA asymptotic behavior at large
%layer separations, even for zero-gap systems, a feature not shared by any of
%the schemes just mentioned.}

A relatively successful approach is the full numerical (direct) Random Phase Approximation ((d)RPA) correlation
energy method \cite{DobsonWangPRL99,RPAEnergyBN_MariniRubioPRL2006,RPABulkPropertiesHarlKressePRL09,GraphiteDispIntPRL2010LebegueEtal,Lu:09}, henceforth denoted simply "RPA".
This method and its extensions \cite{SOSEXtestsGruneis2009},
\cite{OlsenThygesenEnergiesFrFxc_PRL2014}
have been found to give the best
and most consistent treatment to date 
%{\blue ( Seb: I think that we should be %careful here: Kresse and Thygesen are doing beyond RPA calculations now)} 
of energy differences in layered
systems at all layer spacings provided electromagnetic retardation and higher many-body effects 
\cite{HowMBTAffectsGraphenevdW} are are unimportant.
  \cite{AreWevdWreadyBjorkmanJPCM2012,RPASolidsvdWCoval_OlsenThygesenPRB2013}. In particular, the RPA approach
correctly describes the very long-wavelength charge fluctuations not
captured in pairwise theories, but responsible for anomalous long-ranged van
der Waals interactions \cite{AsyDispIntPRL2006JFDWhiteRubio}. These
differences from the pairwise predictions are manifested in both the
dependence of the energy on layer spacing $D$ \cite{AsyDispIntPRL2006JFDWhiteRubio,HChainsLiuAngyanDobsonJCP2011} and,
for finite systems, on the number of atoms $N$ \cite{GobreTkatvdWscalingNatComm2013,vdWFullerenesDefyConvWisdomRuzsEtalPRL2012}. 
Furthermore, the RPA (without extensions) has been shown 
\cite{GraphiteDispIntPRL2010LebegueEtal} to give simultaneously a good description of the layer spacing, the breathing elastic constant $C_{33}$ and the layer binding energy of graphite, a task that proved elusive with other methods. 
The RPA correlation  approach,
 however, comes at a high computational cost that makes the analysis of complex systems difficult or impossible.

\section{Layer Response Theory}

In the present work we aim to produce a simplified, computationally
efficient theory of the outer parts of the RPA correlation energy curves
involving nano-thin layers. We then introduce a simple extension allowing
analysis right down to intimate contact between layers. 

For macrospically
thick systems (slabs) there is already a very good theory of this type for
the outer parts of the energy curve, namely the Lifshitz approach \cite{vdWGenThDzyaloshinskiilLifshitzPitaevski1961,ParsegianvdW,vdWBooks3}  
which describes non-contacting systems but, unlike the pure
RPA, is not applicable to intimately contacting cases. The Lifshitz
approach, rooted in quantum electrodynamics, has been further elaborated
recently by a number of authors, e.g. refs \onlinecite{ScattThCasimirRahiEmigEtalPRD09,ParsegianvdW,MacroQEDLifshitzScheelBuhmannSlovakJ2008,BuhmannDispersionFrcBk1,BuhmannDispersionFrcBk2}. It
uses the long-wavelength dielectric properties of the infinite solid as
input (i.e. these are not predicted by Lifshitz theory, and must be obtained from experiment or independent calculations). The
electromagnetically non-retarded limit of Lifshitz theory can alternatively
be obtained by using an RPA-style correlation energy treatment \cite{CalcDispersionEnergiesDobsonJPCM2012} to predict the van der Waals
interaction between two semi-infinite solids. The RPA\ correlation energy
approach, and our simplification of it described here, have a distinct advantage over
traditional Lifshitz theory, however, in that they both remain accurate at
small inter-layer separations right down to the binding distance.

%{\blue (Seb: I think that these two sentence should be merged.) 
The present theory is nevertheless somewhat similar in spirit to Lifshitz
theory, in that it evaluates semi-analytically the outer parts of the
energy-versus distance curves, using the long-wavength dielectric properties
of a single layer as input. An important aspect of our approach, however, is that, rather than attempting to treat a monolayer as a very thin 3D macroscopic 
dielectric as one might do in Lifshitz theory, we  
carefully evaluate the realistic dielectric properties of a single 2D layer at the RPA level to the lowest three orders in a wavenumber expansion. We obtain this data
using RPA-level calculation of the \emph{long-wavelength}
\emph{macroscopic} dielectric function of a \emph{slightly stretched} layered solid made
from the individual layers under study. Such a calculation is now
routinely available with great computational efficiency in modern
ab-initio 3D codes (in our case VASP \cite{VASP1,VASP2}). In particular, we do not
require expensive calculations with large vacuum spaces between the layers and
consequently large unit cells, because we show below how to deconvolute the
Coulomb interactions between the layers analytically (see also the recent
work of Nazarov \cite{LayerResponseQuantitiesUsefulExptNazarovNJP2015} and
of \ Andersen et al. \cite{DielectricGenomevdWHetAndersenThygesenNanolett2015}).

We take the long-wavelength $O(Q^{0})$ layer polarizability
data so obtained and augment it via analytic reasoning and a fitted
parameter to obtain the $O( Q^{1}) $ and 
$O(Q^{2})$ contributions to the polarizability respectively, for
each species of monolayer.\ The $O(Q^{2})$ term puts our
theory well beyond the usual Lifshitz theory, which assumes strictly local
dielectric properties. It allows us to obtain the correlation energy
semi-analytically for systems approaching the contact configuration.
We term our approach the ''Layer Response Theory''\ of interactions in our
target systems. \ We first introduce the electronic response functions of a
single layer to external longitudinal fields, in a particularly convenient
form, namely reflection and (differential) transmission coefficients $%
R(Q,\omega )$ and $T\left( Q,\omega \right) $, and then relate
these to screened layer polarizabilities $\alpha ^{2D,scr}$. The
reflection/transmission (more generally, scattering) approach is in fact
 the popular modern way to
implement Lifshitz theory \cite{ScattThCasimirRahiEmigEtalPRD09}.

\section{Reflection and transmission of electrostatic fields by an isolated
layer}

\begin{figure}[tbp]
% currently Fig 1
\includegraphics[width=\linewidth]{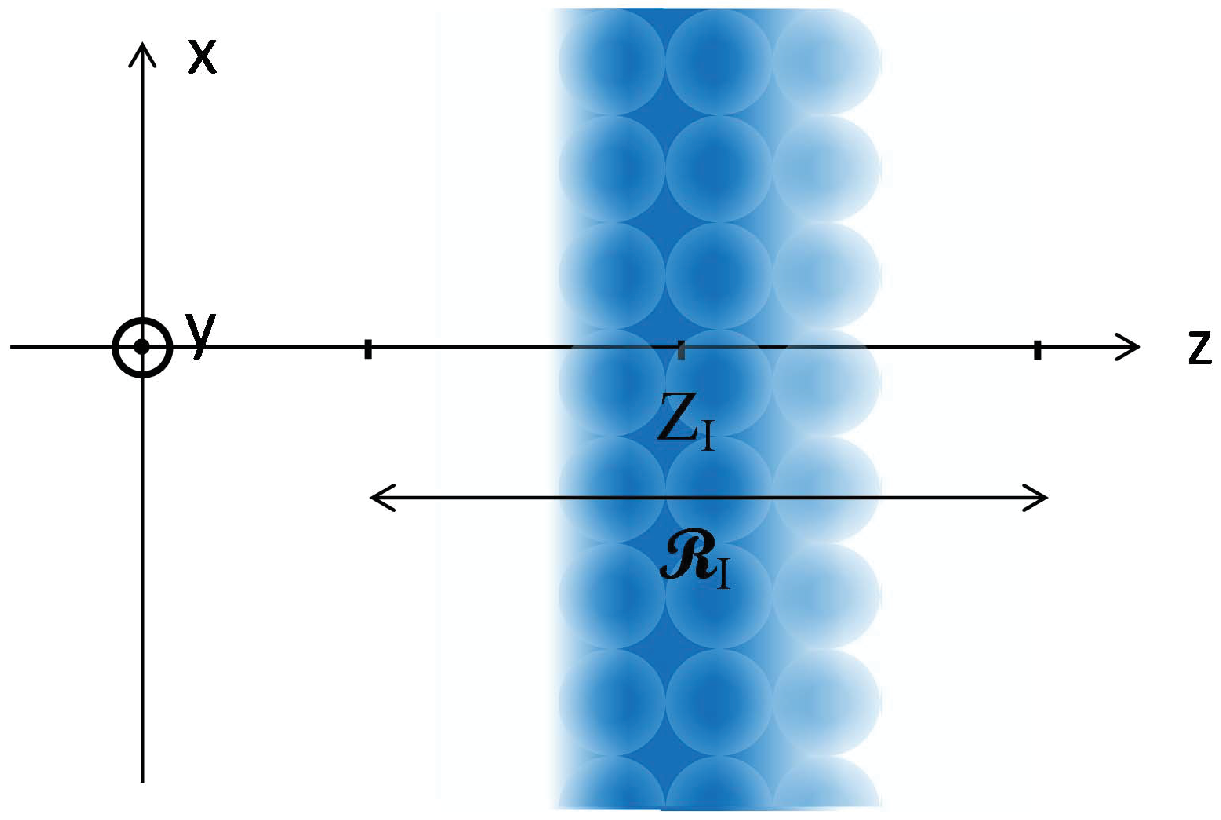}
\caption{%
Scheme for division of the $z$ axis  
\label{fig:LayerDiagram}
}
\end{figure}

We define a \textquotedblright layer\textquotedblright , labelled
\textquotedblright $I$\textquotedblright , \ of a material \ that has a
periodic structure in the $x$ and $y$ spatial directions and is confined to a
finite region $\mathcal{R}_{I}$ of the $z$ axis, containing a reference
point $Z_{I}\in \mathcal{R}_{I}$.(See Fig {\ref{fig:LayerDiagram}).  By \textquotedblright
confined\textquotedblright\ we mean that the groundstate electronic number
density $n_{0I}\left( \vec{r}\right) $ and any linear time-dependent density
perturbations $\delta n_{I}$ to it are assumed to vanish outside $\mathcal{R}%
_{I}$ . In a perturbed situation, we can decompose the time-dependent
density perturbations $\delta n_{I}$ into components varying as $\exp \left(
i\left( \vec{Q}+\vec{G}\right) .\vec{r}\right) \exp \left(
ut\right) \delta n_I\left( \vec{Q}+\vec{G},z,iu\right) $. Here 
$\vec{G}= G_x \vec{i} + G_y \vec{j}$ is a reciprocal lattice vector of the 2D periodic lattice, and 
 $\vec{Q} = q_x \vec{i} + q_y \vec{j}$ is a 2D wavenumber lying inside the corresponding 2D Brillouin
zone. Such a density perturbation will set up potentials $\delta W$ outside the
layer given by%
\begin{equation}
\delta W_{due\,to\,I}\left( x,y,z\right) =\left\{ 
\begin{array}{c}
m_{I}^{+}\exp \left( i\left( \vec{Q}+\vec{G}\right) .\vec{r}%
\right) \frac{2\pi e^2 }{\left\vert \vec{Q}+\vec{G}\right\vert }%
e^{-\left\vert \vec{Q}+\vec{G}\right\vert \left( z-Z_{I}\right)
},\;\;\;z\;to\;right\;of\,\mathcal{R}_{I} \\ 
m_{I}^{-}\exp \left( i\left( \vec{Q}+\vec{G}\right) .\vec{r}%
\right) \frac{2\pi e^2 }{\left\vert \vec{Q}+\vec{G}\right\vert }%
e^{+\left\vert \vec{Q}+\vec{G}\right\vert \left( z-Z_{I}\right)
},\;\;\;z\;to\;left\;of\,\mathcal{R}_{I}%
\end{array}%
\right.  \label{PotlDueTolayerI}
\end{equation}%
Here the two moments 
\begin{eqnarray*}
m_{I}^{+}(\vec{Q},\vec{G},iu) &=&\int_{\mathcal{R}_{I}}e^{\left\vert 
\vec{Q}+\vec{G}\right\vert \left( z-Z_{I}\right) }
%\deltan_{I}\left( z-Z_{I}\right) dz \\
\delta n_I\left( \vec{Q},\vec{G},z,iu\right) dz\\
m_{I}^{-}(\vec{Q},\vec{G},iu) &=&\int_{\mathcal{R}_{I}}e^{-\left\vert 
\vec{Q}+\vec{G}\right\vert \left( z-Z_{I}\right) }
%\delta n_{I}\left( z-Z_{I}\right) dz
\delta n_I\left( \vec{Q},\vec{G},z,iu\right) dz
\end{eqnarray*}%
completely specify (via Eq (\ref{PotlDueTolayerI})) the effects of layer $I$
on non-overlapping objects such as other layers or physisorbed atoms. From
here onwards we assume initially that the matter in layer $I$ is sufficiently well
separated from the edges of its region $\mathcal{R}_{I}$ that the rapidly
(exponentially) decaying $\vec{G}\neq \vec{0}$ components in Eq (\ref{PotlDueTolayerI})
are negligible outside $\mathcal{R}_{I}$.  See Fig \ref{fig:LayerDiagram}. (We do modify this assumption for our
 final theory, however) Thus initially we only need the $\vec{G}%
=\vec{0}$ moments, which we shall denote 
%{\blue Seb: The quantity on the %right depends	 on $Z_{I}$ but not the quantity %on the left. In fact I think %that we integrate over $z-Z_{I}$ ???}
\begin{equation}
m_{I}^{\pm }\equiv m_{I}^{\pm }(\vec{Q},\vec{G}=\vec{0},iu)=\int_{%
\mathcal{R}_{I}}e^{\pm \left\vert \vec{Q}\right\vert \left(z-Z_{I}\right) }
%\delta n_{I}\left( z-Z_{I}\right) dz\;\;.
%\delta n\left( z\right) dz\;\;.
\delta n_I\left( \vec{Q},\vec{G}=\vec{0},z,iu\right) dz
\label{DefMplusMinus}
\end{equation}

The potential $V_{I}(z)\exp \left( i\vec{Q}.\vec{r}\right) exp(ut)$
due to sources outside layer $I$ and acting on the matter inside layer $I$,
is assumed to lack the rapidly (exponentially)  damped $G \neq 0$ components because of
the isolation of \ the matter deep enough inside layer $I$.  
This
potential has no sources inside layer $I$ and hence obeys Poisson's equation 
$\nabla ^{2}V_{I}=\left( -Q^{2}+d^{2}/dz^{2}\right) V_{I}=0$ for $z\in $
$\mathcal{R}_{I}.$ Thus it can be completely specified by two amplitudes $%
V_{I}^{\pm }$: 
\begin{equation}
V^{ext}\left( z\right) =V_{I}^{+}e^{\left\vert \vec{Q}\right\vert
 z }+V_{I}^{-}e^{-\left\vert \vec{Q}\right\vert
 z },\ \ \ \ z\epsilon \mathcal{R}_{I}  \label{VextI}
\end{equation}

Suppose that a small external field impinges on layer $I$ from the right, so
that $V_{I}^{-}=0$ in (\ref{VextI}). This potential gives rise to a density
perturbation inside layer $I$ 
\begin{equation}
\delta n_{I}\left( z\right) =\int_{\mathcal{R}_{I}}dz^{\prime }\chi
_{I}\left( Q,\vec{G}=\vec{G}^{\prime }=\vec{0}%
,z,z^{\prime },iu\right) V_{I}^{+}e^{\left| \vec{Q}\right|
\left( z^{\prime }-Z_{I}\right) }  \label{DeltaNLayerIFromChiI}
\end{equation}%
where the finite-frequency density-density response $\chi _{I}$ of the layer has to include
all Coulomb screening and local field effects inside layer $I$. 
%{\blue (Seb: I %suggest to remove this sentence to keep the discussion focus) Later we
%will show how to extract the needed integrals of $\chi _{I}$ for $%
%Q\rightarrow 0$ from inexpensive macroscopic dielectric function
%calculations for a slightly stretched layered solid using 
%%a package such as
%the dielectric tensor capabilities of off-the-shelf codes for three-dimensional crystals.}
%

The density (\ref{DeltaNLayerIFromChiI}) in turn creates a reflected field
to the right of $\mathcal{R}_{I}$ of form%
\begin{equation}
V^{refl,r}=R_{I}^{right}V_{I}^{+}\exp \left( i\vec{Q}.\vec{r}\right)
\exp \left( -Q\left( z-Z_{I}\right) \right) ,\;\;\;z\text{ to right of
\ }\mathcal{R}_{I}  \label{FormOfVreflRight}
\end{equation}%
where, from (\ref{PotlDueTolayerI}) and (\ref{DeltaNLayerIFromChiI}), the
dimensionless \textquotedblright right hand reflection
coefficient\textquotedblright\ $R_{I}^{right}$ is%
\begin{equation}
R_{I}^{right}=\frac{2\pi e^{2}}{Q}\int_{%
\mathcal{R}_{I}}dzdz^{\prime }e^{ Q  \left(
z+z^{\prime }-2Z_{I}\right) }\chi _{I}\left( \vec{Q},z,z^{\prime
},iu\right)  \label{RRightFromChi}
\end{equation}

Similarly, still with a potential $V_{I}^{+}$ impinging from the right,
layer $I$ also creates an additional \textquotedblright
transmitted\textquotedblright\ potential to its left 
\begin{equation}
V^{tr,right}=T_{I}^{right}V_{I}^{+}\exp \left( i\vec{Q}.\vec{r}\right)
\exp \left( Q\left( z-Z_{I}\right) \right) ,\;\;\;z\text{ to left of \ }%
\mathcal{R}_{I}  \label{FormOfVTrasmRight}
\end{equation}%
where the dimensionless \textquotedblright right (differential) transmission
coefficient\textquotedblright\ is 
\begin{equation}
T_{I}^{right}=\frac{2\pi e^{2}}{ Q}\int_{%
\mathcal{R}_{I}}dzdz^{\prime }e^{ Q \left(
-z+z^{\prime }\right) }\chi _{I}\left( \vec{Q},z,z^{\prime
}iu\right)  \label{TrightFromChi}
\end{equation}%
Similarly for fields impinging on layer $I$\ from the left we have
\textquotedblright left\textquotedblright\ reflection and transmission
coefficients%
\[
R_{I}^{left}=\frac{2\pi e^{2}}{Q}%
\int_{\mathcal{R}_{I}}dzdz^{\prime }e^{-Q \left(
z+z^{\prime }-2Z_{I}\right) }\chi _{I}\left( \vec{Q},z,z^{\prime
},iu\right) 
\]%
\[
T_{I}^{left}=\frac{2\pi e^{2}}{Q}\int_{%
\mathcal{R}_{I}}dzdz^{\prime }e^{Q \left(z-z^{\prime }\right) }
\chi _{I}\left( \vec{Q},z,z^{\prime},iu\right) 
\]%`
For a layer with inversion symmetry in the $z$ direction, the distinction
between fields impinging from left or right disappears, and we just have a
reflection coefficient $R_{I}$ and and a transmission coefficient $T_{I}$
given by (\ref{RRightFromChi}) and (\ref{TrightFromChi}).

It is useful to re-express these reflection and transmission coefficients in
terms of an interacting or screened polarizability $a_{ij}^{scr}$ giving the
change in polarization $p_{i}$ (dipole moment per unit volume) per unit
applied external electric field\ $E_{j}^{ext}$. By contrast the above
quantity $\chi $ gives the response of the number density $\delta n$ to a
unit external electronic potential energy perturbation $\delta V^{ext}$. \
Since $\delta n=\left| e\right| ^{-1}\vec{\nabla}.\vec{p}$ and $\vec{E}%
^{ext}=\left| e\right| ^{-1}\vec{\nabla}V^{ext}$ it follows that in general 
\cite{JFDDintvdW96},\cite{DobsonBrisvdWChap} 
\begin{equation}
\chi \left( \vec{r},\vec{r}~^{\prime },\omega \right) =-e^{-2}\partial
_{i}\partial _{j^{\prime }}\alpha _{ij}^{scr}\left( \vec{r},\vec{r}%
~^{\prime }\omega \right)   \label{Chi3DFromAlphaScr3D}
\end{equation}%
Note that this ''screened''\ or ''interacting ''\ polarizability $\alpha
^{scr}$ (denoted $ -e^{-2} F$ in ref. \onlinecite{JFDDintvdW96}) gives the response of the
polarization to the \emph{external} field whereas the usual polarizability $%
\alpha \equiv \left( \varepsilon -1\right) /4\pi $ gives the response to the \emph{%
total} field. $\ $Thus $\alpha $ is an ''irreducible''\ quantity that does
not describe the RPA-level screening effects within a layer, in contrast to $%
\alpha^{scr}$. Applying (\ref{Chi3DFromAlphaScr3D}) to (\ref{RRightFromChi}) and
(\ref{TrightFromChi}) we find via integration by parts (where henceforth we assume a maximally
symmetric layer where the polarizability decouples into an 
$xx\equiv yy \equiv \parallel$
component and a $zz \equiv \perp$ component)%
\begin{eqnarray}
R_{I}\left( \vec{Q},\omega \right)  &=&-2\pi Q
\left( \alpha _{xx}^{2D,scr,+}\left( Q,\omega \right) 
+\alpha_{zz}^{2D,scr,+}\left( Q,\omega \right) \right) 
\label{RFromAlphaScr2D} \\
T_{I}\left( \vec{Q},\omega \right)  &=&-2\pi Q%
\left( \alpha _{xx}^{2D,scr,-}\left( Q,\omega \right) -\alpha
_{zz}^{2D,scr,-}\left( Q,\omega \right) \right) 
\label{TFromAlphaScr2D}
\end{eqnarray}

where the screened layer polarizability of layer $I$ is 
\begin{equation}
\alpha _{ii}^{2D,scr,\pm }\left( Q,\omega \right) =\int_{\mathcal{R}%
_{I}}dzdz^{\prime }e^{Q\left( \left(z-Z_{I}\right) 
\pm \left( z^{\prime }-Z_{I}\right) \right) }\alpha
_{ii}^{scr}\left( \vec{Q},z,z^{\prime },iu\right)
\label{DefAlphaScrLayerPM}
\end{equation}%
Here the four quantities $\alpha _{ii}^{2D,scr,\pm }\left( Q,\omega\right) $
 for $i=x$ or $z$ are the lumped screened polarizabilities of a
single layer, and in the limit $Q\rightarrow 0$ they all reduce to the
polarization per unit area of the layer per unit \emph{external} electric
field $E^{ext}$. \ For a layer with inversion symmetry in the $z$ direction, 
with $Z_I$ chosen at the symmetry point,
an expansion in powers of $Q$ shows that the \textquotedblright
+\textquotedblright\ and \textquotedblright -\textquotedblright\ versions 
$\alpha _{ii}^{2D,scr,\pm }\left( Q,\omega \right) $ are the same
through $O\left( Q\right) $. Thus in (\ref{RFromAlphaScr2D}) and (\ref%
{TFromAlphaScr2D}) we can drop the superscripts+ and - on the
polarizabilities $\alpha $ in many of the applications below.

An advantage of using (\ref{RFromAlphaScr2D}) and (\ref{TFromAlphaScr2D})
is that, \emph{for an insulating/semiconducting layer}, $\alpha
_{ii}^{2D,scr\pm }$ approach finite nonzero constants $\alpha
_{xx}^{scr}\left( \omega \right) $ and $a_{zz}^{scr}\left( \omega \right) $
as $Q\rightarrow 0$, so that $R$ and $T$ $\ $vanish linearly with $Q$. 

For a metallic layer or graphene, 
$\alpha _{xx}^{scr}\left( \omega \right)$ diverges as $\omega \rightarrow 0$ 
and special care must be taken.

\section{Moment response of a single layer}

%{\blue (Seb: It seems that equations (1) and (2) are not needed to derive Eqs %(3) to 
%(12). Shouldn't we put Eqs. (1) and (2) in this paragraph rather than in the% %previous one ?)}

When an external potential of form (\ref{VextI}) is applied to layer $I$, we
find from (\ref{DeltaNLayerIFromChiI}) that the moments defined in (\ref%
{DefMplusMinus}) are 
\begin{eqnarray*}
m_{I}^{+} &=&\frac{Q}{2\pi e^{2}}\left(
R_{I}^{right}V_{I}^{+}+T_{I}^{left}V_{I}^{-}\right) \\
m_{I}^{-} &=&\frac{Q}{2\pi e^{2}}\left(
T_{I}^{right}V_{I}^{+}+R_{I}^{left}V_{I}^{-}\right)
\end{eqnarray*}

This can be written%
\begin{equation}
\vec{m}_{I}=\frac{Q}{2\pi e^{2}}\mathbf{c}_{0I}\vec{V}_{I}=\frac{Q%
}{2\pi e^{2}}\sum_{J}\mathbf{c}_{0IJ}\vec{V}_{J}  \label{MvecEqC0Vvec}
\end{equation}%
where $\vec{m}_{I}=\left( m_{I}^{+},m_{I}^{-}\right) ^{T}$ and $\vec{V}%
_{I}=\left( V_{I}^{+},V_{I}^{-}\right) ^{T}$ are 2D column vectors and 
\begin{equation}
\mathbf{c}_{0I}=\left( 
\begin{array}{cc}
R_{I}^{right} & T_{I}^{left} \\ 
T_{I}^{right} & R_{I}^{left}%
\end{array} \right) ,
\;\;\mathbf{c}_{0IJ}=\delta _{IJ}\left( 
\begin{array}{cc}
R_{I}^{right} & T_{I}^{left} \\ 
T_{I}^{right} & R_{I}^{left}%
\end{array}%
\right) \;  \label{DefC0asymm}
\end{equation}%
is a dimensionless response function for the single isolated layer. \ We
note that even for a non-symmetric layer the general reciprocity relation $%
\chi (\vec{r},\vec{r}\,^{\prime },iu)=\chi (\vec{r}\,^{\prime },\vec{r}%
\,,iu) $ implies that $T^{left}\left( \vec{Q},iu\right) =T^{right\ast
}\left( Q,iu\right) $ so that $\mathbf{c}_{0I}$ is a hermitian
matrix. (But there is in general no simple relation between $R^{right}$ and $%
R$ $^{^{left}}$ for a non-symmetric layer$).$

\section{RPA equation for response of interacting layers}

Here the electron-electron Coulomb interactions \emph{within} a layer are
assumed to be included already in the layer response quantities $R_{I}$ and $%
T_{I}$. We now treat the Coulomb interactions \emph{between} the layers in
the Random Phase Approximation (RPA, time-dependent Hartree approximation).
This means that the potential coefficients $V_{I}^{\pm }$ acting on layer $I$
in (\ref{MvecEqC0Vvec}) consist of an externally imposed potential $%
V_{I,ext}^{\pm }$ and an internal potential $V_{I,int}^{\pm }$ that is
generated by the dynamic charge moments $m_{J}^{\pm }$, $J\neq I$, existing
on the other layers, as in (\ref{DefMplusMinus}):%
\[
V_{I,int}^{+}=\sum_{J>I}\int_{z^{\prime }\in R_{J}}dz^{\prime }\frac{2\pi
e^{2}}{Q}e^{-Q\left\vert Z_{I}-(Z_{J}+z^{\prime })\right\vert
}\delta n_{J}\left( z^{\prime }\right) =\frac{2\pi e^{2}}{Q}%
\sum_{J}v_{IJ}^{+-}m_{J}^{-} 
\]%
and similarly $V_{I,int}^{-}$ is sourced from layers to the left, with $J<I.$
These two cases can be summarized by 
\[
\vec{V}_{I,int}=\frac{2\pi e^{2}}{Q}\sum_{J}\mathbf{v}_{IJ}\vec{m}_{J} 
\]%
where we have defined a $2\times 2$ dimensionless interlayer Coulomb matrix
by%
\begin{equation}
\mathbf{v}_{IJ}=\equiv \left( 
\begin{array}{cc}
v_{IJ}^{++} & v_{IJ}^{+-} \\ 
v_{IJ}^{-+} & v_{IJ}^{--}%
\end{array}%
\right) =e^{-Q\left\vert Z_{I}-Z_{J}\right\vert }\left( 
\begin{array}{cc}
0 & \bar{\theta} _{J>I} \\ 
\bar{\theta} _{I>J} & 0%
\end{array}%
\right)  \label{DefuIJ}
\end{equation}%
where $\bar{\theta}_{cond}=1$ if the condition $cond$ is satisfied, and equals $0$ otherwise. Then
the RPA equations are (with a Coulomb reduction factor $\lambda $ that
equals $1$ for the real system)

\begin{equation}
\vec{m}_{I}=\vec{m}_{I}^{ext}+\mathbf{c}_{0I}\lambda \sum_{J}\mathbf{v}_{IJ}%
\vec{m}_{J}\;\;.  \label{RPALayerEqWithGeneralSource}
\end{equation}%
Here, in the source term 
\begin{equation}
m_{I}^{\pm ext}=\int \exp \left( \pm Qz\right) \chi _{I}\left(
Q,z,z^{\prime },iu\right) V^{ext}\left( z^{\prime }\right) dzdz^{\prime}\;\;,
\label{Gemeralsource}
\end{equation}%
we have assumed no particular $z$ dependence of the external potential $%
V^{ext}\left( z^{\prime }\right) \exp \left( \vec{Q}.\vec{r}%
--i\omega t\right) $. If indeed the external potential is of the form (%
\ref{VextI}) then $\vec{m}_{I}^{ext}=\frac{Q}{2\pi e^{2}}\mathbf{c}_{0I}%
\vec{V}_{I}^{ext}$ and so \ 
\begin{equation}
\vec{m}_{I}=\frac{Q}{2\pi e^{2}}\mathbf{c}_{0I}\vec{V}_{I}^{tot}=%
\mathbf{c}_{0I}\left( \frac{Q}{2\pi e^{2}}\vec{V}_{I}^{ext}+\lambda
\sum_{J}\mathbf{v}_{IJ}\vec{m}_{J}\right)  \label{CoupledLayersRPAEqIJ}
\end{equation}%
The equations are off-diagonal in the layer index $I$ and have the formal
solution 
\begin{equation}
\vec{m}_{I}=\frac{Q}{2\pi e^{2}}\sum_{J}\mathbf{c}_{\lambda IJ}\vec{V}%
_{J}^{ext}  \label{MIFromCIJ}
\end{equation}%
\begin{equation}
\mathbf{c}_{\lambda }=(\mathbf{I}-\lambda \mathbf{c}_{0}\mathbf{v)}^{-1}%
\mathbf{c}_{0}  \label{InteractingCDimless}
\end{equation}%
where the inverse is over both the $I,J$ (layer) and $+-$ (reflection /
transmision) indices. \ ( $\mathbf{c}_{\lambda }$ is the response of the
multi-layer system with both intra- and inter-layer Coulomb interactions
included).

\section{Practical determination of $\protect\alpha ^{2D,scr}$for $%
Q\rightarrow 0$}

The long-wavelength 2D screened polarizability $\alpha ^{2D,scr}$ of a
single isolated layer from (\ref{DefAlphaScrLayerPM}) will be used 
as input for much of what follows. While
there exist intrinsically 2D codes (such as ADF\ Band) and other approaches
 \cite{RPAGrapheneOnMedtalsDespojaPRB2014} that can
calculate $\alpha ^{2D,scr}$ (or the correponding density response) directly, most condensed matter
calculations use 3D plane-wave codes such as ABINIT or VASP \cite{VASP1},%
\cite{VASP2}. We show here that we can obtain $\alpha ^{2D,scr}\left(
Q\rightarrow 0,i\omega \right) $ from such 3D codes by finding the 3D macroscopic
dielectric function of an infinite stack of parallel layers, with a layer
spacing $D$ that only modestly exeeds the equilibrium spacing, $D\geq 2D_{0}.$
Of course the single-layer properties could be deduced by choosing a very
large $D$ value, but such calculations are numerically prohibitive because
of the large unit cell involved. Instead we show how to deconvolute the
layer-layer Coulomb interactions, so the result for $\alpha ^{2D,scr}$
converges as soon as $D$ is large enough to avoid overlap of layers.

\subsection{Response of a homogeneous stack of layers to a uniform field
directed parallel to the layers.}

Consider an infinite stack of identical symmetric non-overlapping layers
with an equal interlayer spacing $D\neq D_{0}$ that is not necessarily equal
to the equilibrium crystal spacing $D_{0}$. We aim to relate the screened
polarizability $\alpha _{xx}^{2D}\left( Q,\omega \right) $ of a single
isolated layer to the macroscopic dielectric properties of the stack. To
simulate linear response to a uniform field $E_{0\text{ }}$ in the $x$
direction, we choose 
\begin{equation}
V^{ext}\left( \vec{r}\right) =+E_{0}\left| e\right| x=E_{0}\left| e\right|
\lim_{Q\rightarrow 0}\frac{\sin \left( Qx\right) }{Q}%
=E_{0}\left| e\right| \lim_{Q\rightarrow 0}
\frac{e^{iQx}-e^{-iQx}}{2iQ}  \label{ExtPotlUnifFieldX}
\end{equation}%
Since there is no component of the external field in the z direction, this
is independent of z. Since for a symmetric layer such as graphene or BN the
long-wavelength density response is even in $\vec{Q}$, this
field induces a density of the same sinusoidal form, plus more rapidly
varying local-field contributions: 
\begin{equation}
\delta n\left( \vec{r}\right) =E_{0}\left| e\right| \frac{\sin \left(
Qx\right) }{Q}\lim_{Q\rightarrow 0}\delta n\left(
Q,z\right) +\left\{\vec{G} \neq \vec{0}\ terms\right\} 
\label{InducedDenswLocalFieldComps}
\end{equation}%
where $\delta n\left( Q,z\right) =\int \tilde{\chi}\left( \vec{Q}%
,z,z^{\prime }\right) dz^{\prime }$ is the response to a unit potential 
$\exp \left( i\vec{Q}.\vec{r}\right)$ and $\tilde{\chi}$ is here the
density response with all layer-layer interactions included. Because of the
sum rules for density response \cite{JFDDintvdW96}, $\tilde{\chi}$ can be
derived from a well-behaved polarizability $\tilde{\alpha}$ as per Eq (\ref%
{Chi3DFromAlphaScr3D}). The dipole moment $\vec{p}$ per unit volume, of the
stack, in the $x$ direction can be found by looking at the polarization,
specifically the dipole moment induced in a spatial region one
half-wavelength $\lambda /2=\pi /Q$ long in the $x$ direction, one
layer (say the $I=0$ layer) wide in the z direction, and of unit width in
the $y$ direction. The polarization induced by a strictly uniform field is a
tricky concept and in reality is related to surface charges. Its calculation
from the density perturbation in a finite region inside the crystal is well
known to be subject to ambiguities: the answer depends on choice of the
''edges''\ of the region. Equivalently, the net dipole moment depends
strongly on the edge positions via the high-wavenumber ($\vec{G}\neq \vec{0}$%
) components of the induced electron density. 

However, for an external field
with a small but finite wavenumber $Q=2\pi /\lambda $, considering a
region $-\lambda /4<x<\lambda /4$, 
we find that 
the field $E_{0}\cos \left(
Qx\right) $ from (\ref{ExtPotlUnifFieldX}) vanishes at the edges.
Correspondingly, explicit integration over this region shows that the dipole
moment from the $\vec{G}\neq \vec{0}$ components of the induced density (\ref%
{InducedDenswLocalFieldComps}) contributes a zero fraction of the total moment
 as $Q$ $\rightarrow 0$. Thus
the dipole moment $\vec{p}$ per unit volume can be calculated from just the $%
\vec{G}=\vec{0}$ density components, as we do here in deriving Eqs (\ref%
{AlphaLayerScrxxFromEpsMacroxx}) and (\ref{Aplha2DScrzzFromEpsMacro}) below.

Then the dipole moment per unit volume is 
\begin{eqnarray}
\vec{p} &=&\frac{1}{1\left( \lambda /2\right) D}\sum_{I}\int_{-\lambda
/4}^{+\lambda /4}dxx\int_{-D/2}^{+D/2}dz\left( -\left| e\right| \delta
n\left( \vec{r}\right) \right)   \notag \\
&=&\frac{-E_{0}\left| e\right| ^{2}}{1\left( \pi /Q\right) D}\left(
\int_{-\pi /(2Q)}^{\pi /(2Q)} x\frac{\sin \left( Qx\right) }{%
Q}dx\right) \int dzdz^{\prime }\tilde{\chi}\left( Q,z,z^{\prime
},\omega \right)   \label{pFromChiTilde}
\end{eqnarray}%
From (\ref{Chi3DFromAlphaScr3D}) the moment integral here is 
\begin{eqnarray}
\int dzdz^{\prime }\tilde{\chi}\left( Q,z,z^{\prime },\omega \right) 
&=&\int \left( Q^{2}\tilde{\alpha}_{^{||}}\left( Q,z,z^{\prime
},\omega \right) -\frac{\partial ^{2}}{\partial z\partial z^{\prime }}\tilde{%
\alpha}_{\perp }\left( Q,z,z^{\prime },\omega \right) \right)
dzdz^{\prime }  \notag \\
&=&Q^{2}\int \tilde{\alpha}_{^{||}}\left( Q,z,z^{\prime },\omega
\right) dzdz^{\prime }  \label{RequiredMoment}
\end{eqnarray}%
and this is not identical to the moments $m^{\pm }$ introduced in (\ref%
{DefMplusMinus}). \ However for small finite $Q$ the combination 
\begin{eqnarray*}
\frac{1}{2}\left( m^{+}+m^{-}\right)  &=&\int \cosh \left( Qz\right)
\delta n\left( Q,z,\omega \right) dz \\
&=&\int \left( 1+\frac{1}{2}Q^{2}z^{2}+...\right) \tilde{\chi}\left(
Q,z,z^{\prime },\omega \right) dzdz^{\prime } \\
&=&\int \left( 1+\frac{1}{2}Q^{2}z^{2}+...\right) \left( Q^{2}%
\tilde{\alpha}_{^{||}}\left( Q,z,z^{\prime },\omega \right) -\frac{%
\partial ^{2}}{\partial z\partial z^{\prime }}\tilde{\alpha}_{\perp }\left(
Q,z,z^{\prime },\omega \right) \right) dzdz^{\prime } \\
&=&Q^{2}\int \tilde{\alpha}_{^{||}}\left( Q,z,z^{\prime },\omega
\right) dzdz^{\prime }+0+\frac{1}{2}Q^{4}\int z^{2}\tilde{\alpha}%
_{^{||}}\left( Q,z,z^{\prime },\omega \right) +0
\end{eqnarray*}%
gives the required moment (\ref{RequiredMoment}) to lowest order in $Q$.%
Then (\ref{pFromChiTilde}) becomes to leading order in $Q$ 
\begin{equation}
\vec{p}=\frac{-E_{0}e^{2}\hat{x}}{1\left( \pi /Q\right) D}\left(
2/Q^{3}\right) \frac{1}{2}\left( m^{+}+m^{-}\right) \;\;\;.
\label{pFrom(mplus+mminus)}
\end{equation}

Here we used $\int_{-\pi /2}^{\pi /2}x\sin xdx=2$ so$~\int_{-\pi /q}^{\pi
/q}q^{-1}x\sin \left( qx\right) dx=2/q^{3}$:

For this external potential (\ref{ExtPotlUnifFieldX}) the external source
term (\ref{Gemeralsource}) in (\ref{RPALayerEqWithGeneralSource}) is, with $%
\chi _{I}$ the density response of an isolated layer and arguments $Q$, 
$\omega$ suppressed in places for brevity,
\begin{eqnarray*}
m^{\pm ext} &=&\int \exp \left( \pm Qz\right) \chi _{_{I}}\left(
Q,z,z^{\prime },iu\right) dzdz^{\prime } \\
&=&\int \exp \left( \pm Qz\right) \left( -e^{-2}\right) \left(
Q^{2}\alpha _{^{||}}\left( z,z^{\prime }\right) -\frac{\partial ^{2}}{%
\partial z\partial z^{\prime }}\alpha _{\perp }\left( z,z^{\prime }\right)
\right) dzdz^{\prime } \\
&=&-e^{-2}Q^{2}\int \exp \left( \pm Qz\right) \alpha _{||}\left(
z,z^{\prime }\right) dzdz^{\prime }\;\;\;\sin ce\;\;\int \frac{\partial ^{2}%
}{\partial z\partial z^{\prime }}\alpha _{\perp }\left( z,z^{\prime }\right)
dz^{\prime }\equiv 0 \\
&=&-e^{-2}Q^{2}\alpha _{||}^{2D,scr}(Q=0)+O\left(
Q^{4}\right) \;\;for\;insulating\;layers\;\;\;.
\end{eqnarray*}%
For this z-independent source potential the density perturbation $m_{I}^{\pm 
\text{ }}$and the response matrix $\mathbf{c}_{0}$ are the same for any
layer, $m_{I}^{\pm }=m^{\pm }$ and then (\ref{RPALayerEqWithGeneralSource})
becomes, to lowest order in $Q$%
\begin{eqnarray}
\vec{m} &=&-e^{-2}Q^{2}\alpha _{||}^{2D,scr}(Q=0)\left( 
\begin{array}{c}
1 \\ 
1%
\end{array}%
\right) +\mathbf{c}_{0}\sum_{J}\mathbf{v}\left( I-J\right) \vec{m}  \notag \\
\vec{m} &=&-e^{-2}Q^{2}\alpha _{||}^{2D,scr}(Q=0)\left( 
\begin{array}{c}
1 \\ 
1%
\end{array}%
\right) +\left( 
\begin{array}{cc}
Tv & Rv \\ 
Rv & Tv%
\end{array}%
\right) \vec{m}  \label{MIntermed}
\end{eqnarray}%
where%
\begin{equation*}
v=\sum_{K=1}^{\infty }e^{-KQD}=\sum_{K=-1}^{-\infty }e^{KQD}=%
\frac{1}{e^{QD}-1}\approx \frac{1}{QD}\;\,for\;\;Q\rightarrow
0\;.
\end{equation*}%
Defining $m=\frac{1}{2}\left( m^{+}+m^{-}\right) $ and adding the $+$ and $-$
components of (\ref{MIntermed}) we find%
\begin{eqnarray}
2m &=&-2e^{-2}Q^{2}\alpha _{||}^{2D,scr}(Q=0)+2\left( T+R\right) vm
\notag \\
m &=&\frac{-e^{-2}Q^{2}\alpha _{||}^{2D,scr}(Q=0)}{1-\left(
T+R\right) \frac{1}{QD}}=\frac{-e^{-2}Q^{2}\alpha
_{||}^{2D,scr}(Q=0)}{1+4\pi D^{-1}\alpha _{||}^{2D,scr}\left(
Q=0\right) }  \label{MForUnifEx}
\end{eqnarray}%
The dipole moment per unit volume from (\ref{pFrom(mplus+mminus)}) is then%
\begin{equation}
\vec{p}=\frac{2}{\pi }\frac{E_{0}\hat{x}}{D}\frac{\alpha
_{||}^{2D,scr}(Q=0)}{1+4\pi D^{-1}\alpha _{||}^{2D,scr}\left(
Q=0\right) }  \label{DipMomX}
\end{equation}%
We denote by ''3D\ macroscopic polarizability''\ $\alpha ^{3D,macro}$ the
macroscopic dipole moment per unit volume induced per unit \emph{external}
field, $\alpha ^{3D,macro}=p/\bar{E}$. Since the average external field in
our chosen volume is $\bar{E}=\int_{-\pi /\left( 2Q\right) }^{\pi
/\left( 2Q\right) }E_{0}\cos \left( Qx\right) dx/\int_{-\pi
/\left( 2Q\right) }^{\pi /\left( 2Q\right) }dx=2E_{0}/\pi $ we
have from (\ref{DipMomX}) 
\begin{equation}
\alpha ^{3D,macro}=\frac{p}{\bar{E}}=\frac{D^{-1}\alpha
_{||}^{2D,scr}(Q=0)}{1+4\pi D^{-1}\alpha _{||}^{2D,scr}\left(
Q=0\right) }  \label{AlphaMacro3DFromAlphaScr2D}
\end{equation}%
However the 3D macroscopic polarizability can also be expressed in terms of
the (3D) macroscopic dielectric function $\varepsilon ^{macro\text{ }}$of
the stack , via $\alpha ^{3D,macro}=\left( 4\pi \right) ^{-1}\left(
1-1/\varepsilon ^{macro}\right) $. By equating this to (\ref%
{AlphaMacro3DFromAlphaScr2D}) we find an expression for the polarizability
of an isolated layer, valid when the layers do not overlap.%
\begin{equation}
\alpha _{||}^{2D,scr}\left( Q=0,\omega \right) =\frac{D}{4\pi }\left(
\varepsilon _{||}^{macro}\left( D,\omega \right) -1\right) 
\label{AlphaLayerScrxxFromEpsMacroxx}
\end{equation}%
This equation can alternatively be obtained via the type of analysis leading
to  
%(\ref{cfromc0}) and 
(\ref{vMatrix}) below, with $q_{z}=0$ exactly and $Q$
small but finite. Eq (\ref{AlphaLayerScrxxFromEpsMacroxx}) is also implicit in
the work of Nazarov \cite{LayerResponseQuantitiesUsefulExptNazarovNJP2015}.

\subsection{Response of a homogeneous stack of layers to a uniform field
directed perpendicular to the layers.}

For the case of a layer with inversion symmetry in the $xy$ directions, a
uniform field in the $z$ direction (i.e. with $Q=0$) can only move the $%
\vec{G}=\vec{0}$ component of charge in each layer of the stack in the $z$
direction, creating a perturbation charge on each layer that sums to zero
and is zero outside the layer. By Gauss's theorem, such a charge
distribution produces zero electric field outside the layer. Therefore the
layers do not interact in this geometry provided that the spacing $D$ is
stretched far enough from equilibrium that the layer charge densities do not
overlap. Therefore the polarization per unit volume in the bulk is just
obtained by adding the polarizations of the layers, traeated as
independent, to the same external field. \ The polarization per unit volume
in bulk in terms of the macroscopic dielectric function is 
\begin{equation*}
a_{zz}^{3D,scr}=\frac{1}{4\pi }\left( 1-\left( \varepsilon _{rpa}^{-1}\left(
q\hat{z},q\rightarrow 0\right) \right) _{00}\right) =\frac{1}{4\pi }\left( 1-%
\frac{1}{\varepsilon _{zz}^{macro}}\right) 
\end{equation*}%
and for mutually non-interacting layers this must equal 
$\alpha_{zz}^{2D,scr}/D$. Thus 
\begin{equation}
\alpha _{zz}^{2D,scr}\left( Q=0,\omega \right) =\frac{D}{4\pi }\left( 1-%
\frac{1}{\varepsilon _{zz}^{macro}\left( D,\omega \right) }\right)
\label{Aplha2DScrzzFromEpsMacro}
\end{equation}

Contrast this with the parallel-field case (\ref{AlphaLayerScrxxFromEpsMacroxx}%
) where the macroscopic dielectric constant $\varepsilon $ appears in
different way because of interactions between the layers in the
parallel-field case. The result (\ref{Aplha2DScrzzFromEpsMacro}) can be
derived more formally along the same lines as for (\ref%
{AlphaLayerScrxxFromEpsMacroxx}), by using the type of analysis leading to 
%(\ref{cfromc0})and 
(\ref{vMatrix}), then setting $Q=0$ exactly while  $q_{z}$
is small but finite.  Eq (\ref{Aplha2DScrzzFromEpsMacro}) can also be obtained from 
equations in ref. \onlinecite{LayerResponseQuantitiesUsefulExptNazarovNJP2015} and is related to the
work of Andersen et al \cite{DielectricGenomevdWHetAndersenThygesenNanolett2015}.

As an example of the formulae (\ref{Aplha2DScrzzFromEpsMacro}) and (\ref%
{AlphaLayerScrxxFromEpsMacroxx}), Figs \ref{fig:alphaxxFromepsMacro}(a) and %
\ref{fig:alphazzFromEpsMacro}(a) show the parallel and perpendicular layer
polarizabilities $Im\{\alpha _{xx}^{2D,scr}\left( \omega +i0\right)
\}$ and $Im\left\{ \alpha _{zz}^{2D,scr}\left( \omega +i0\right)
\right\} $ of a BN monolayer, deduced from VASP calculations of the
macroscopic dielectric functions of solid BN that is only slightly
stretched. The agreement for layer spacings $D=0.8~$nm and $D=1.2$ nm shows
that, with the analytic interlayer-descreening implied in (\ref%
{AlphaLayerScrxxFromEpsMacroxx}), large layer spacings are not needed to
extract well-converged properties of an isolated layer. On the other hand,
if the ''noninteracting''\ formula (\ref{Aplha2DScrzzFromEpsMacro}) is
inappropriately applied to the parallel polarizability, the the results are
strongly $D$-dependent (see Fig \ref{fig:alphaxxFromepsMacro}(b) )
and will only converge to the correct layer polarizability when $%
D\rightarrow \infty$.

As a further test we note that the BN binding energies and asymptotics that we obtain
below are stable to $1\%$ when we change from $D=0.8$ \thinspace to $%
D=1.2\,nm$ for the reference solid, in calculations of $\varepsilon ^{macro}$.

\section{Inter-layer correlation energy in ACFD-RPA}

For a collection of parallel non-overlapping layers we can separate the
Coulomb interaction into intra-layer and inter-layer parts:

\begin{equation}
V=V_{intra}+\lambda V_{inter}  \label{VintraPlusVInter}
\end{equation}%
with $0 \leq \lambda \leq 1$:$\,$\ $\lambda =1$ defines the physical system, while
for $\lambda =0$ we have a collection of isolated systems. We start from the
full interaction $V_{intra}$ inside each layer, and turn on the interlayer
interactions by varying $\lambda $ from $0$ to $1$. By evaluating the work
to increase from $\lambda $ to $\lambda +d\lambda $ we find the interlayer
correlation energy in the Adiabatic Connection Fluctuation 
(ACFD) approach \cite{CalcDispersionEnergiesDobsonJPCM2012}:%
\[
E^{interlayer}=\frac{\hbar }{2\pi }\int_{0}^{\infty }du
\int_{r,r^{\prime}\textrm{ in different layers}}
d\vec{r}d\vec{r}~^{\prime }\int_{0}^{1}d\lambda
\chi _{\lambda }(\vec{r},\vec{r}~^{\prime },iu)\frac{e^{2}}{\left| \vec{r}-%
\vec{r}~^{\prime }\right| } 
\]%
Here $\chi _{\lambda }$ is the density-density response with the interaction
(\ref{VintraPlusVInter}) present. Fourier transformation of the $x$ and $y$
coordinates, with neglect of the exponentially decaying $\vec{G}\neq 
\vec{0}$ terms, gives the energy per unit area as

\begin{eqnarray}
\frac{E^{interlayer}}{A} &=&\frac{\hbar }{\left( 2\pi \right) ^{3}}%
\int_{0}^{\infty }du\int d^{2}Q\sum_{I\neq J}e^{-Q\left|
Z_{I}-Z_{J}\right| }\frac{2\pi e^{2}}{Q}\int_{0}^{1}d\lambda  \nonumber
\\
&&\times \int dZdZ^{\prime }e^{\theta (Z_{J}-Z_{I})Q\left( Z-Z^{\prime
}\right) }\chi _{\lambda }\left( Q,Z_{I}+Z,Z_{J}+Z^{\prime },iu\right)
\label{EInterlayerFromIJ}
\end{eqnarray}%
where $A$ is the area of one layer and $\theta $ is the Heaviside function.
Here for $J>I$ the $\int dZdZ^{\prime }$ can be recognized as the moment $%
m_{I}^{+}$ (Eq (\ref{DefMplusMinus})) induced on layer $I$ by a perturbing
external potential component $V_{J}^{-}$ of unit strength (see Eq (\ref%
{VextI})) applied to layer $J$ only. This moment would be zero if the layers
did not interact. For interacting layers, $\vec{m}$ is obtained from
solution of (\ref{CoupledLayersRPAEqIJ}) with $\vec{V}_{I}^{ext}=(\delta
_{IJ}$ , $0).$ Similarly for $J<I~$the integral $\int dzdz^{\prime }$ \
gives $m_{I}^{-}$ for a unit source $V^{-\text{ }}$in layer $J$,  so that 
$\vec{V}_{I}^{ext}=(0,\delta _{IJ})$. Thus from (\ref{MIFromCIJ}) 
\[
\int dzdz^{\prime }e^{\theta (J-I)Q\left( z-z^{\prime }\right) }\chi
\left( Q,Z_{I}+z,Z_{J}+z^{\prime },iu\right) =\frac{Q}{2\pi e^{2}}%
\left( c_{IJ}^{+-}\theta \left( J-I\right) +c_{IJ}^{-+}\theta \left(
I-J\right) \right) 
\]%
and then (\ref{EInterlayerFromIJ}) with (\ref{DefuIJ} \ ) becomes 
\begin{eqnarray}
\frac{E^{interlayer}}{A} &=&\frac{\hbar }{\left( 2\pi \right) ^{3}}%
\int_{0}^{\infty }du\int d^{2}Q\sum_{IJ}\int_{0}^{1}d\lambda \left(
c_{\lambda IJ}^{+-}v_{JI}^{-+}+c_{\lambda IJ}^{+-}v_{JI}^{-+}\right) 
\nonumber \\
&=&\frac{\hbar }{\left( 2\pi \right) ^{3}}\int_{0}^{\infty }du\int
d^{2}Q\sum_{IJ}\int_{0}^{1}d\lambda Tr_{\pm }\left( \mathbf{c}_{\lambda
IJ}\mathbf{v}_{JI}\right)  \label{EInterlayerasTracePm}
\end{eqnarray}%
%&=&\frac{\hbar }{\left( 2\pi \right) ^{3}}\int_{0}^{\infty }du\int
%d^{2}Q\sum_{IJ}\int_{0}^{1}d\lambda \mathbf{Tr}_{\pm }\left( \mathbf{c}_{\lambda
%IJ}\mathbf{v}_{JI}\right)  \label{EInterlayerasTracePm}

where $\mathbf{v}_{IJ}$ is defined in (\ref{DefuIJ}). Here the trace $%
{\mathbf Tr}_{\pm }$ is over a $2\times 2$ matrix but the layer sum in (\ref%
{EInterlayerasTracePm}) can also be recognized as a trace of the product $%
\mathbf{cv}$ over the layer index $I$. Expanding (\ref{InteractingCDimless})
to zero order in the interlayer interaction $u$ and noting $\mathbf{c}%
_{0}=\delta _{IJ}\mathbf{c}_{0II}$ we find that the lowest-order
contribution to (\ref{EInterlayerasTracePm}) is zero, being proportonal to $%
{\mathbf Tr}_{\pm }\sum_{I}\mathbf{c}_{0II}v_{II}=0$ since $\mathbf{v}_{II}=\mathbf{0}$
from (\ref{DefuIJ}). The expansion of (\ref{EInterlayerasTracePm}) therefore
starts at second order in $\mathbf{v}$. The second order term is
proportional to a sum of interactions between pairs $I,J,$ of layers, $%
\sum_{IJ}\int \lambda d\lambda Tr\left( \mathbf{c}_{0II}\mathbf{v}_{IJ}%
\mathbf{c}_{0JJ}\mathbf{v}_{JI}\right) .$ Similar arguments show that
expansion terms in (\ref{EInterlayerasTracePm}) containing solely $v^{+}$
factors, or solely $v^{-}$ factors, are also zero - these terms couple only
layers to the right(left) of each other and so cannot contribute to the
same-layer components that make up the trace.

\section{Correlation energy between two parallel layers distant $D$}

Here we set the layer spacing $Z_{2}-Z_{1}=D.$ and (\ref{DefuIJ}) becomes 
\[
\mathbf{v}_{\mathbf{12}}=\left( 
\begin{array}{cc}
0 & \exp \left( -QD\right) \\ 
0 & 0%
\end{array}%
\right) =\mathbf{v}_{\mathbf{21}}^{T} 
\]%
Here it is appropriate to drive with $V_{1}^{-}=V_{2}^{+}=0$, so that only 
two of the four scalar RPA equations (\ref{CoupledLayersRPAEqIJ}) are coupled:
\begin{equation}
\vec{m}_{I}=\frac{Q}{2\pi e^{2}}\mathbf{c}_{0I}\vec{V}_{I}^{tot}=%
\mathbf{c}_{0I}\left( \frac{Q}{2\pi e^{2}}\vec{V}_{I}^{ext}+\lambda
\sum_{J}\mathbf{v}_{IJ}\vec{m}_{J}\right)
\end{equation}%
\begin{eqnarray*}
m_{1}^{+} &=&c_{01}^{++}\left( \frac{Q}{2\pi e^{2}}V_{1}^{+}+\lambda
e^{-QD}m_{2}^{-}\right) \\
m_{2}^{-} &=&c_{02}^{--}\left( \frac{Q}{2\pi e^{2}}V_{2}^{-}+\lambda
e^{-QD}m_{1}^{+}\right)
\end{eqnarray*}%
with solution%
\[
m_{1}^{+}=\frac{Q}{2\pi e^{2}}\frac{c_{01}^{++}V_{1}^{+}+\lambda
e^{-QD}c_{01}^{++}c_{02}^{--}V_{2}^{-}}{\left( 1-\lambda
^{2}e^{-2QD}c_{01}^{++}c_{02}^{--}\right) }=\frac{Q}{2\pi e^{2}}\left(
c_{11}^{++}V_{1}^{+}+c_{12}^{+-}V_{2}^{-}\right) 
\]%
so that the interacting response function between the layers is 
\[
c_{12}^{\pm }=\frac{\lambda e^{-QD}c_{01}^{++}c_{02}^{--}}{%
1-\lambda ^{2}e^{-2QD}c_{01}^{++}c_{02}^{--}}=c_{21}^{\mp } 
\]%
Then the first line of (\ref{EInterlayerasTracePm}) becomes%
\[
\frac{E^{interlayer}}{A}=\frac{\hbar }{\left( 2\pi \right) ^{3}}%
\int_{0}^{\infty }du\int d^{2}Q\int_{0}^{1}d\lambda \left( c_{\lambda
IJ}^{+-}v_{JI}^{-+}+c_{\lambda IJ}^{-+}v_{JI}^{+-}\right) 
\]

\begin{eqnarray}
\frac{E^{interlayer}}{A} &=&\frac{\hbar }{\left( 2\pi \right) ^{3}}%
\int_{0}^{\infty }du\int d^{2}Q\int_{0}^{1}d\lambda \frac{2\lambda
c_{02}^{--}c_{01}^{++}e^{-2QD}}{1-\lambda
^{2}e^{-2QD}c_{01}^{++}c_{02}^{--}}  \nonumber \\
&=&\frac{\hbar }{\left( 2\pi \right) ^{3}}\int_{0}^{\infty }du\int
d^{2}Q\ln \left( 1-e^{-2QD}c_{01}^{++}c_{02}^{--}\right)
\label{vdW2LayersFromcs} \\
&=&\frac{\hbar }{\left( 2\pi \right) ^{3}}\int_{0}^{\infty }du\int
d^{2}Q\ln \left( 1-e^{-2QD}R_{1}^{right}R_{2}^{left}\right)
\label{Evdw2LayersFromRs}
\end{eqnarray}%
where (\ref{DefC0asymm}) was used.

We note in passing that (\ref{Evdw2LayersFromRs}) is valid even for infinitely thick parallel slabs separated by distance $D$, where we choose 
the regions $\mathcal{R}_{1}$  and $\mathcal{R}_{2}$ to be $z<0$ and $z>0$ respectively, with the reference points $Z_1$ and $z_2$ at 
$\pm D/2$ (i.e. at the edges of the two slabs). See Fig \ref{fig:LayerDiagram}. By evaluating the slab reflection coefficients $R_1$ and $R_2$ in the small-q limit (see for example \cite{CalcDispersionEnergiesDobsonJPCM2012}) one can then re-derive the electromagnetically non-retarded limit of the original Lifshitz result for non-overlapping slabs. Here however we consider nano-thin layers.

 For isotropic inversion-symmetric nano-layers ,
use of (\ref{RFromAlphaScr2D}) gives%
\begin{equation}
\frac{E^{interlayer}}{A}=\frac{\hbar }{\left( 2\pi \right) ^{3}}%
\int_{0}^{\infty }du\int d^{2}Q\ln \left( 1-e^{-2QD}\left( -2\pi
Q\right) ^{2}\alpha _{1tot} \alpha _{2tot}\right)
\label{EvdW2LayersFromAlphas}
\end{equation}%
where for each layer species $\alpha_{tot}=\alpha _{xx}^{2D,scr}(\vec{Q},iu)+\alpha
_{zz}^{2D,scr}\left( \vec{Q},iu\right) $.

For large $D$ (at least for insulating layers) we can expand the logarithm in 
(\ref{EvdW2LayersFromAlphas}) to second order, giving%
\begin{equation}
\frac{E^{interlayer}}{A}=-\frac{\hbar }{\left( 2\pi \right) ^{3}}%
\int_{0}^{\infty }du\int \left( 2\pi Q\right)
^{2}d^{2}Qe^{-2QD}\alpha _{1tot}\alpha _{2tot}
\label{EcTwoLaersLogExpanded}
\end{equation}

For the case of a BN bilayer our numerical work (see below) shows that the
maximum value of $\alpha_{tot}\left( iu\right) $ is about $0.1\,nm$, and
since the maximum value of $Q{}^{2}\exp \left( -2QD\right) $ is $%
e^{-2}D^{-2}$ , the nontrivial term under the logarithm in (\ref%
{Evdw2LayersFromRs}) is \ always less than about $\left( 2\pi \right)
^{2}10^{-2}e^{-2}\left( D/nm\right) ^{-2}=5.3\times 10^{-2}\left(
D/nm\right) ^{-2}$.  \ Thus for any $D>2D_{0}=0.7nm\;$\ this term is less
than $\approx 5.3\times 10^{-2}\left( .7\right) ^{-2}=\allowbreak 0.11\,$and
so $\exp $ansion of the logarithm is justified \emph{a priori} for BN layers except
very close to contact.\ \ In practice we found the expansion is suitable
for correlation energy calculations even near to contact.

\subsection{Asymptotic vdW interaction of two insulating layers}

For two insulating layers $\alpha _{1tot}$ and $\alpha _{2tot}$ are
both constant as $Q\rightarrow 0.$ The asymptotic ($D\rightarrow \infty$,
implying $Q\rightarrow 0$) \ vdW interaction (\ref%
{EvdW2LayersFromAlphas}) is then evaluated by taking each $\alpha_{itot}$ as
independent of $Q$, and Taylor-expanding the logarithm in (\ref{EvdW2LayersFromAlphas}%
). We this obtain \ the asymptotic $(D\rightarrow \infty )$ form

\[
\frac{E^{interlayer}}{A}\left( 2\;insulators)\right) =-\frac{3\hbar }{8D^{4}}%
\int_{0}^{\infty }\alpha _{1tot}(Q=0,iu)\alpha
_{2ot}(Q=0,iu)du\ \ \ . 
\]

This is total correlation energy per unit area of one side of the double
layer. The correlation energy per unit area per layer is then half of this:%
\begin{equation}
\frac{E^{interlayer}\left( 2\;ins\right) }{NA}=-\frac{3\hbar }{16D^{4}}%
\int_{0}^{\infty }\alpha _{1tot}(Q=0,iu)\alpha
_{2tot}(Q=0,iu)du  \label{EPerAreaPerLayerAsyInsulator2Layer}
\end{equation}%
where $N=2$ \ and $\alpha_{tot}$ is defined just below Eq (\ref%
{EvdW2LayersFromAlphas}).

\section{Inter-layer RPA interactions in an infinite uniform stack of
identical symmetric layers}

Here all layers are identically constructed. With reference to 
Fig \ref{fig:LayerDiagram},
 the $Ith$ layer is centred on the reference point 
\[
Z_{I}=ID\;\;,I=0,\pm 1,\pm 2,... 
\]%
and the region $\mathcal{R}_{I}$ containing layer $I$\thinspace 
%(referred to above)
 is 
\[
\mathcal{R}_{I}=\left\{ z:\;\;\left( I-1/2\right)D <z<\left( I+1/2\right)D
\right\} . 
\]%
Here $D$ is the common spacing between adjacent layers, which does not have
to be the equilibrium spacing (e.g. the layered crystal could be stretched).

For this equally-spaced case all two-index layer functions such as $\mathbf{c%
}_{IJ}$ are functions of $I-J$ only. The inverse over the layer indices $I,J$
in (\ref{InteractingCDimless}) then becomes analytic via a reverse Fourier
Series transformation to a wavenumber variable $q_{z}$, instead of layer
index $I$. For an arbitrary function $f_{I}$ we define

\begin{equation}
f\left( q_{z}\right) =\sum_{J}e^{-iq_{z}DJ}f_{J},\ \ \ f_{I}=\frac{D}{2\pi }%
\int_{-\pi /D}^{\pi /D}dq_{z}f\left( q_{z}\right) e^{iq_{z}DI}
\label{FourierSeriesTransf}
\end{equation}%
and a convolution \ $X_{\Delta I}\equiv X_{I-J}=\sum_{K}F_{I-K}G_{K-J}$\ has
a transform $X\left( q_{z}\right) =F\left( q_{z}\right) G\left( q_{z}\right) 
$.

Then (\ref{InteractingCDimless}) becomes%
\[
\label{cfromc0}
%\label{InteractingcMatrixStack}
\mathbf{c}\left( Q,q_{z}\right) =\left( I-\mathbf{c}_{0}\left(
Q,q_{z}\right) \mathbf{v}\left( Q,q_{z}\right) \right) ^{-1}%
\mathbf{c}_{0}\left( Q,q_{z}\right) \
\]%
where the products and inverse are solely over $2\times 2$ matrices. Here
from (\ref{DefC0asymm}) and (\ref{FourierSeriesTransf}) 
\begin{equation}
\mathbf{c}_{0}\left( Q,q_{z}\right) =\left( 
\begin{array}{cc}
R\left( Q,\omega \right) & T\left( Q,\omega \right) \\ 
T\left( Q,\omega \right) & R\left( Q,\omega \right)%
\end{array}%
\right)  \label{c0Matrix}
\end{equation}%
is independent of $q_{z}$. Similarly from (\ref{DefuIJ}) and (\ref%
{FourierSeriesTransf}), 
\begin{equation}
\mathbf{v}\left( Q,q_{z}\right) =\left( 
\begin{array}{cc}
0 & v^{+}\left( Q,q_{z}\right) \\ 
v^{+}\left( Q,q_{z}\right) ^{\ast } & 0%
\end{array}%
\right)  \label{vMatrix}
\end{equation}%
where 
\begin{equation}
v^{+}\left( Q,q_{z}\right) =\sum_{I>0}\exp \left( \left(
-Q+iq_{Z}\right) DI\right) =\frac{e^{iq_{z}D}e^{-QD}}{%
1-e^{iq_{z}D}e^{-QD}}=\left( v^{-}\right) ^{\ast }  \label{vPlusComplex}
\end{equation}

Then the interlayer correlation energy per unit area, per layer of the
infinite stack ($N\rightarrow \infty $) from (\ref{EInterlayerasTracePm}) is
a convolution on $J$: 
\begin{eqnarray}
\frac{E_{c}}{NA} &=& \frac{\hbar }{\left( 2\pi \right) ^{3}}\int_{0}^{\infty
}du\int d^{2}Q\frac{1}{N}\sum_{I}\int_{0}^{1}d\lambda \mathbf{Tr}_{\pm }\left( 
\mathbf{c}_{\lambda }\mathbf{v}\right) _{II}  \nonumber \\
&=&\frac{\hbar }{\left( 2\pi \right) ^{3}}\int_{0}^{\infty }du\int
d^{2}Q\int_{0}^{1}d\lambda \frac{D}{2\pi }\int dq_{z}\mathbf{Tr}_{\pm }\left[
\left( 1-\lambda \mathbf{c}_{0}\mathbf{v}\right) ^{-1}\mathbf{c}_{0}\mathbf{v%
}\right] _{Q,q_{z}.iu}  \nonumber \\
&=&\frac{\hbar }{\left( 2\pi \right) ^{3}}{\mathbf Tr}_{\vec{q}}\mathbf{Tr} _{\pm
}\int_{0}^{\infty }\hbar du\ln _{\pm }\left( \mathbf{I-c}_{0}\mathbf{v}%
\right)  \nonumber \\
&=&\int_{0}^{\infty }\hbar du {\mathbf Tr}_{\vec{q}}\ln \det_{\pm }\left( \mathbf{I-c}%
_{0}\mathbf{v}\right) \\
&=&\left( 2\pi \right) ^{-3}\mathbf{Tr}_{\vec{q}}\int_{0}^{\infty }\hbar du\ln \left(
1-2TRev^{+}+\left( T^{2}-R^{2}\right) \left| v^{+}\right| ^{2}\right)
\label{EperLayerAreaAs3DTrace}
\end{eqnarray}%
Here ${\mathbf Tr}_{\vec{q}}\{\}\equiv \int_{0}^{\infty }d^{2}Q\frac{D}{2\pi }%
\int_{-\pi /D}^{\pi /D}\{\}dq_{z}$, while ${\mathbf Tr}_{\pm }$ and $\ln _{\pm }$
are operations on the $2\times 2$ matrices. In (\ref{EperLayerAreaAs3DTrace}),
the last step used (\ref{c0Matrix}) and (\ref{vMatrix})). From (\ref{vMatrix}) 
we have

\begin{eqnarray}
Rev^{+} &=&\left( \cos \left( q_{z}D\right) -\exp \left(
-QD\right) \right) /b,\;Trv^{+}=\left| \;v^{+}\right| ^{2}=\exp \left(
-QD\right) /(2b).\;  \label{ExplicitvPlusQparQz} \\
\;b &=&\cosh \left( QD\right) -\cos \left( q_{z}D\right)  \nonumber
\end{eqnarray}
Then the inter-layer correlation energy of the infinite stack is 
\begin{equation}
\frac{E_{c}}{NA}=\left( 2\pi \right) ^{-3}{\mathbf Tr}_{\vec{q}}\int_{0}^{\infty
}\hbar du\ln \left( 1+\frac{\left( \exp \left( -QD\right) -\cos \left(
q_{z}D\right) \right) T+\exp \left( -QD\right) \left(
T^{2}-R^{2}\right) }{\cosh \left( QD\right) -\cos \left( q_{z}D\right) }%
\right)  \label{ExplicitEcPerVolUnifStack}
\end{equation}%
where $T\left( \vec{Q},iu\right) $ and $R\left( \vec{Q},iu\right) $
are the reflection and transmission coefficients. 

 This can be written in dimensionless form using  
 $\kappa_{\parallel}=QD$, $\kappa_{z}=q_{z}D$
 and applying 
(\ref{TFromAlphaScr2D}) and (\ref{RFromAlphaScr2D}): 
\begin{eqnarray}
\frac{E_{c}}{NA} &=&\left( 2\pi \right) ^{-3}D^{-2}\int_{\vec{\kappa}%
_{||}}d^{2}\kappa_{||}\frac{1}{2\pi }\int_{-\pi }^{\pi }d\kappa_{z}\int_{0}^{\infty
}\hbar du  \nonumber \\
&&\times \ln 
\left( 
1+
\frac{-2\pi \kappa_{||}D^{-1}
\left( \exp \left(-\kappa_{||}\right) -\cos \left( \kappa_{z}\right) \right)
 \alpha _{diff}
+\left( 2\pi\kappa_{||}\right) ^{2}D^{-2}\exp \left( -\kappa_{||}\right) 
\left( \alpha_{sum}^{2}-\alpha _{diff}^{2}\right) }
{\cosh \left( \kappa_{||}\right) -\cos \left( \kappa_{z}\right) }
\right)
  \label{EInfStackDimensionless}
\end{eqnarray}%
Here $\alpha _{diff}=\alpha _{xx}^{2D,scr,-}\left( \kappa_{||}D^{-1},iu\right)
 - \alpha _{zz}^{2D,scr,-}\left( \kappa_{||}D^{-1},iu\right) ,\;
 \alpha_{sum}=\;\alpha _{xx}^{2D,scr,+}\left( \kappa_{||}D^{-1},iu\right) +\alpha
{zz}^{2D,scr,+}\left(\kappa _{||}D^{-1},iu\right)$.

\subsection{$D\rightarrow \infty $ limit for an infinite stack of insulating
layers}

In the large-$D$ limit (\ref{EInfStackDimensionless}) calls for the
single-layer polarizabilities $\alpha _{sum}$, $\alpha _{diff}$ at very
small parallel wavenumber $Q=\kappa_{||}D^{-1}$ where the cosh function on the
denominator ensures that large $\kappa_{||}$ values are not relevant. \ For
insulators these polarizabilities have finite values as $Q\rightarrow 0$%
, and then (\ref{EInfStackDimensionless}) shows explicitly that the argument
of the logarithm is close to $1$, justifying a Taylor expansion of the log.
In fact the expansion of the logarithm is most transparently done from the
original formula (\ref{EperLayerAreaAs3DTrace}), giving

\[
\frac{E^{(2)}}{NA}=\frac{1}{\left( 2\pi \right) ^{3}}{\mathbf Tr}_{\vec{q}%
}\int_{0}^{\infty }\hbar du\left[ -T(v^{+}+v^{+\ast })+\left(
T^{2}-R^{2}\right) v^{+}v^{+\ast }-\frac{1}{2}T^{2}\left(
v^{+2}+2v^{+}v^{+\ast }+v^{+\ast 2}\right) \right] 
\]

The terms containing only $v^{+}$, or only $v^{+\ast }$, and those
containing $\ $only $v^{+2}$ or $v^{+\ast 2}$are zero as explained following
Eq (\ref{EInterlayerasTracePm}). The terms containing $T$ cancel and the
remaining lowest term involves only the reflection coefficient and is of
second order in $\mathbf{v}$: 
\begin{eqnarray*}
\frac{E^{(2)}}{NA} &=&\frac{-1}{\left( 2\pi \right) ^{3}}{\mathbf Tr}_{\vec{q}}\left(
v^{+}v^{+\ast }\int_{0}^{\infty }\hbar duR^{2}\right) \\
&=&\frac{-1}{\left( 2\pi \right) ^{3}}\int \left( \frac{D}{2\pi }\int
v^{+}v^{+\ast }dq_{z}\right) \left( 2\pi Q\right)
^{2}d^{2}Q\int_{0}^{\infty }\alpha _{tot}^{2}\left( iu\right) \hbar du
\end{eqnarray*}%
where $\alpha _{tot}$ is defined below Eq (\ref{EvdW2LayersFromAlphas}).
%\alpha _{xx}^{2D,scr}(Q,iu)+a_{zz}^{2D,scr}\left(Q,iu\right) $.
 Now the $q_{z}$ integral is a Fourier
coefficient of a product and therefore gives the on-layer value of a layer
convolution,  
\[
\frac{D}{2\pi }\int v^{+}v^{+\ast
}dq_{z}=\sum_{J}v_{IJ}^{+}v_{JI}^{-}=\sum_{K=1}^{\infty }e^{-2KDQ} 
\]%
Then 
\begin{eqnarray}
\frac{E^{\left( 2\right) \inf \,stack}\left( D\right) }{NA} &=&\frac{-1}{%
\left( 2\pi \right) ^{3}}\sum_{K=1}^{\infty }\int
d^{2}Qe^{-2KDQ}\left( 2\pi Q\right) ^{2}\int_{0}^{\infty
}\alpha _{tot}^{2}\left( Q,iu\right) \hbar du  \nonumber \\
&=&2\sum_{K=1}^{\infty }\frac{E^{\left( 2\right) bilayer}\left( KD\right) }{%
2A}  \label{Ec)2)InfStackAsLayerPairSum}
\end{eqnarray}%
where (\ref{EcTwoLaersLogExpanded}) was used. This form, representing a
pairwise sum over layers, results from our expansion of the logarithm in (%
\ref{EperLayerAreaAs3DTrace}). The fully asymptotic D$\rightarrow \infty $
result can be obtained for insulating layers by using the $Q=0$ value
of the layer response. The $Q$ integral can then be done analytically
for each value of $K$, giving the asymptotic (large-$D$) layer binding
energy for a uniform stack of identical insulating layers: $\ \ \ \ $%
\begin{equation}
\frac{E\,}{NA}=-\frac{3}{8}D^{-4}\left( \sum_{K=1}^{\infty }K^{-4}\right)
\int_{0}^{\infty }\hbar du \alpha^2_{tot}(Q=0,iu)
\label{AsymptoticEcInfiniteInsulatorStack}
\end{equation}%
where $\alpha_{tot}$ was defined following Eq (\ref{EvdW2LayersFromAlphas}).
Note $\sum_{K=1}^{\infty }K^{-4}\approx \allowbreak 1.\,\allowbreak 082\,$.
The leading term in the infinite stack result (\ref%
{AsymptoticEcInfiniteInsulatorStack}) is twice the two-layer result (\ref%
{EPerAreaPerLayerAsyInsulator2Layer}), which makes sense because in the
infinite stack each layer has two nearest-neighbour layers.

We can convert (\ref{AsymptoticEcInfiniteInsulatorStack}) to an energy per
atom by multiplying by the area $A_{atom}$ per atom in a layer.

\section{Sub-asymptotic theory: response and energetics for small non-zero $Q$}

The asymptotic results (\ref{EPerAreaPerLayerAsyInsulator2Layer}),  
(\ref{AsymptoticEcInfiniteInsulatorStack}) %
above were based (at least for insulating layers) 
on the layer polarizabilities at $Q=0$. Now we look
at the response of an isolated layer for small but finite $Q$, which
will allow us to consider the interactions of layers with objects at
somewhat closer distances, but still not in the region of strong overlap of
electronic clouds. (i.e. it gives sub-asymptotic results).\ To do this we
need to consider the Coulomb interaction within a layer, which we do at the
RPA\ level. For these intra-layer interactions we cannot \emph{a priori} ignore the 
$\vec{G}\neq \vec{0}$ (local-field) Coulomb interaction effects.

The bare (Kohn-Sham) and interacting density responses $\chi ^{_{\left( 0\right) }}$, $%
\chi $ of a single periodic layer are related by the RPA equation 
\begin{equation}
\chi _{\vec{G}\vec{G}\,^{\prime }}(Q,\omega ,z,z^{\prime })=\chi _{\vec{%
G}\vec{G}\,^{\prime }}^{(0)}(Q,\omega ,z,z^{\prime })+\sum_{\vec{G}\
^{\prime \prime }}\int dz^{\prime \prime } dz^{\prime \prime \prime}\chi _{\vec{G}\vec{G}\,^{\prime
\prime }}^{(0)}(Q,z,z^{\prime \prime },\omega )V_{\vec{G}\,^{\prime
\prime }+\vec{Q}}(z^{\prime \prime },z^{\prime\prime \prime })\chi _{\vec{G}%
\,^{\prime \prime }\vec{G}\,^{\prime }}(Q,\omega ,z^{\prime \prime \prime},z^{\prime })
\label{FullIntraayerScreeningEq}
\end{equation}%
or more compactly $\chi =\chi ^{0}+\chi ^{0}\ast V\ast \chi $ where the 2D
Coulomb potential is $V_{\vec{k}}\left( z,z^{\prime }\right) =2\pi
e^{2}\left| \vec{k}\right| ^{-1}\exp \left( -\left| \vec{k}\right| \left|
z-z^{\prime }\right| \right) $. We break this Coulomb interaction into a
part $W^{ rap }$ that varies rapidly in space even when 
$Q\rightarrow 0$, and a part $W^{\left( slow\right) \text{ }}$that does
not: $V=W^{rap}+W^{slow}$.

The rapid component contains the $\vec{G}\neq \vec{0}$ local-field
terms for $Q=0$, and another part with a discontinuous $z$ derivative: 
%\[
\begin{equation}
\frac{W_{G}^{rap}\left( z,z^{\prime }\right) }{2\pi e^{2}}=\left(
1-\delta _{\vec{G},\vec{0}}\right) \frac{\exp \left( -\left| \vec{G}\right|
\left| z-z^{\prime }\right| \right) }{\left| \vec{G}\right| }
-\delta_{\vec{G},\vec{0}}\left| z-z^{\prime }\right| 
\label{Wrap}
%\]
\end{equation}

The remaining slowly varying part is analytic at small $Q$:%
\begin{eqnarray}
\frac{W_{\vec{G}}^{slow}\left( z,z^{\prime }\right) }{2\pi e^{2}} 
&=&\frac{V-W^{rap}}{2\pi e^{2}}  \nonumber \\
&=&\left( 1-\delta _{\vec{G},\vec{0}}\right) \left( \frac{\exp \left( -\left| \vec{G%
}+\vec{Q}\right| \left| z-z^{\prime }\right| \right) }{\left| \vec{%
G}+\vec{Q}\right| }-\frac{\exp \left( -\left| \vec{G}\right|
\left| z-z^{\prime }\right| \right) }{\left| \vec{G}\right| }\right) \nonumber \\
&&+\delta _{\vec{G}_{00},\vec{0}}\left( \frac{\exp \left( -Q\left|
z-z^{\prime }\right| \right) }{Q}+\left| z-z^{\prime }\right| \right) \nonumber \\
&=&\delta _{\vec{G}_{00},\vec{0}}\left( Q^{-1}+\frac{1}{2}Q\left(
z-z^{\prime }\right) ^{2}\right) +O\left( Q^{2}\right)
\label{WslowExpanded}
\end{eqnarray}

We define a ''rapid'' single-layer response $\chi _{\vec{G}\vec{G}^{\prime
}}^{rap}$ $\left( Q,z,z^{\prime },\omega \right) $ that satisfies (\ref%
{FullIntraayerScreeningEq}) with $W^{rap}$ in place of the full Coulomb
interaction $V$: 
\begin{equation}
\chi ^{rap}=\chi ^{0}+\chi ^{0}\ast W^{rap}\ast \chi ^{rap}
\label{ChiRapSreeningEq}
\end{equation}%
and it then follows that the full RPA response $\chi $ from (\ref%
{FullIntraayerScreeningEq}) exactly satisfies%
\begin{equation}
\chi =\chi ^{rap}+\chi ^{rap}\ast W^{slow}\ast \chi
\label{ScreeningeqChiRPAFRomChiRap}
\end{equation}

To study the subasymptotic interaction of the layer with other layers, we
only need the component $\vec{G}=\vec{G}^{\prime }=\vec{0}$ of $\chi$.
 We consider this component of (\ref{ScreeningeqChiRPAFRomChiRap}).
Expressing $\chi ^{rap}$ and $\chi$ in terms of polarizabilities $\alpha^{rap}$, 
$\alpha^{scr}$ as in (\ref%
{Chi3DFromAlphaScr3D}) we write the $\vec{G}=\vec{G}^{\prime }=\vec{0}$ 
component of the convolution on the RHS of (\ref%
{ScreeningeqChiRPAFRomChiRap}) in the the form%
\begin{eqnarray}
\left( \chi ^{rap}\ast W^{slow}\ast \chi \right) _{\vec{0}\vec{0}}\left(
Q,z,z^{\prime }\right) &=&\int dz^{\prime \prime }dz^{\prime \prime
\prime }\left[ Q^{2}\alpha _{||}^{rap}\left( Q,z,z^{\prime \prime
}\right) -\partial _{z}\partial _{z^{\prime \prime }}\alpha _{\perp
}^{rap}\left( Q,z,z^{\prime \prime }\right) \right]  \nonumber  \\
&&\times \left( Q^{-1}+Q\left( z^{\prime \prime }-z^{\prime \prime
\prime }\right) ^{2}\right)
\left[ Q^{2}\alpha _{||}^{scr}\left( Q,z^{\prime \prime \prime},z^{\prime}\right)
 -\partial _{z^{\prime \prime \prime}}\partial _{z^{\prime}}\alpha^{scr}_{\perp}
\left( Q,z^{\prime \prime \prime},z^{\prime}\right) \right]
%\chi _{\vec{0}\vec{0}}\left( Q,z^{\prime
%\prime \prime },z^{\prime }\right) 
 \nonumber\\
&=&Q\int dz^{\prime \prime }dz^{\prime \prime \prime }\alpha
_{||}^{rap}\left( Q,z,z^{\prime \prime }\right)
Q^2 \alpha^{scr}_{||}\left( Q, z^{\prime \prime \prime},z^{\prime}\right)
 \left( 1+O\left(Q^{2}\right) \right) 
  \label{Inter2}
\end{eqnarray}%
where we used the identities $\int \partial _{z}\partial _{z^{\prime \prime
}}\alpha _{\perp }^{rap}\left( Q,z,z^{\prime \prime }\right) dz^{\prime
\prime }=0 = \int \partial _{z^{\prime \prime \prime}}\partial _{z^{\prime}}
\alpha _{\perp}^{scr}\left( Q,z^{\prime \prime \prime },z^{\prime }\right) dz^{\prime
\prime \prime }$. 
We take the $\exp \left( Q\left( z\pm z^{\prime
}\right) \right) dzdz^{\prime }$ moments of (\ref{ScreeningeqChiRPAFRomChiRap}%
) using (\ref{Inter2}) and also the definition (\ref{DefAlphaScrLayerPM}), finding%
\begin{eqnarray}
Q^{2}\left( \alpha _{||}^{2D,scr}(Q,iu)\pm \alpha _{\perp
}^{2D,scr}  (Q,iu)\right) &=&Q^{2}\left( \alpha
_{||}^{2D,rap}(Q,iu)\pm \alpha _{\perp }^{2D,rap}(Q,iu)\right) 
\nonumber \\
&&+Q\alpha _{||}^{2D,rap} (Q,iu)Q^{2}\alpha
_{||}^{2D,scr}(Q,iu)+O\left( Q^{4}\right)
\label{PmMomentsAlphaToNextOrder}
\end{eqnarray}

Here we have noted that the moments 
$\int \exp \left( Q\left(z \pm z^{\prime }\right) \right)
 \alpha _{ii}\left( Q,z,z^{\prime}\right)  dz dz^{\prime}$
, %
$\int \exp \left(\pm Qz\right) 
\alpha _{ii}\left(Qz,z^{\prime}\right)%
 dzdz^{\prime}$
 are equal to to lowest order 
($O( Q^{0})$), and also to $O(Q^{1})$
for symmetric layers, which we consider exclusively here.
Thus we were able to drop the $\pm$ superscript on the $\alpha^{2D}$ terms (see also the
commentary following Eq (\ref{DefAlphaScrLayerPM})). The same is not true
for the equivalent moments of $\chi $. Adding and subtracting the two
equations (\ref{PmMomentsAlphaToNextOrder}) and cancelling $Q^{2}$
throughout, we find for a symmetric layer%
\begin{equation}
\alpha _{||}^{2D,scr}(Q,iu)=\alpha _{||}^{2D,rap}(Q,iu)+2\pi
Q\alpha _{||}^{2D,rap}(Q,iu)\alpha _{||}^{2D,scr}(Q,iu)+O\left(
Q^{2}\right)  \label{O(Qpar)ScreeningAlphaPar}
\end{equation}

\begin{equation}
\alpha _{\perp }^{2D,scr}(Q,iu)=\alpha _{\perp
}^{2D,rap}(Q,iu)+O\left( Q^{2}\right)
\label{O(Qpar)ScreeninAlphaPerp}
\end{equation}

These screening equations are deceptively simple, as they include all
one-electron local-field effects to the order stated. \ Furthermore by setting $Q\rightarrow 0$ in (\ref{O(Qpar)ScreeninAlphaPerp}) and (\ref%
{O(Qpar)ScreeningAlphaPar}) we find that the response $\alpha%
^{2D,rap}$ is identical to the interacting layer response $\mathbf{\alpha }%
^{2D,scr}$ in the $Q\rightarrow 0$ limit. \ Also, using the facts that $%
W^{rap}$ is $Q$-independent (see (\ref{Wrap})) and $\chi ^{0}\left( Q,iu\right) =\chi
^{0}\left( Q\rightarrow 0\right) \left( 1+O\left( Q^{2}\right)
\right)$, we can show from (\ref{ChiRapSreeningEq}) that ${\alpha }%
^{2D,rap}\left( Q,iu\right) =\mathbf{\alpha }^{2D,rap}\left(
Q=0,iu\right) +O\left( Q^{2}\right)$. 
Thus (\ref{O(Qpar)ScreeningAlphaPar}) and (\ref{O(Qpar)ScreeninAlphaPerp}) 
can be written, for a symmetric layer at least%
\begin{eqnarray}
\alpha _{xx} &\equiv &\alpha _{||}^{2D,scr}(Q,iu)=\frac{\alpha
_{||}^{2D,scr}(Q\rightarrow 0,iu)}{1+2\pi Q\alpha
_{||}^{2D,scr}(Q\rightarrow 0,iu)}+O\left( Q^{2}\right) ,\ \ 
\label{1stOrderScrndAlphaxx} \\
\alpha _{zz} &\equiv &\alpha _{\perp }^{2D,scr}(Q,iu)=\alpha _{\perp
}^{2D,scr}(Q,\rightarrow 0,iu)+O\left( Q^{2}\right) \ \ 
\label{1stOrderScreenedAplhazz}
\end{eqnarray}%
where the ${\alpha _{ii} }^{2D,scr}(Q\rightarrow 0,iu)$ are readily
obtained from macroscopic dielectric function calculations as in 
(\ref{AlphaLayerScrxxFromEpsMacroxx}) and (\ref{Aplha2DScrzzFromEpsMacro}).

We can use \ (\ref{1stOrderScrndAlphaxx}) and (\ref{1stOrderScreenedAplhazz}%
) in two ways, a formal next-order correction, and a less systematic
approach that is more useful in practice.

\subsection{Formal next-order correction to inter-layer correlation energy}

We can expand (\ref{1stOrderScrndAlphaxx}) formally to 1st order, obtaining $%
\alpha _{||}^{2D,scr}(Q,iu)=\alpha _{||}^{2D,scr}\left( 0,iu\right) -2\pi
Q\left[ \alpha _{||}^{2D,scr}\left( 0,iu\right) \right] ^{2}$. Putting
this into (\ref{EcTwoLaersLogExpanded}), for example, we obtain a
sub-asymptotic correction to (\ref{EPerAreaPerLayerAsyInsulator2Layer}) for
the cross-correlation energy of two identical insulating layers

\begin{equation}
E^{c,next}=\frac{1}{D^{5}}\frac{3\pi }{2}\hbar \int \left( {\alpha _{||}}%
\right) ^{2}{\alpha }_{tot}du  \label{EcNext2InsultingLayers}
\end{equation}%
$\ \allowbreak $where $\alpha _{tot}=\alpha _{xx}^{2D,scr}\left(
Q=0,iu\right) +\alpha _{zz}^{2D,scr}\left( Q=0,iu\right)$ and ${%
\alpha _{||}}=\alpha _{xx}^{2D,scr}\left( Q=0,iu\right) $. Eq (\ref%
{EcNext2InsultingLayers}) is to be added to the leading $D^{-4}$asymptotic
term from (\ref{EPerAreaPerLayerAsyInsulator2Layer}). For an infinite stack,
according to (\ref{Ec)2)InfStackAsLayerPairSum}), (\ref%
{EcNext2InsultingLayers}) should be multiplied by $2\sum_{J=1}^{\infty
}J^{-5}=2.0738$.\ \ Surprisingly, we found that for BN layers, (\ref%
{EcNext2InsultingLayers}) becomes comparable to the leading asymptotic term (%
\ref{EPerAreaPerLayerAsyInsulator2Layer}) for $D<1\;nm$ \ For separations as
large as $1$ nm the interaction is \ $O\left( meV/atom\right) $, too small to
be found accurately by a full numerical correlation energy calculation using
(e.g.) VASP. \ Thus for this system there is NO range of separations $D$
where we can meaningfully test the full calculation against these
asymptotics! \ Our sub-asymptotic approach in the next Section is more
useful in this regard.

\subsection{More useful sub-asymptotic form}

Another way to use (\ref{1stOrderScrndAlphaxx}) and (\ref%
{1stOrderScreenedAplhazz}) is to leave the screening denominator un-expanded
in (\ref{1stOrderScrndAlphaxx}). While we include some $O(Q^{2})$
terms by doing this, and so should formally be including
all $O( Q^{2})$ terms, we have estimated that the
denominator in (\ref{1stOrderScrndAlphaxx}) can be far from unity at
intermediate layer separations $D$, where the other $O(Q^{2})$ terms
are still small. With this approach the $Q$ integration can be done
analytically (assuming an isotropic layer response in the plane), giving 
\begin{eqnarray}
&&\frac{E^{2Layers(sub-asy)}\left( D\right) }{2A}=-\frac{1}{2}\left( \frac{1%
}{2\pi }\right) ^{4}\int_{0}^{\infty }\hbar du\left[ \alpha _{||}\left(
iu\right) \right] ^{-4}  \label{Ec2SuAsy2Laer} \\
&&\times \left( \alpha _{||}^{2}\left( iu\right) f_{2}\left( \xi \right)
+2\alpha _{||}\left( iu\right) \alpha _{\perp }\left( iu\right) f_{1}\left(
\xi \right) +\alpha _{\perp }^{2}\left( iu\right) f_{0}\left( \xi \right)
\right) \;
\end{eqnarray}%
where $\xi \equiv \frac{D}{\pi \alpha _{||}\left( iu\right) }$ and%
\begin{eqnarray}
f_{n}\left( \xi \right) &=&\int_{0}^{\infty }q^{3}\frac{e^{-\left| \xi
\right| q}}{\left( 1+q\right) ^{n}}dq\approx 6\xi ^{-4}\;for\;\xi
\rightarrow \infty  \label{DefFn(Xi)} \\
f_{0} &=&6\xi ^{-4}  \nonumber \\
f_{1} &=&-e^{\xi }E_{1}\left( \xi \right) +\xi ^{-1}-\xi ^{-2}+2\xi ^{-3} 
\nonumber \\
f_{2} &=&\left( 3+\xi \right) e^{\xi }E_{1}\left( \xi \right) +\xi
^{-2}-2\xi ^{-1}-1  \label{EvalF3(Xi)}
\end{eqnarray}%
Here $E_{1}\left( \xi \right) \equiv Ei\left( 1,\xi \right) $ is the
exponential integral function, item 5.1.1 in Abramowitz and Stegun \cite%
{AbramowitzStegun}. Alternatively the $Q$ integration in (\ref%
{EcTwoLaersLogExpanded}) can be done numerically, which avoids the use of
the $E_{1}$ function.

For the infinite stack of uniformly spaced layers (still within a second
order expansion of the logarithm) the correlation energy according to (\ref%
{Ec)2)InfStackAsLayerPairSum}) is

\begin{equation}
\frac{E^{\infty stack(sub-asy)}\left( D\right) }{NA}=2\sum_{K=1}^{\infty }%
\frac{E^{2Layers(sub-asy)}\left( KD\right) }{2A}
\label{EInfHomStackFrom2LayerCalc}
\end{equation}%
The sum over layer separation index $K$ converges rapidly, like 
$\sum_{K}K^{-4}$.

In the above arguments, analytic evaluation of the $Q$ integrations
depended on perturbative expansion of the logarithm in the full RPA energy
expressions, a procedure that we showed to be`valid asymptotically for
insulating 2D layers, essentially because 2D intra-layer RPA screening is
ineffective as $Q\rightarrow 0$ (This contrasts with thick layers,
where conventional Lifshitz theory applies, 3D screening remains signifant
as $Q\rightarrow 0$, and the second order expansion of the logarithm is
known to make errors of up to 20\% in the vdW energy \cite%
{vdWGenThDzyaloshinskiilLifshitzPitaevski1961}). \ However one can easily
revert to non-expanded logarithmic expressions such as (\ref%
{EInfStackDimensionless}) or (\ref{EvdW2LayersFromAlphas}). \ Numerically
exact evaluation for symmetric layers is easily done as a two-dimensional
numerical integral over $Q\equiv \left| \vec{Q}\right| $ and $u$
for 2 layers, or a 3D integration on $Q,q_{z}$ and $u$ for the case of
an infinite stack. This obtains the correlation energy in seconds to minutes
on a laptop machine, in contrast to full RPA energy calculations with a
packaged plane-wave code, which takes hours, days or even weeks on a large
parallel cluster, even for simple geometries. \ 

\section{Including $O(Q^{2})$ corrections in the layer response}

As shown above, the polarizability of a symmetric insulating layer through 
$O(Q)$ can be determined directly from the macroscopic dielectric
function $\varepsilon ^{macro}\left( Q=0,\omega +i0\right) $ of a
slightly stretched stack of layers, without further input. \
However to predict layer interactions at shorter distances one needs the
layer response to $O( Q^{2})$. At this order many
effects come in:
(i) local-field ($\vec{G}\neq \vec{0}$) effects within layers
(see (\ref{WslowExpanded},\ref{O(Qpar)ScreeningAlphaPar},
\ref{O(Qpar)ScreeninAlphaPerp}}));
(ii) local-field effects between layers; 
(iii) effects of finite layer width (consider expansion of the exponential in 
(\ref{DefAlphaScrLayerPM}) for a symmetric layer);
(iv) finite spatial extent of any adsorbates; 
(v) electron pressure (diffusion) effects in the layer response 
(for example, the long-wavelenth bare density
 response of a metallic layer is of form 
 $(\chi_{0}(Q,iu))^{-1}\propto u^2 +\gamma^2 Q^2$ where the diffusion term 
  involves a mean-square velocity spread $\gamma^2$). 
It would be too difficult to account for all these effects analytically 
(or even numerically without complexity comparable to a full RPA 
correlation energy calculation),  so we now introduce an empirical 
$O(Q^{2})$ correction both to the layer response $\alpha ^{2D,scr}$
and (treated elsewhere) to the polarizability $A\left( iu\right)$ of any adsorbate 
where applicable. Accordingly we write for the layer response through
$O(Q^{2})$
\begin{eqnarray}
\alpha _{xx}^{2D,scr}(Q,iu) &=&\frac{\alpha _{xx}^{2D,scr}(Q=0,iu)%
}{1+2\pi Q\alpha _{xx}^{2D,scr}(Q=0,iu)}\sqrt{1+B^{2}Q^{2}}%
\exp \left( -\left( \beta Q\right) ^{6}/2\right)  \label{OqsquaredAlpxx}
\\
\alpha _{zz}^{2D,scr}(Q,iu) &=&\alpha _{xx}^{2D,scr}(Q=0,iu)\sqrt{%
1+B^{2}Q^{2}}\exp \left( -\left( \beta Q\right) ^{6}/2\right)
\label{OqsquaredAlpzz}
\end{eqnarray}%
where $B$ has dimensions of length and can be regarded as a sort of
effective layer width governing polarization properties, and the final exponential
factor is discussed below. 
The square root form in (\ref{OqsquaredAlpxx},\ref{OqsquaredAlpzz}) is not unique but it allows 
for the required $O(Q^2)$ correction for small $Q$, while avoiding over-emphasis of 
large $Q$ values where our theory is at any rate not accurate.
\ Moment expansion
of the exponential in (\ref{DefAlphaScrLayerPM}) suggests that $B$ is real,
i.e. that the correction is positive, and this is borne out by the fits to
the RPA correlation energy that we obtain below. There really should be
separate values $B_{xx}$ and $B_{zz}$ \ to fit the parallel and
perpendicular responses, but here we assume for simplicity that $%
B_{xx}=B_{zz}=B$. The constant $B$ for each species of layer was fixed by
fitting one point on the curve of RPA correlation energy vs layer spacing $D$
of an infinite stack of that particular layer, the fit being done at a point that
is near to the equilibrium spacing $D_{0}$. \ Our theory is then used for
all other $D$ values, and we shall see that this reproduces the entire interlayer
correlation energy curve $E_{c}\left( D\right) $ of a stack of layers very
well, including $D$ values near to and far from the equilibrium layer spacing $D_{0}$ of
the solid. In doing this, we needed to consider one more point, as follows.

%The $\vec{Q}$ integrations arein principle limited to the first Brillouin zone, but our analytic 
%expressions are at any rate not accurate near the zone boundary.  To preserve the simnplicity of
% our scheme, rather than integrating of such a geometrically inconveniennt region 
%we achieve a similar result by integrating over all $\vec{Q}$ but using the exponential factor in  of approximatoinas The additional exponential factor in (\ref{OqsquaredAlpzz}) was introduced 
%to provide smooth damping the response for for $Q$ lying outside the 2D Brillouin zone. foreoutside  the response 
%In addition to the $O(Q^2)$ correction introduced to
%reproduce sub-asymptotic physics, we also needed to introduce a further
%exponential factor  into equations (71) and (72) to extend our scheme all the way to
%"contact". 
At moderate inter-layer distances $D$ the inverse exponential
$e^{-2QD}$, from the Coulomb potential in correlation energy expressions such as 
(\ref{EPerAreaPerLayerAsyInsulator2Layer}), creates a natural cutoff ensuring that 
the unphysical large-$Q$ behaviour of our small-$Q$ resonse expressions 
(\ref{OqsquaredAlpxx},\ref{OqsquaredAlpzz}) does not contribute significantly to the energy.
 However, as the two layers
approach their natural binding distance $D_0$ this exponential decay is
insufficient for the task, and we need to recognize that the $\vec{Q}$ 
integrations should at least be restricted to the first Brillouin zone. To this end we 
include an additional
damping factor in (\ref{OqsquaredAlpxx},\ref{OqsquaredAlpzz}) of form $e^{-(\beta Q)^6}$. The value of $\beta$ 
is chosen to satisfy a normalisation condition based on the area $A_{BZ}$ of 
the 2D Brillouin zone (BZ) of the layer's crystal structure:
\[
\int \exp \left( -\left( \beta Q\right) ^{6}\right)
d^{2}Q=A_{BZ} 
\]
This ensures that the energy integration only samples values of $\vec{Q}$ lying
roughly within the BZ, even though we formally include larger  $Q$ values in our 
$Q$  integration. This smooth damping
factor avoids the numerical difficulties posed by a sharp BZ cutoff in
$Q$, but contributes only to $O(Q^6)$ for
$Q \rightarrow 0$ and thus does not spoil the desirable small-wavenumber
physics of the polarizability model (\ref{OqsquaredAlpxx},\ref{OqsquaredAlpzz}) that we have created. 
Apart from this consideration, the use of the 6th power is not unique.

\section{Numerical results for BN layers}

%\subsection{ Stretched bulk BN}

\begin{figure}[tbp]
% currently Fig 4
%\includegraphics{CEnergyBN.eps}
%\includegraphics{CEnergyBN_6_238.eps}
%\includegraphics{CEnergyBN_6_2.38.eps}  seb
\includegraphics{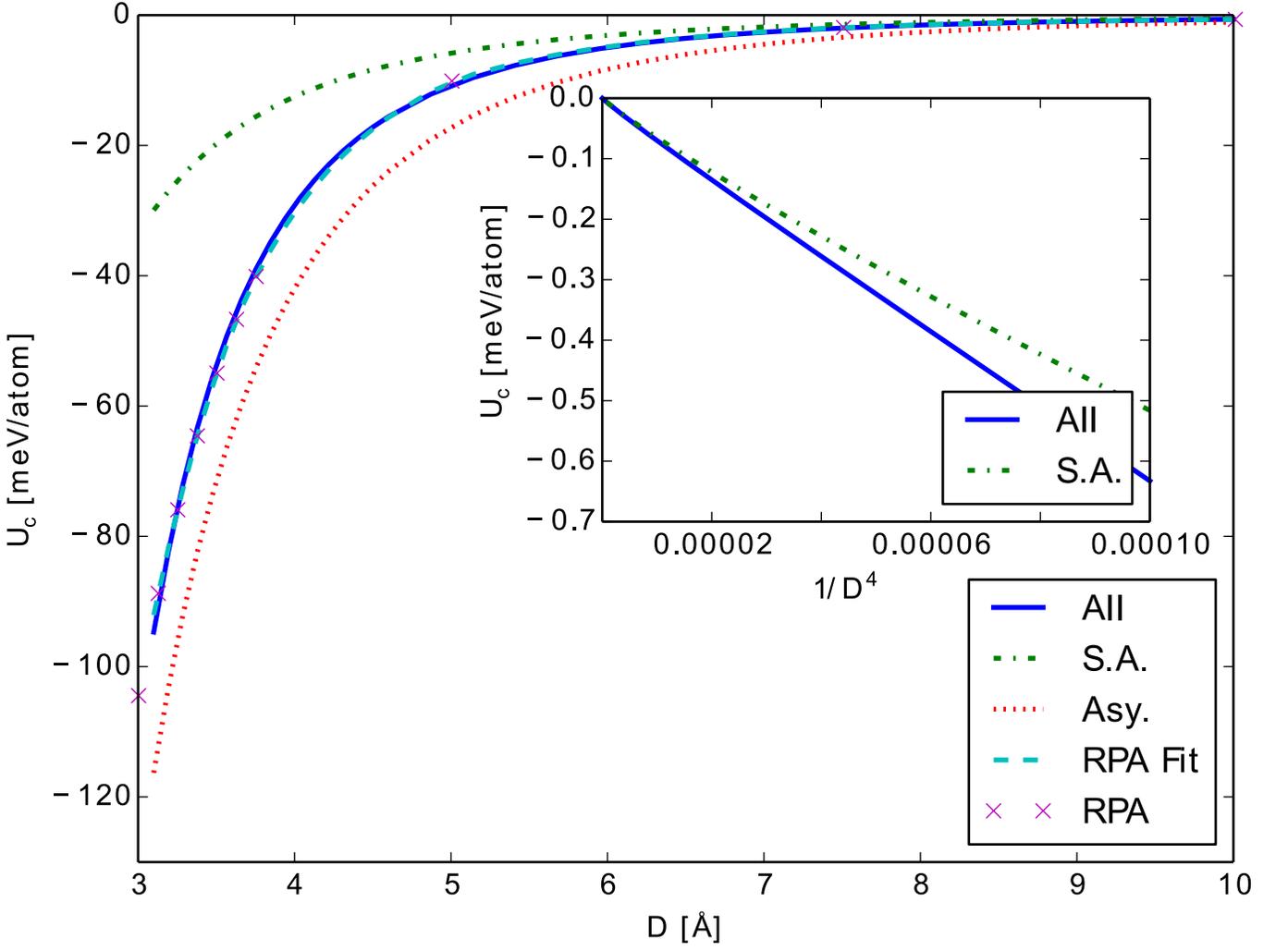}
\caption{Inter-layer correlation energy $U_c \equiv E_c(D)$ per atom (meV) of stretched bulk BN with 
layer spacing $D$. Short green dashes with pale blue crosses: RPA correlation energy from VASP. 
Red dots: asymptotic $D^{-4}$ result from Eq  (\ref{AsymptoticEcInfiniteInsulatorStack}).
Dark green dash-dots: sub-asymptotic energy from Eqs (\ref{Ec2SuAsy2Laer}) and (\ref{Ec)2)InfStackAsLayerPairSum}). 
Solid blue line: our semi-analytic result ($O(Q)$ sub-asymptotics plus $O(Q^2)$ 
correction) from (\ref{EcTwoLaersLogExpanded}),
(\ref{Ec)2)InfStackAsLayerPairSum}), (\ref{OqsquaredAlpxx}) and 
(\ref{OqsquaredAlpzz}).
Insert: $E_c$ vs $D^{-4}$ (Angstrom $ ^{-4}$) for larger $D$ values, showing approximate
proportionality $E_c \approx KD^{-4}$.
\label{fig:EcBulkBN}
}
\end{figure}

\begin{figure}[tbp]
% currently Fig 5
%\includegraphics{CEnergyxxBN_6_2.38.eps}
%\includegraphics{CEnergyxxBN_6_2.38.eps}  SEB
\includegraphics{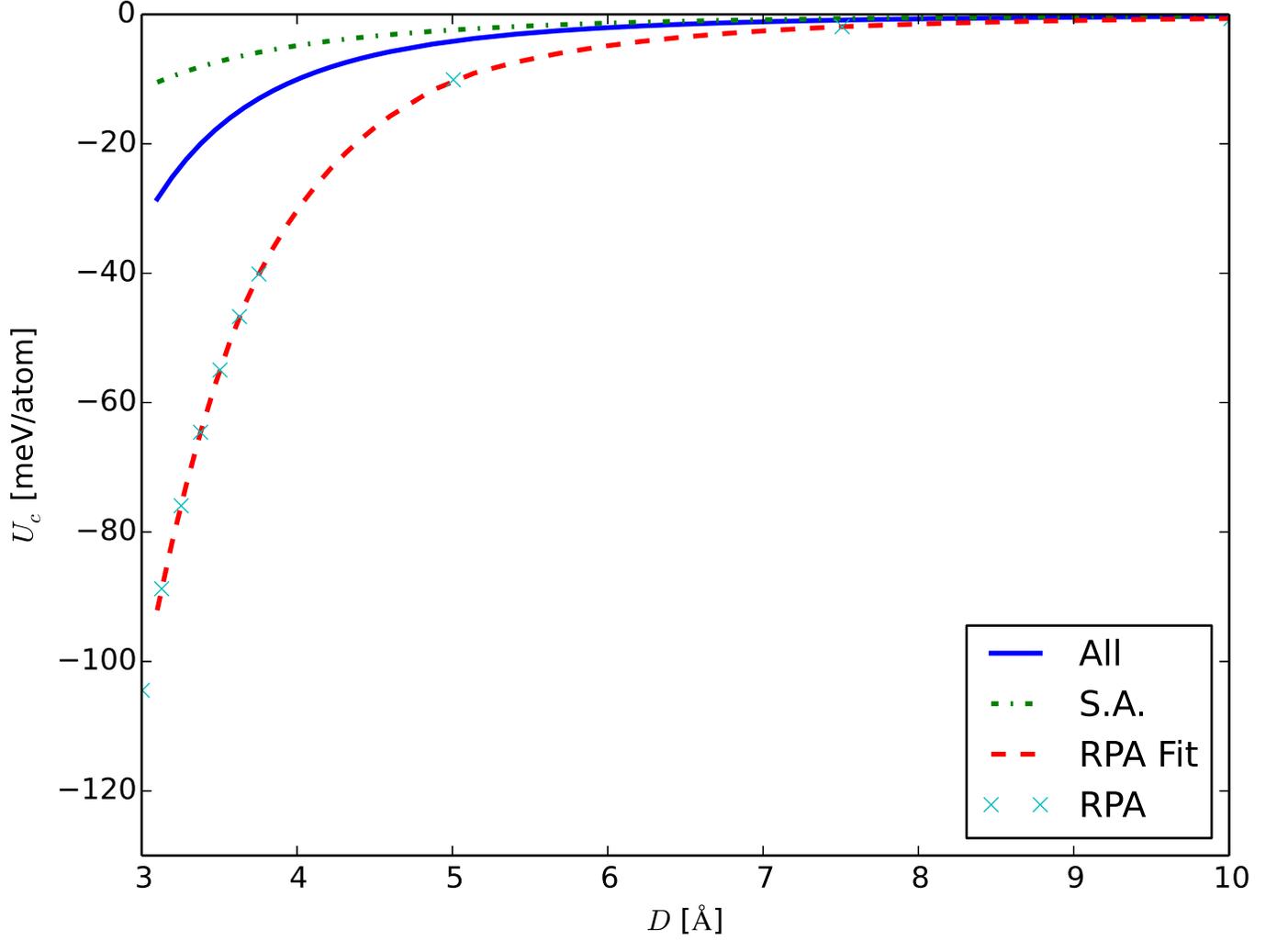}
\caption{
Interlayer correlation energy $U_c \equiv E_c(D)$ (meV/atom) of stretched bulk BN using only the parallel 
response, showing the importance of including both perpendicular and parallel layer polarizability 
in the analytic work. Red dashed line (with light blue crosses at the numerical data points): 
Full $E_c(D)$ from numerical RPA using VASP. 
Solid blue line: Our best analytic fit, but with the perpendicular polarizability $\alpha_{zz}^{2D,scr}$ 
set to zero. 
Green dash-dots: Our sub-asymptotic energy (\ref{Ec2SuAsy2Laer}) with 
$\alpha_{zz}^{2D,scr}$ set to zero. 
 Note the poor fit of our best theory to full RPA data here when $\alpha_{zz}^{2D,scr}$ set to zero, 
 in contrast to the excellent fit in 
Fig \ref{fig:EcBulkBN} where both 
$\alpha_{zz}^{2D,scr}$ and $\alpha_{xx}^{2D,scr}$ are included.
\label{fig:EcBulkBNwithoutAlphazz}
}
\end{figure}

The results of our procedure for a stretched h-BN crystal are shown in Fig 
\ref{fig:EcBulkBN}. \ The fully asymptotic energy prediction (\ref%
{AsymptoticEcInfiniteInsulatorStack}) substantially over-estimates the RPA
correlation energy at smaller separations, and so cannot be corrected by a
positive $O(Q^{2}$) term of the form motivated above on grounds of
layer width and other physics. The sub-asymptotic $O\left( Q\right) $
form of layer polarizability (\ref{Ec2SuAsy2Laer}), (\ref%
{EInfHomStackFrom2LayerCalc}) somewhat under-estimates $E_{c}$ and so is
eligible for physically meaningful $O( Q^{2}) $ correction.
By fitting the RPA layer correlation energy at a single point near the
equilbrium spacing $D=D_{0}$ we found optimal values $B=0.238\,nm$ for
monolayer BN and $B=0.265\,nm$ for monolayer graphene (see 
 (\ref{OqsquaredAlpxx}) and (\ref{OqsquaredAlpzz})). These choices led to
excellent fits to the available RPA interlayer correlation energy at all
layer separations $D$ down to $D=D_{0}$, as seen in Fig. \ref{fig:EcBulkBN}.
\ \ In Fig \ref{fig:EcBulkBNwithoutAlphazz} we also show the same quantity
with the perpendicular polarizability $a_{zz}^{2D,scr}$ set to zero. \ This
demonstrates that neglect of either one of $a_{zz}^{2D,scr}\,$or $%
a_{xx}^{2D,scr}$yields large errors in the inter-layer correlation energy. \ 

\begin{figure}[tbp]
% currently fig 6
%\includegraphics{TimsEnergyBN.eps}
%\includegraphics{EnergyBN_6_2.38.eps}
\includegraphics{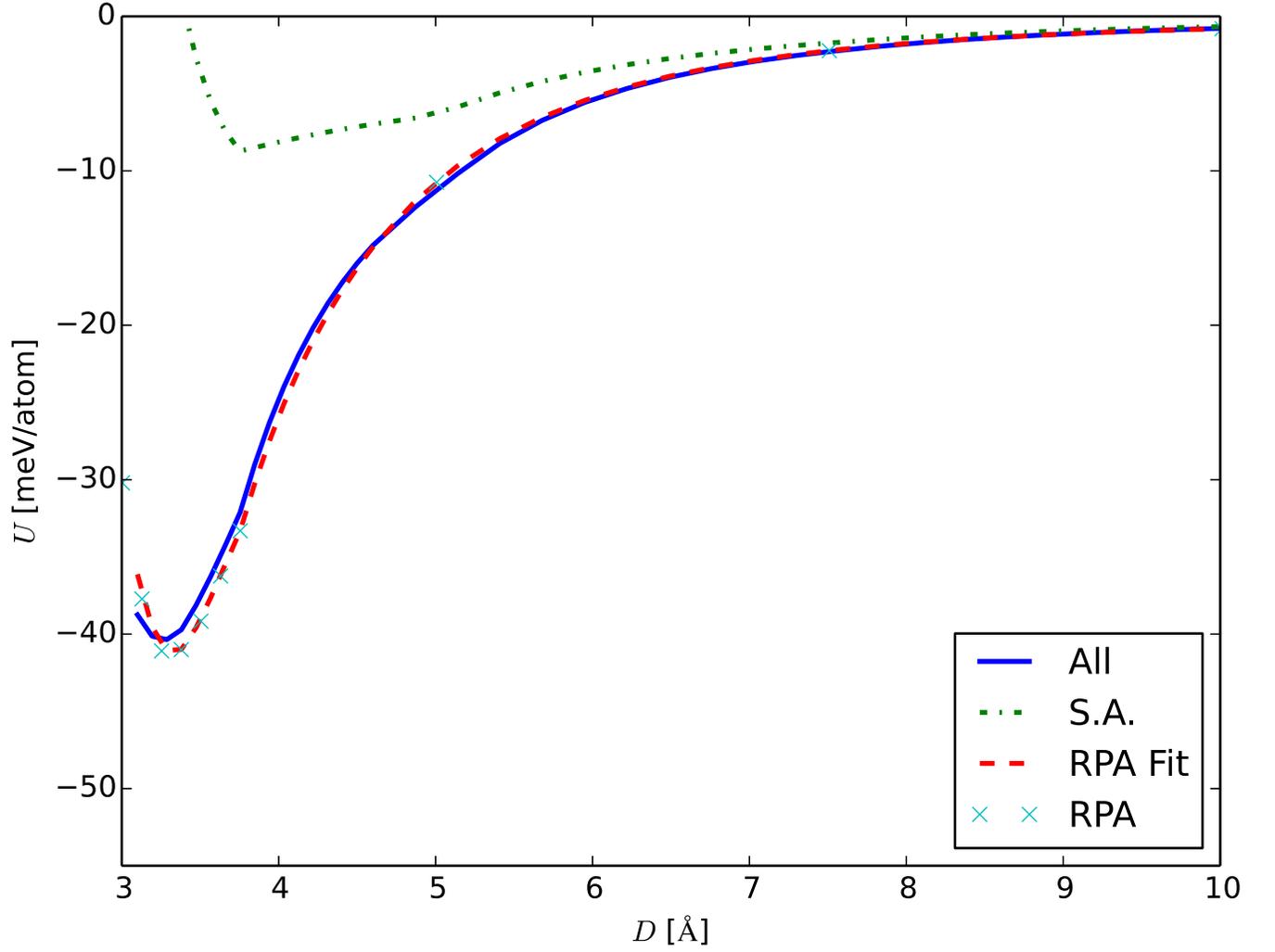}
\caption{
Total interlayer energy $U \equiv E_tot(D)$ of stretched bulk BN (meV/atom), including exact exchange energy 
fom VASP in all cases. 
Red dashes with blue-green crosses: with umerically exact RPA correlation energy from VASP. 
 Solid blue line: with correlation energy from our best semi-analytic theory (see also 
Fig \ref{fig:EcBulkBN}).  
Dark green dash-dots: with sub-asymptotic correlatiom energy from (\ref{Ec2SuAsy2Laer}) and (\ref{Ec)2)InfStackAsLayerPairSum}).  
(\ref{AsymptoticEcInfiniteInsulatorStack}). 
\label{fig:EtotBulkBN}
}
\end{figure}

Since our prediction of the interlayer RPA\ correlation energy alone is very
good, in order to reproduce the total RPA energy $E\left( D\right) $ we
should add the exact Hartree and exchange energies. This gives a very good
account of the total RPA binding energy curve $E_{tot}\left( D\right) $ as
shown in Fig.\ref{fig:EtotBulkBN}. The convergence parameters of the full 
RPA energy calculation were as follows: a $12\times 12 \times 6$ k-space grid
 for the exchange calculation, and an $8\times 8 \times 3$ grid for the 
 correlation energy calculation, reducing to $8\times 8 \times 1$ for large
 inter-layer separations. The cut-off for the polarisability matrices was 
 300 eV and the cut-off of the wavefunction was 700 eV.The energies agree well
 with previous RPA calculations on 
 h-BN \cite{RPAEnergyBN_MariniRubioPRL2006},\cite{AreWevdWreadyBjorkmanJPCM2012}.

\section{The special case of graphene \ }

Graphene is not an insulator and its polarizability is singular as $%
Q,u\rightarrow 0,0$. Graphene is also a special case because it is very
difficult to include enough $\vec{k}$ points, in standard numerical
plane-wave codes, to capture the delicate physics of gapless electronic
transitions near to the Dirac point, responsible for the graphene's unusual
electronic and van der Waals properties. In calculating the layer
polarizability for graphene from the plane wave code VASP \cite{VASP1},\cite%
{VASP2} we therefore excised from the numerics any electronic transitions
with an energy less than $\varepsilon _{c}=1.25eV$, leading to a 2D
polarizability $\alpha ^{(ins)}$ that is insulator-like - i.e. nonsingular
as $u\rightarrow 0$. The remaining gapless $\pi _{z}$ transitions give rise
to the bare graphenic response from a truncated conical Bloch band \cite%
{GraphiteDispIntPRL2010LebegueEtal}. This response is singular as $%
u\rightarrow 0$ but is known analytically \cite%
{GouldDobsongrapheneConesPRB2013},\cite%
{DispInductInteractGrapheneNanostrDobsonSurfSci2011}. Adding the numerical
and analytic contributions we obtain the ''rapid''\ polarizabilities $\alpha
^{rap}$corresponding to $\chi ^{rap}$ as defined in (\ref{ChiRapSreeningEq}%
). We note that the $\pi _{z}$ response contains no local field terms and
does not contribute to the perpendicular polarizability $\alpha _{zz}$ \ 
\begin{equation}
\alpha _{xx}^{2D,rap,gr}(Q,iu)=\alpha _{xx}^{2D,rap,ins,2D}\left(
Q,iu\right) 
+\alpha^{cone}(Q,iu) + O\left( Q^{2}\right) 
\label{AlphaRapxxGraphene}
\end{equation}%

\begin{equation}
\alpha^{cone} = e^{-2}\left( 4\hbar \right) ^{-1} \left(
u^{2}+v_{0}^{2}Q^{2}\right) ^{-1/2}\theta
\end{equation}%

Here $v_{0}$ is the characteristic velocity of the graphene electronic
bandstructure, and 
\[
\theta =\frac{2}{\pi }\arctan \left( \tilde{\varepsilon}/\sqrt{%
u^{2}+v_{0}^{2}Q^2}\right) ,\ \ \ \ \ \ \ \tilde{\varepsilon}=\sqrt{%
\text{max}\left( \varepsilon _{c}^{2} - v_{0}^{2}Q,0\right) }\rightarrow
\varepsilon _{c}\text{ as }Q\rightarrow 0\,. 
\]%
Then using (\ref{ScreeningeqChiRPAFRomChiRap}) to calculate the interacting 
$\chi ^{2D,scr}$ we obtain the layer
polarizability at small but finite surface-parallel wavenumber in a form including 
an $O(Q^2)$ correction as in (\ref{OqsquaredAlpxx}) : 
\begin{equation}
\alpha _{xx}^{2D,scr,graphene}\left(Q,iu\right) =\frac{\alpha
_{||}^{2D,rap,gr}\left( Q,iu\right) }{1+2\pi e^{2}Q\alpha
_{xx}^{2Drap,gr}\left( Q,iu\right) }\sqrt{1+BQ^{2}}\exp \left(
-\left( \beta Q\right) ^{6}/2\right) +O\left( Q^{3}\right)
\label{RForGraphene}
\end{equation}%
where $\alpha _{xx}^{2D,rap,gr}$is given by (\ref{AlphaRapxxGraphene}
), while $\alpha _{zz}^{2D,scr}\left( Q,iu\right) =\alpha
_{zz}^{rap}\left( Q,iu\right) \sqrt{1+BQ^{2}}\exp \left( -\left(
\beta Q\right) ^{6}/2\right) +O\left( Q^{3}\right) $ just as for
an insulating layer. \ The cross-correlation energy \ for two graphene
layers is then found by putting (\ref{RForGraphene}) into (\ref%
{EvdW2LayersFromAlphas}) or (\ref{EcTwoLaersLogExpanded}). \ For an infinite
stack of graphenes we put (\ref{RForGraphene}) into (\ref%
{EInfStackDimensionless}) or (\ref{Ec)2)InfStackAsLayerPairSum}).

\begin{figure}
% currently Fig 7
%\includegraphics{CEnergyGrCut1.eps}
%\includegraphics{CEnergyGrCut1.25_6_2.65.eps}
\includegraphics{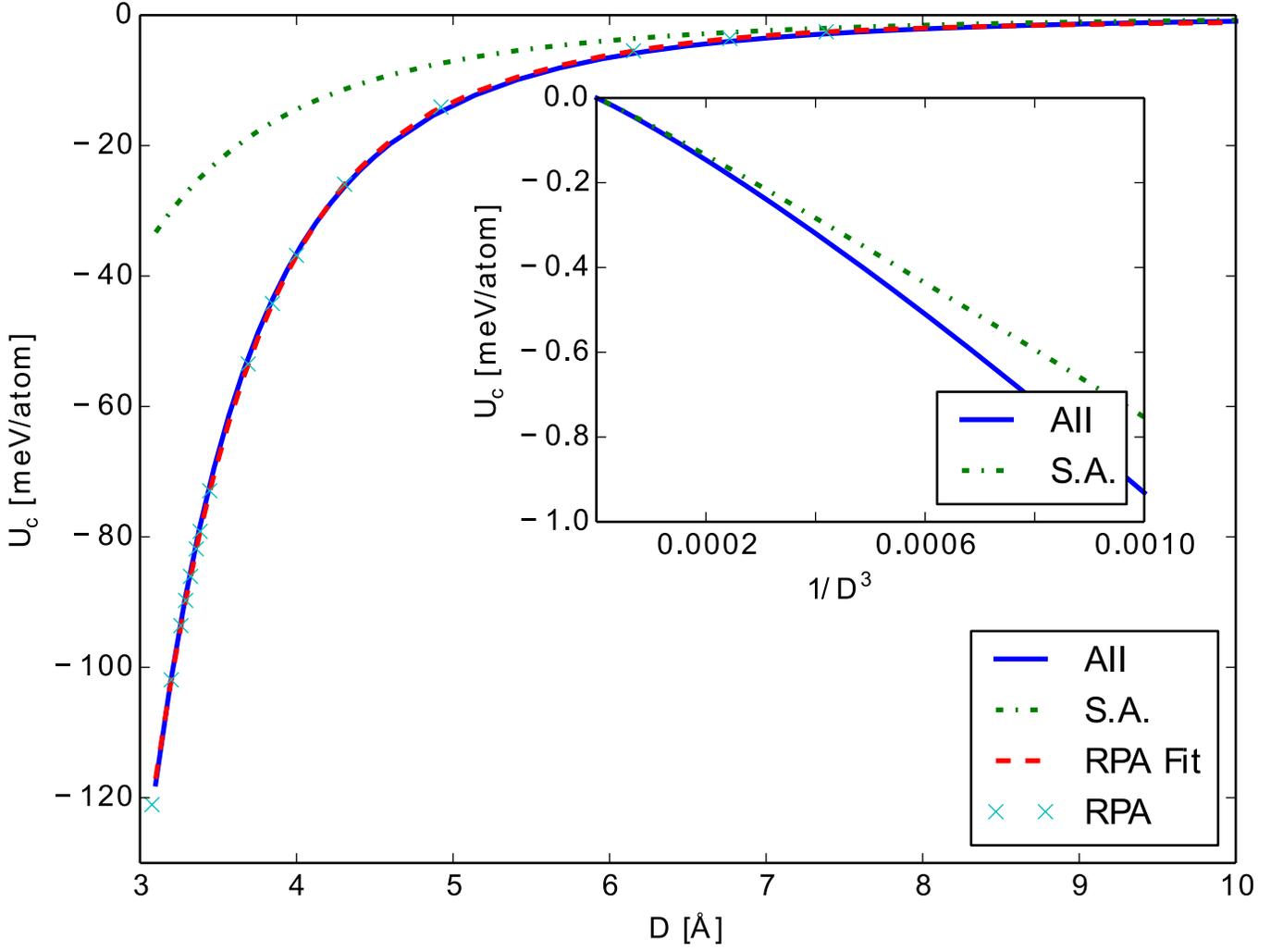}
\caption{
Interlayer correlation energy $E_c(D)$ of stretched graphite (meV/atom). 
Red dashes: RPA correlation energy from VASP, 
with individual data points shown by crosses.  
Blue solid line: our best semi-analytic (sub-asymptotic + 
$O(Q^2)$) theory from (\ref{RForGraphene}), 
(\ref{EcTwoLaersLogExpanded}) and (\ref{Ec)2)InfStackAsLayerPairSum}).  
Green dash-dots: our sub-asymptotic theory.  .
Insert: energy vs $D^{-3}$ for $D > 1$ nm. See also ref. \onlinecite{GouldDobsongrapheneConesPRB2013}.  
\label{fig:Ec(D)stretchedGraphite}
}
\end{figure}

Figure \ref{fig:Ec(D)stretchedGraphite} shows the inter-layer correlation
energy $E_{c}\left( D\right) $ of stretched graphite as a function of layer
separation $D$ in the  sub-asymptotic approximation, and in our full semianalytic $O(Q^{2})$
 theory, as well as $E_{c}\left( D\right) $ from a full
RPA\ correlation energy calculation using VASP.   Agreement of our analytic 
$O(Q^2)$ theory with the VASP results is excellent over the whole range
 of $D$ values covered in the figure.

For relatively large layer separations $D$ not covered in Fig \ref{fig:Ec(D)stretchedGraphite}
 (in practice for $D >10 nm$), graphene layers are known \cite{AsyDispIntPRL2006JFDWhiteRubio},
 \cite{GouldDobsongrapheneConesPRB2013} %
to have an unusual van der Waals energy that falls off more slowly than
that of insulating layers, as $D^{-3}$ instead of $D^{-4}$. 

This asymptotic result can be obtained as follows. For the case of
two zero-temperature undoped graphene layers at sufficiently large distance $%
D$ so that only very small $Q$ values are sampled, 
the electronic $\pi _{z}$ response dominates, so 
$\alpha_{1tot}=\alpha_{2tot}=e^{-2}Q^{-2}\chi ,$ $\chi =\chi _{0}/\left( 1-2\pi
e^{2}Q^{-1}\chi _{0}\right) ,$ $\chi _{0}=\left( -Q^{2}/4\hbar
\right) \left( u^{2}+v_{F}^{2}Q^{2}\right) ^{-1/2}$. Introducing new
dimensionless integration variables $\kappa =QD$, $U=u/\left(
v_{F}Q\right) $ we reduce (\ref{EvdW2LayersFromAlphas}) to a $D^{-3}$
 power law, which has been known for some time \cite%
{AsyDispIntPRL2006JFDWhiteRubio}:

\begin{equation}
\frac{E^{interlayer}}{A}\left( 2\;graphenes)\right) =-\frac{\hbar v_{F}}{%
\left( 2\pi \right) ^{2}}D^{-3}\int_{0}^{\infty }Q^{2}dQ\int_{0}^{\infty
}dU\ln \left( 1-e^{-2Q}\left( \mu \sqrt{U^{2}+1}+1\right) ^{-2}\right)
\label{DMinus3} 
\end{equation}

where $\mu =2\hbar v_{F}/\left( \pi e^{2}\right)$. Note however that for 
zero-temperature, undoped graphene layers, in practice this formula is
accurate only for layer separations $D$ exceeding about $10\;nm$: for
smaller separations the insulator-like contributions to the layer response,
from non-$\pi _{z}$ electronic transitions, plus finite-Dirac-cone effects,, 
must be included as we have done here, and 
the $D^{-3}$ dependence is masked by larger terms.
 The infinite-stack case can be obtained from the layer-pair result as in Eq
(\ref{Ec)2)InfStackAsLayerPairSum}):\ there is evidence that expanding the
logarithm in the RPA correlation energy formula is adequate here (see Table
III of \cite{ValidityComparisonvdWDobsonJCompThNanosci2009}). 

The insert to Fig \ref{fig:Ec(D)stretchedGraphite} plots the layer binding energy 
$E$ of stretched graphite vs. $D^{-3}$ using 
our Layer Response Theory, for large separations $D > 10$ Angstrom where the full 
RPA energy cannot be obtained meaningfully because of numerical noise. The approximate linear 
dependence of $E$ on $D^{-3}$ (see  (\ref{DMinus3})) is apparent for large $D$ 
(left-hand part of the insert).  So our LRT has treated all regimes of separation correctly.

For even larger separations the RPA itself is suspect for graphene, and an 
approximate many-body treatment \cite{HowMBTAffectsGraphenevdW} 
suggests that (\ref{DMinus3}) is further modified
by a logarithmic factor or even a change to the power of $D$.  This is not 
relevant to the energetics at the $D$ values investigated nuimerically here,
however.

\begin{figure}[tbp]
% currently Fig 8
%\includegraphics{EnergyGrCut1.eps}
%\includegraphics{EnergyGrCut1.25_6_2.65.eps}
\includegraphics{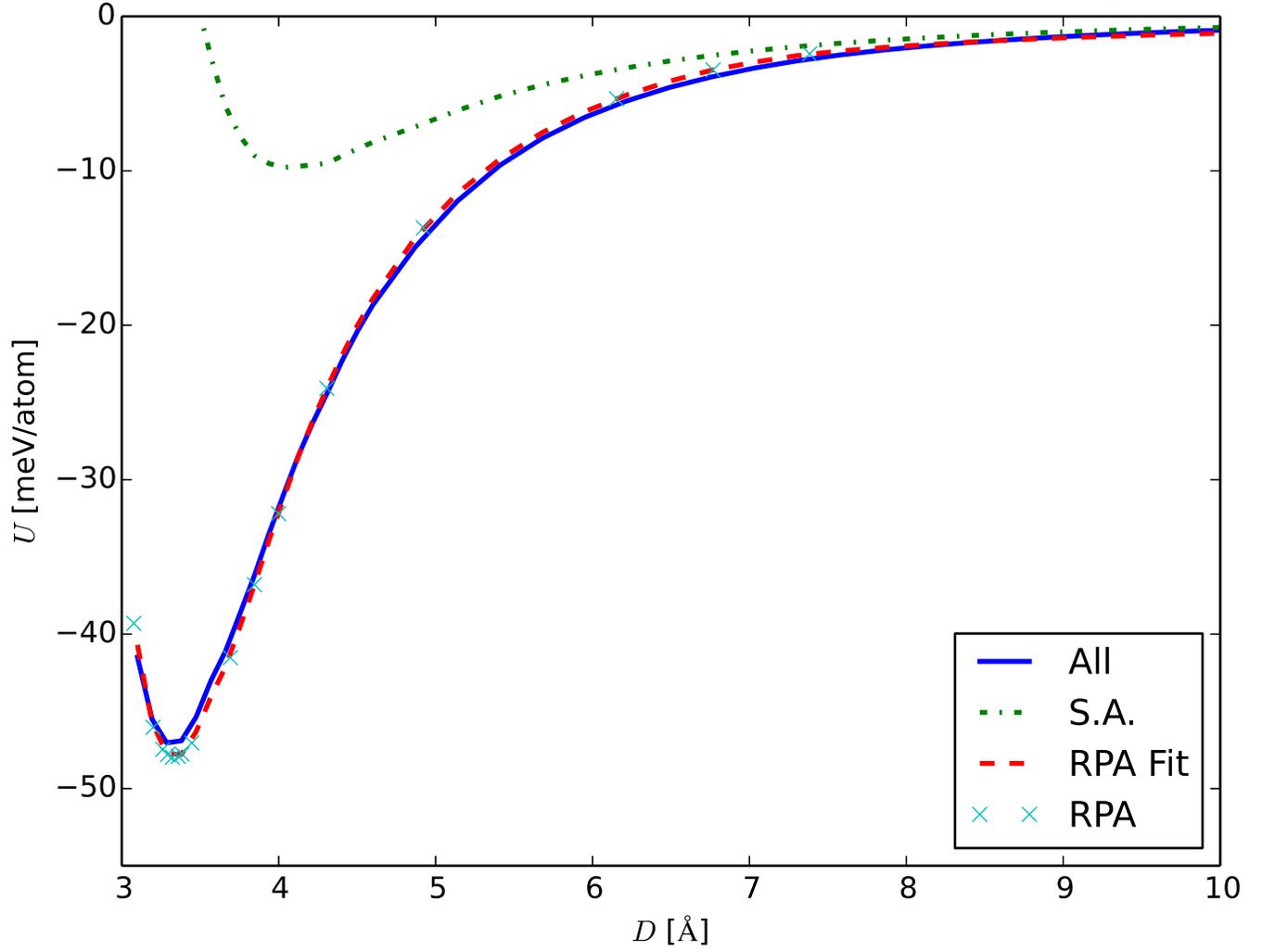}
\caption{
Total interlayer energy $U \equiv E_{tot}(D)$ of stretched graphite (meV/atom).  
Red dashes with crosses: numerically exact RPA energy including exact exchange energy, from VASP. 
Solid blue line: using correlation energy from our best semi-analytic theory plus Hartree and exact 
exchange energy from VASP. 
Green dash-dots: our sub-asymptotic approximation. 
\label{fig:Etot(D)Graphite}
}
\end{figure}

Figure \ref{fig:Etot(D)Graphite} shows the total interlayer energy $%
E_{c}\left( D\right) $ of bulk graphite. We include the full RPA\
correlation energy from a large VASP calculation, and our approximation
using (\ref{Ec)2)InfStackAsLayerPairSum}) and (\ref{RForGraphene}) together with
the numerically exact exchange energy from VASP.  
The convergence parameters for the full RPA energy calculation for graphite were as 
for Ref \onlinecite{GraphiteDispIntPRL2010LebegueEtal}, resulting in an estimated 
1meV/atom numerical uncertainty.
 Our LRT energies agree with full RPA to about this level. Thus
once more the agreement of our extremely efficient analytic $O(Q^2)$ theory 
with the large costly VASP calculation is excellent.

\subsection{Example: Dispersion interaction in an 
infinite stack of alternating graphene and BN layers}

For the infinite stack -BN-Gr-BN-Gr-BN-Gr- we use the sum of layer-pair
interactions as in (\ref{Ec)2)InfStackAsLayerPairSum}). \ Partial
justification of use of this log-expanded approximation for graphene layers
comes from \cite{ValidityComparisonvdWDobsonJCompThNanosci2009}. 
The inter-layer correlation energy of a graphene-BN pair is obtained from 
 (\ref{EcTwoLaersLogExpanded}) with the layer polarizabilities $\alpha_1$ 
and $\alpha _2$ for BN and graphene taken from the working above, with %
effective layer-width parameters $B^{BN}$ and $B^{gr}$ already determined 
by the previous fits to pure BN and pure graphite. The Brillouin zone 
cutoff parameter $\beta$ was taken as the avarage of those of BN and graphene, 
since the 2D lattice parameters are so similar. 
We also chose the effective layer width parameter $B$ to be the average of 
that for graphene and that for BN: 
\begin{eqnarray}
\alpha _{1tot} &=&\left( \frac{\alpha _{xx}^{BN}\left( iu\right) }{1+2\pi
Q\alpha _{xx}^{BN}\left( iu\right) }+\alpha _{zz}^{BN}\left( iu\right)
\right) f \label{Alpha1TotBNgr} \\
%\sqrt{1+B^{BN}Q^{2}}\exp \left( -\left( \beta Q\right)
%^{6}/2\right)  
\alpha _{2tot} &=&\left( \frac{\alpha _{xx}^{gr,cut}\left( iu\right)
+\alpha ^{cone}\left( Q,iu\right) }{1+2\pi Q\left( \alpha
_{xx}^{gr,cut}\left( iu\right) +\alpha ^{cone}\left( Q,iu\right)
\right) }+\alpha _{zz}^{gr,cut}\left( iu\right) \right) f
\label{Alpha2TotBNgr} 
\end{eqnarray}
%&&\times \sqrt{1+B^{gr}Q^{2}}\exp \left( -\left( \beta Q\right)
%^{6}/2\right) \\
where 
\begin{equation}
f= \sqrt{1+(B_{gr}+B_{BN})^2Q^{2}/4}\exp \left( -\left( \beta Q\right)^6/2\right)
\end{equation}

\begin{figure}[tbp]
% currently Fig 9
%\includegraphics{CEnergyBoth_6_2.51.eps}  SEB
\includegraphics{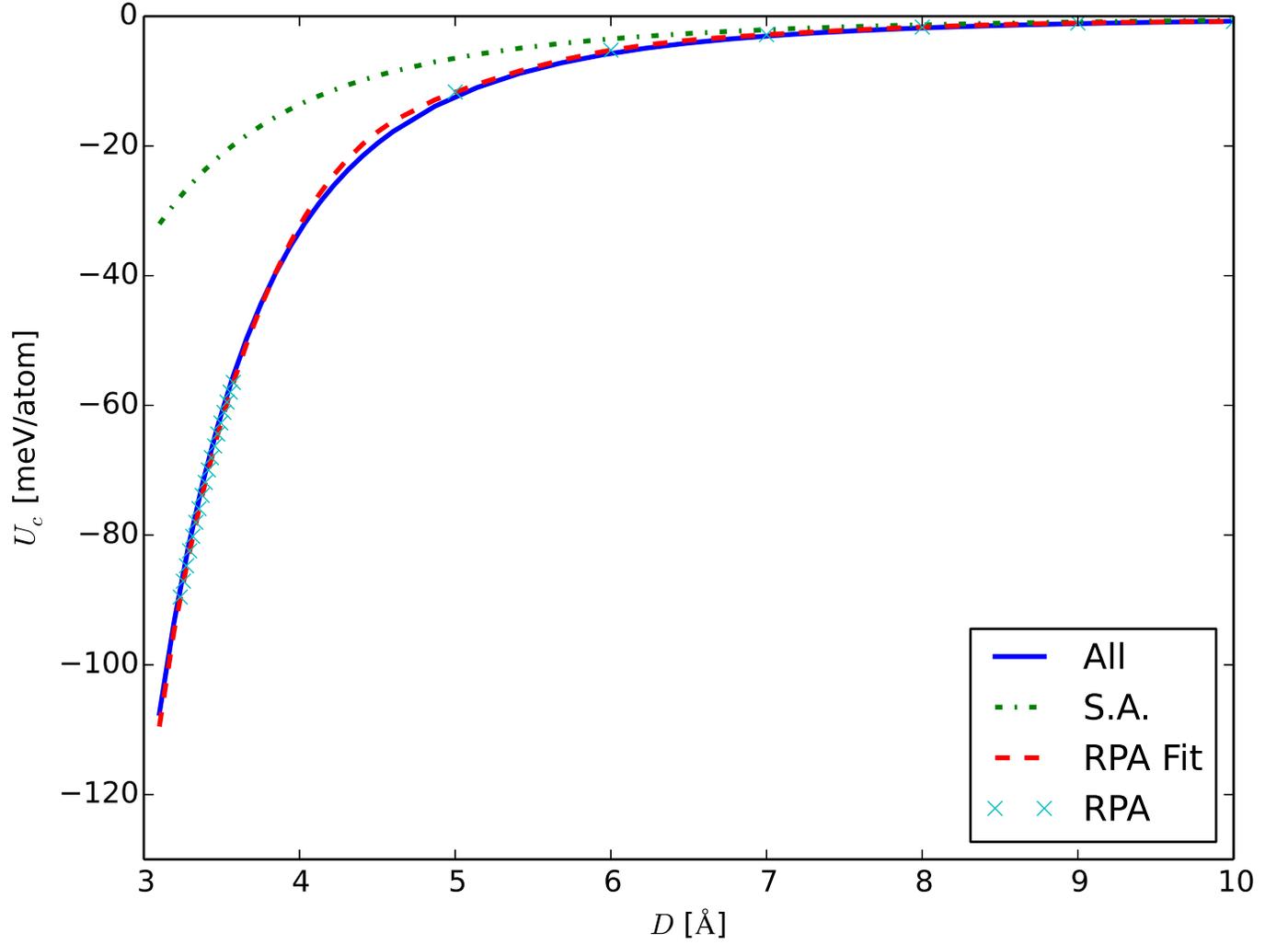}
\caption{
Interlayer correlation energy $U_c \equiv E_c(D)$ from an infinite stretched stack ...BN-gr-BN-gr...of
 alternating graphene and BN layers. Red dashes with crosses: RPA correlation energy from VASP. 
Green dash-dots: our sub-asymptotic formula. 
 Blue solid line: our best Layer Theory prediction through $O(Q^2)$ from
 (\ref{Ec)2)InfStackAsLayerPairSum}), 
(\ref{EcTwoLaersLogExpanded}), (\ref{1stOrderScrndAlphaxx}) and 
(\ref{RForGraphene}). 
\label{fig:Ec(D)BNGrapheneStack}
}
\end{figure}

In Fig. \ref{fig:Ec(D)BNGrapheneStack} our interlayer correlation energy
predictions are compared with those from a full VASP calculation for the hetero-stack.
Our semi-analytic $O(Q^2)$ theory gives an excellent fit to the VASP 
data, without the use of any new fitting parameters specific to this 
particular hetero-structure.  For the full RPA energy calculations on the 
hetero-stack the convergence parameters were similar to those quoted above for 
the pure BN calculations.

\begin{figure}[tbp]
% currently Fig 10
%\includegraphics{EnergyBoth_6_2.51.eps}
\includegraphics{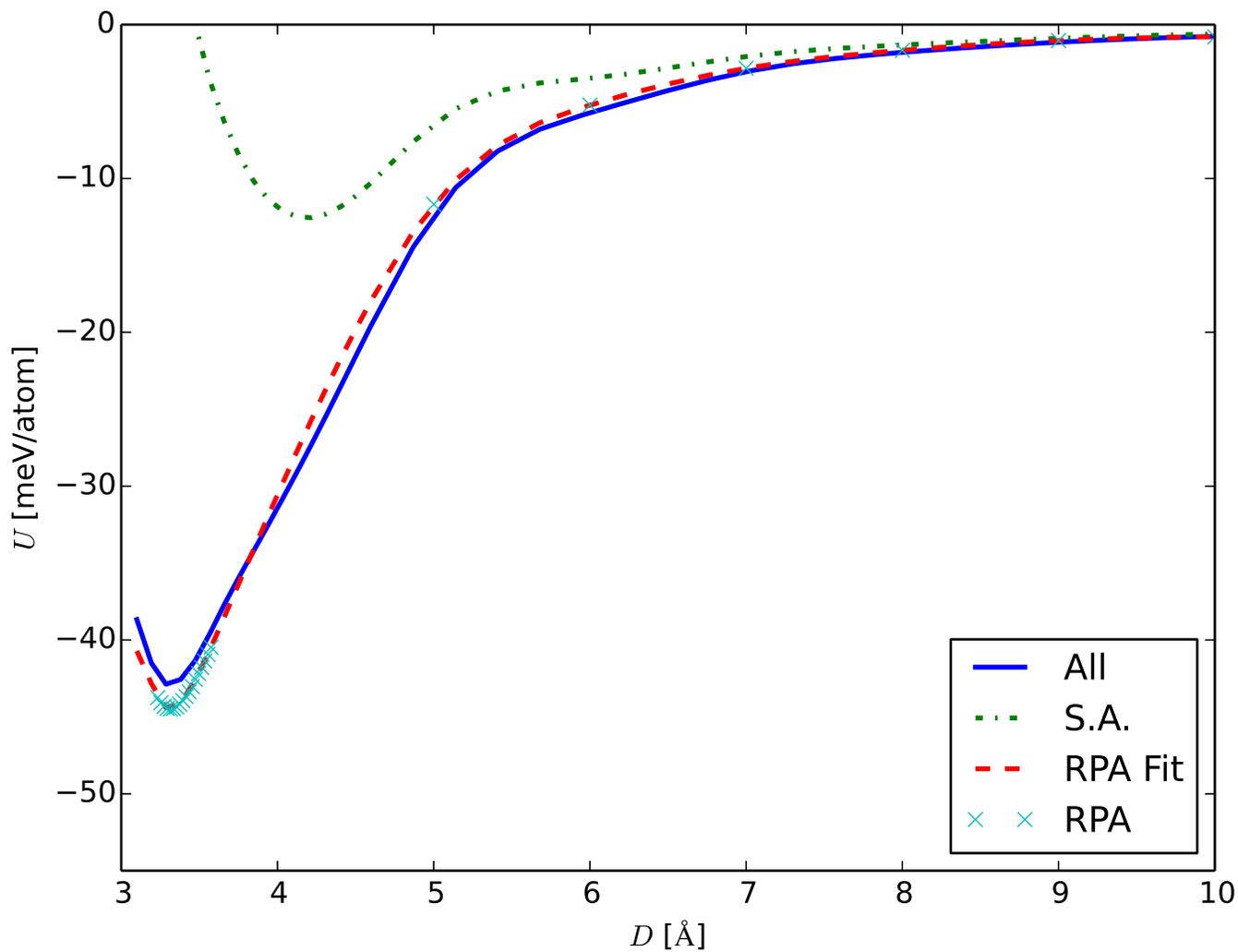}
\caption{
Total interlayer energy $U \equiv E_{tot}(D)$ from an infinite stretched stack ...BN-gr-BN-gr...of alternating graphene and BN layers (meV/atom), including exact exchange energy fom VASP. Color scheme of graph as for Fig. 
\ref{fig:Ec(D)BNGrapheneStack}.
\label{fig:Etot(D)BNGrapheneStack}
}
\end{figure}

In Fig \ref{fig:Etot(D)BNGrapheneStack} we show the total interlayer energy
for the BN-gr stack using our Layer Response Theory plus the exact exchange
contribution from VASP. Again our results are in excellent agreement with 
the total interlayer energy from VASP.

\section{Summary}

Our aim here was to provide the necessary semi-analytic correlation energy theory for prediction
of RPA-level cohesive energetics on nano-thin layer systems without very
large computations. We began by obtaining the long-wavelength
imaginary-frequency screened polarizablities $\alpha _{xx}^{2D,scr}\left(
Q\rightarrow 0,\omega =iu\right) $, $\alpha _{zz}^{2D,scr}\left(
Q\rightarrow 0,\omega =iu\right) $ of each isolated monolayer species.
We did this using efficient calculations of the macroscopic dielectric function $%
\varepsilon ^{macro}\left( \omega +i0\right) $ for\ an infinite stack of the
relevant layer, with a modestly stretched interlayer spacing $D\approx 2D_{0}
$. This was possible without the use of highly stretched layer spacings
because we treated the long-wavelength inter-layer Coulomb screening
analytically. \ See \ Eqs (\ref{AlphaLayerScrxxFromEpsMacroxx}) and (\ref%
{Aplha2DScrzzFromEpsMacro}).

\ We then used microscopic analytic theory to obtain the finite-wavenumber
$O( Q)$ corrections to $\alpha_{ii}^{2D,scr}(Q,iu)$ in the
presence of all intra-layer local-field ($\vec{G}\neq \vec{0})$
effects, without further numerical input. See eqs (\ref{1stOrderScrndAlphaxx}%
) and (\ref{1stOrderScreenedAplhazz}).

For each type of monolayer, we then obtained further 
$O(Q^{2}) $ corrections (see \ Eqs (\ref{OqsquaredAlpxx}) and (\ref%
{OqsquaredAlpzz})) to $\alpha _{ii}^{2D,scr}$ by fitting to a single
point $D=D_{00}\approx D_{0}$ on the curve of the interlayer correlation
energy $E_{c}\left( D\right) $of an infinite stack of the same layer
species, from RPA energy calculations using VASP. These large numerical RPA
energy calculations were tractable because only a small unit cell is needed to
represent such a homogeneous periodic stack. They need to be performed only
once for each monolayer species, and are then used for all subsequent
calculations involving that species with multiple layer types and other
species. \ This spatially nonlocal physics represented in the 
$O(Q^{2}) $ correction puts the theory well outside the normal scope
of the traditional Lifshitz approach to dispersion forces, and in particular
we obtained good energetics right down to equilibrium binding geometry.

We then used our layer polarizabiltes through $O(Q^{2})$ to evaluate the 
inter-layer correlation energies of layered systems (including the dispersion
 energy) within the RPA. This could be done semi-analytically because of the 
simplified $O(Q^{2})$ form of the response functions.  See 
(\ref{EInterlayerasTracePm}), (\ref{Evdw2LayersFromRs}), 
(\ref{EvdW2LayersFromAlphas}), (\ref{EPerAreaPerLayerAsyInsulator2Layer}),
(\ref{ExplicitEcPerVolUnifStack}) and 
(\ref{Ec)2)InfStackAsLayerPairSum}).

We found that for layer systems of the types considered, the "standard" \emph{single-power} $D^{-p}$ asymptotic forms 
of interaction energy (see (\ref{EPerAreaPerLayerAsyInsulator2Layer}), (\ref{DMinus3})) are reasonably correct 
only for $D > 10$ nm and are certainly not valid 
at inter-layer separations  $D$ less than $1\,nm$ where the correlation energy per atom is of $O(1\, meV)$ or more. 
That is, they are invalid in the regime where codes such as VASP can give meaningful correlation energy results, 
given practical numerical error considerations. For this regime of relatively small separations, 
it was necessary to consider the response of each layer through second order in the wavenumber component $Q$ 
parallel to the layer. This reflects the weak Coulomb screening within a 2D layer as  $Q\rightarrow 0$, 
and the consequent strong $Q$ dependence (spatially non-local character) of the screened layer response.  
This is in contrast to thick layers, where dielectric prpperties are spatially local in the $Q=0$ limit, 
and where Lifshitz theory is valid down to moderate non-contacting separations.      

We gave the name "Layer Response Theory" (LRT)  to this approach that 
obtains semi-analytically the correlation component of RPA-level interactions
involving monolayers, by determining  
the polarizabilies $a_{ii}^{2D,scr}\left( Q,iu\right)$ of each
monolayer species through $O( Q^{2})$.
  As first examples of our LRT method,
here we calculated the following quantities semi-analytically, finding
excellent agreement with much larger full RPA correlation calculations.

(i) The interlayer correlation energy curve $E_{c}\left( D\right) $ for
stretched bulk BN, obtaining excellent agreement at all $D$ values with
large numerical RPA calculations (see eqs (\ref{Ec)2)InfStackAsLayerPairSum}%
),(\ref{EcTwoLaersLogExpanded}), (\ref{OqsquaredAlpxx}), (\ref{OqsquaredAlpzz}) and Fig \ref{fig:EcBulkBN});

(ii) As in (i) but with neglect of the layer-perpendicular polarizability
component $\alpha _{zz}^{2D,scr}$ (see Fig . 
\ref{fig:EcBulkBNwithoutAlphazz}). This figure shows that both the parallel and
perpendicular polarizabilities of BN  layers contribute significantly to the
interlayer correlation energy. We found a similar situation for
graphene (not shown here).

(iii) The total interlayer energy $E_{tot}\left( D\right) $ for stretched
bulk BN, using the exact exchange energy \ from a large VASP\ calculation.
This shows that our semi-analytic correlation energy is good enough for
quantitative predictions even near to the equilibrium spacing. (See Fig \ref%
{fig:EtotBulkBN})

(iv) The interlayer correlation energy of an infinite stack as in (i), but
for graphite instead of BN. See Eqs (\ref{Ec)2)InfStackAsLayerPairSum}) and (\ref%
{RForGraphene}) and Fig \ref{fig:Ec(D)stretchedGraphite}. Agreement with large 
numerical RPA calculations is excellent.

(v) The total interlayer energy of stretched graphite using our Layer
Response Theory for correlation, plus exact exchange from VASP: see Fig \ref%
{fig:Etot(D)Graphite}.  Agreement with large VASP total RPA energy calculations  
is excellent.

(vi) The correlation and total interlayer \ energies for an infinite
BN-gr-BN-gr-BN-gr...hetero-stack (Eqs (\ref{Alpha1TotBNgr}), 
(\ref{Alpha2TotBNgr}),(\ref{Ec)2)InfStackAsLayerPairSum}), 
(\ref{EcTwoLaersLogExpanded}) and Figs \ref{fig:Ec(D)BNGrapheneStack}, \ref%
{fig:Etot(D)BNGrapheneStack}).  Once more agreement with a large VASP 
RPA energy calculation was excellent, and here we did not use any external 
input data for this specific heterostructure. All data (one parameter each) 
came from VASP calculations of the pure graphene and pure BN stacks.

The energies of physisorbed molecular species on graphene constitute another application that
we will discuss elsewhere. We will report successful semi-analytic
calculations for this case also.

Our experience so far suggests that the correlation energies are remarkably
smooth functions of layer separation $D$ and not strongly dependent on layer-layer registry. A similar conclusion was reached by Loncaric and Despoja 
\cite{RPAGrapheneOnMedtalsDespojaPRB2014} where a metal surface near to a graphene sheet was successfully treated in the jellium model. This smoothness is less true for the Hartree
and exchange contributions, which contain most of the dependence on the
registration of atoms on adjacent layers.  \ Semilocal energy functionals \
can give a reasonable account of such registry energies near to contact
separations. With some empiricism, it is possible to meld our accurate
semi-analytic correlation energies with results from semi-local theories of
exchange and correlation. See for example \cite%
{SimpleModelGrapheneBindingGould2013}.  We will pursue this elsewhere.

Another promising direction is the inclusion of beyond-RPA effects in the
correlation energy via simplified exchange-correlation kernels. See for
example \cite{RenormalizedKernelFxcOlsenThygesenPRB2013},\cite%
{OlsenThygesenEnergiesFrFxc_PRL2014},\cite{BeyondRPAOnTHeCheapTGouldJCP2012}.

\section{Acknowledgments}

We thank Prof. A. Rubio, Dr. T. Bjorkman, Prof. K. Thygesen, Prof. V. Despoja and Prof E. M. Gray for discussions.  
The authors benefited from 
Griffith University funding in the form of a Short Term Visiting 
Research Fellowship (2015). Earlier financial support came from Australian-French 
government funding under the FAST scheme. We also acknowledge financial support from CNRS (Centre National de la Recherche Scientifique) through
 the PICS program "2DvdW".  JFD acknowledges support from CNRS and University of Lorraine during visits to Nancy.

%\bibliographystyle{prsty}  seb
%bibliography{LayerTh19Dec,seb}
%\bibliography{LayerTh25Dec,seb}
%\bibliography{LayerTh30Dec,seb}
%\bibliography{LayerTh,seb}

\begin{thebibliography}{61}
\expandafter\ifx\csname natexlab\endcsname\relax\def\natexlab#1{#1}\fi
\expandafter\ifx\csname bibnamefont\endcsname\relax
  \def\bibnamefont#1{#1}\fi
\expandafter\ifx\csname bibfnamefont\endcsname\relax
  \def\bibfnamefont#1{#1}\fi
\expandafter\ifx\csname citenamefont\endcsname\relax
  \def\citenamefont#1{#1}\fi
\expandafter\ifx\csname url\endcsname\relax
  \def\url#1{\texttt{#1}}\fi
\expandafter\ifx\csname urlprefix\endcsname\relax\def\urlprefix{URL }\fi
\providecommand{\bibinfo}[2]{#2}
\providecommand{\eprint}[2][]{\url{#2}}

\bibitem[{\citenamefont{Novoselov et~al.}(2004)\citenamefont{Novoselov, Geim,
  Morozov, Jiang, Zhang, Dubonos, Grigorieva, and
  Firsov}}]{novoselov_electric_2004}
\bibinfo{author}{\bibfnamefont{K.~S.} \bibnamefont{Novoselov}},
  \bibinfo{author}{\bibfnamefont{A.~K.} \bibnamefont{Geim}},
  \bibinfo{author}{\bibfnamefont{S.~V.} \bibnamefont{Morozov}},
  \bibinfo{author}{\bibfnamefont{D.}~\bibnamefont{Jiang}},
  \bibinfo{author}{\bibfnamefont{Y.}~\bibnamefont{Zhang}},
  \bibinfo{author}{\bibfnamefont{S.~V.} \bibnamefont{Dubonos}},
  \bibinfo{author}{\bibfnamefont{I.~V.} \bibnamefont{Grigorieva}},
  \bibnamefont{and} \bibinfo{author}{\bibfnamefont{A.~A.}
  \bibnamefont{Firsov}}, \bibinfo{journal}{Science}
  \textbf{\bibinfo{volume}{306}}, \bibinfo{pages}{666} (\bibinfo{year}{2004}).

\bibitem[{\citenamefont{Novoselov et~al.}(2005)\citenamefont{Novoselov, Jiang,
  Schedin, Booth, Khotkevich, Morozov, and
  Geim}}]{novoselov_two-dimensional_2005}
\bibinfo{author}{\bibfnamefont{K.~S.} \bibnamefont{Novoselov}},
  \bibinfo{author}{\bibfnamefont{D.}~\bibnamefont{Jiang}},
  \bibinfo{author}{\bibfnamefont{F.}~\bibnamefont{Schedin}},
  \bibinfo{author}{\bibfnamefont{T.~J.} \bibnamefont{Booth}},
  \bibinfo{author}{\bibfnamefont{V.~V.} \bibnamefont{Khotkevich}},
  \bibinfo{author}{\bibfnamefont{S.~V.} \bibnamefont{Morozov}},
  \bibnamefont{and} \bibinfo{author}{\bibfnamefont{A.~K.} \bibnamefont{Geim}},
  \bibinfo{journal}{Proceedings of the National Academy of Sciences of the
  United States of America} \textbf{\bibinfo{volume}{102}},
  \bibinfo{pages}{10451} (\bibinfo{year}{2005}).

\bibitem[{\citenamefont{Coleman et~al.}(2011)\citenamefont{Coleman, Lotya,
  O'Neill, Bergin, King, Khan, Young, Gaucher, De, Smith
  et~al.}}]{coleman_two-dimensional_2011}
\bibinfo{author}{\bibfnamefont{J.~N.} \bibnamefont{Coleman}},
  \bibinfo{author}{\bibfnamefont{M.}~\bibnamefont{Lotya}},
  \bibinfo{author}{\bibfnamefont{A.}~\bibnamefont{O'Neill}},
  \bibinfo{author}{\bibfnamefont{S.~D.} \bibnamefont{Bergin}},
  \bibinfo{author}{\bibfnamefont{P.~J.} \bibnamefont{King}},
  \bibinfo{author}{\bibfnamefont{U.}~\bibnamefont{Khan}},
  \bibinfo{author}{\bibfnamefont{K.}~\bibnamefont{Young}},
  \bibinfo{author}{\bibfnamefont{A.}~\bibnamefont{Gaucher}},
  \bibinfo{author}{\bibfnamefont{S.}~\bibnamefont{De}},
  \bibinfo{author}{\bibfnamefont{R.~J.} \bibnamefont{Smith}},
  \bibnamefont{et~al.}, \bibinfo{journal}{Science}
  \textbf{\bibinfo{volume}{331}}, \bibinfo{pages}{568} (\bibinfo{year}{2011}).

\bibitem[{\citenamefont{Mak et~al.}(2010)\citenamefont{Mak, Lee, Hone, Shan,
  and Heinz}}]{mak_atomically_2010}
\bibinfo{author}{\bibfnamefont{K.~F.} \bibnamefont{Mak}},
  \bibinfo{author}{\bibfnamefont{C.}~\bibnamefont{Lee}},
  \bibinfo{author}{\bibfnamefont{J.}~\bibnamefont{Hone}},
  \bibinfo{author}{\bibfnamefont{J.}~\bibnamefont{Shan}}, \bibnamefont{and}
  \bibinfo{author}{\bibfnamefont{T.~F.} \bibnamefont{Heinz}},
  \bibinfo{journal}{Physical Review Letters} \textbf{\bibinfo{volume}{105}},
  \bibinfo{pages}{136805} (\bibinfo{year}{2010}).

\bibitem[{\citenamefont{Geim and Novoselov}(2007)}]{geim_rise_2007}
\bibinfo{author}{\bibfnamefont{A.~K.} \bibnamefont{Geim}} \bibnamefont{and}
  \bibinfo{author}{\bibfnamefont{K.~S.} \bibnamefont{Novoselov}},
  \bibinfo{journal}{Nature Materials} \textbf{\bibinfo{volume}{6}},
  \bibinfo{pages}{183} (\bibinfo{year}{2007}).

\bibitem[{\citenamefont{Castro~Neto et~al.}(2009)\citenamefont{Castro~Neto,
  Guinea, Peres, Novoselov, and Geim}}]{castro_neto_electronic_2009}
\bibinfo{author}{\bibfnamefont{A.~H.} \bibnamefont{Castro~Neto}},
  \bibinfo{author}{\bibfnamefont{F.}~\bibnamefont{Guinea}},
  \bibinfo{author}{\bibfnamefont{N.~M.~R.} \bibnamefont{Peres}},
  \bibinfo{author}{\bibfnamefont{K.~S.} \bibnamefont{Novoselov}},
  \bibnamefont{and} \bibinfo{author}{\bibfnamefont{A.~K.} \bibnamefont{Geim}},
  \bibinfo{journal}{Reviews of Modern Physics} \textbf{\bibinfo{volume}{81}},
  \bibinfo{pages}{109} (\bibinfo{year}{2009}).

\bibitem[{\citenamefont{Xu et~al.}(2013)\citenamefont{Xu, Liang, Shi, and
  Chen}}]{xu_graphene-like_2013}
\bibinfo{author}{\bibfnamefont{M.}~\bibnamefont{Xu}},
  \bibinfo{author}{\bibfnamefont{T.}~\bibnamefont{Liang}},
  \bibinfo{author}{\bibfnamefont{M.}~\bibnamefont{Shi}}, \bibnamefont{and}
  \bibinfo{author}{\bibfnamefont{H.}~\bibnamefont{Chen}},
  \bibinfo{journal}{Chemical Reviews} \textbf{\bibinfo{volume}{113}},
  \bibinfo{pages}{3766} (\bibinfo{year}{2013}).

\bibitem[{\citenamefont{Liao et~al.}(2014)\citenamefont{Liao, Peng, and
  Liu}}]{liao_chemistry_2014}
\bibinfo{author}{\bibfnamefont{L.}~\bibnamefont{Liao}},
  \bibinfo{author}{\bibfnamefont{H.}~\bibnamefont{Peng}}, \bibnamefont{and}
  \bibinfo{author}{\bibfnamefont{Z.}~\bibnamefont{Liu}},
  \bibinfo{journal}{Journal of the American Chemical Society}
  \textbf{\bibinfo{volume}{136}}, \bibinfo{pages}{12194}
  (\bibinfo{year}{2014}).

\bibitem[{\citenamefont{Geim and Grigorieva}(2013)}]{geim_van_2013}
\bibinfo{author}{\bibfnamefont{A.~K.} \bibnamefont{Geim}} \bibnamefont{and}
  \bibinfo{author}{\bibfnamefont{I.~V.} \bibnamefont{Grigorieva}},
  \bibinfo{journal}{Nature} \textbf{\bibinfo{volume}{499}},
  \bibinfo{pages}{419} (\bibinfo{year}{2013}).

\bibitem[{\citenamefont{Amft et~al.}(2011)\citenamefont{Amft, Leb\`egue,
  Eriksson, and Skorodumova}}]{amft}
\bibinfo{author}{\bibfnamefont{M.}~\bibnamefont{Amft}},
  \bibinfo{author}{\bibfnamefont{S.}~\bibnamefont{Leb\`egue}},
  \bibinfo{author}{\bibfnamefont{O.}~\bibnamefont{Eriksson}}, \bibnamefont{and}
  \bibinfo{author}{\bibfnamefont{N.~V.} \bibnamefont{Skorodumova}},
  \bibinfo{journal}{J. Phys. Cond. Mat.} \textbf{\bibinfo{volume}{23}},
  \bibinfo{pages}{395001} (\bibinfo{year}{2011}).

\bibitem[{\citenamefont{Javaid et~al.}(2013)\citenamefont{Javaid, Leb\`egue,
  Detlefs, Ibrahim, Djeghloul, Bowen, Boukari, Miyamachi, Arabski, Spor
  et~al.}}]{mebphthalo}
\bibinfo{author}{\bibfnamefont{S.}~\bibnamefont{Javaid}},
  \bibinfo{author}{\bibfnamefont{S.}~\bibnamefont{Leb\`egue}},
  \bibinfo{author}{\bibfnamefont{B.}~\bibnamefont{Detlefs}},
  \bibinfo{author}{\bibfnamefont{F.}~\bibnamefont{Ibrahim}},
  \bibinfo{author}{\bibfnamefont{F.}~\bibnamefont{Djeghloul}},
  \bibinfo{author}{\bibfnamefont{M.}~\bibnamefont{Bowen}},
  \bibinfo{author}{\bibfnamefont{S.}~\bibnamefont{Boukari}},
  \bibinfo{author}{\bibfnamefont{T.}~\bibnamefont{Miyamachi}},
  \bibinfo{author}{\bibfnamefont{J.}~\bibnamefont{Arabski}},
  \bibinfo{author}{\bibfnamefont{D.}~\bibnamefont{Spor}}, \bibnamefont{et~al.},
  \bibinfo{journal}{Phys. Rev. B} \textbf{\bibinfo{volume}{87}},
  \bibinfo{pages}{155418} (\bibinfo{year}{2013}).

\bibitem[{\citenamefont{Liu et~al.}(2012)\citenamefont{Liu, Carrasco, Santra,
  Michaelides, Scheffler, and Tkatchenko}}]{liu_benzene_2012}
\bibinfo{author}{\bibfnamefont{W.}~\bibnamefont{Liu}},
  \bibinfo{author}{\bibfnamefont{J.}~\bibnamefont{Carrasco}},
  \bibinfo{author}{\bibfnamefont{B.}~\bibnamefont{Santra}},
  \bibinfo{author}{\bibfnamefont{A.}~\bibnamefont{Michaelides}},
  \bibinfo{author}{\bibfnamefont{M.}~\bibnamefont{Scheffler}},
  \bibnamefont{and}
  \bibinfo{author}{\bibfnamefont{A.}~\bibnamefont{Tkatchenko}},
  \bibinfo{journal}{Physical Review B} \textbf{\bibinfo{volume}{86}},
  \bibinfo{pages}{245405} (\bibinfo{year}{2012}).

\bibitem[{\citenamefont{Loncaric and
  Despoja}(2014)}]{RPAGrapheneOnMedtalsDespojaPRB2014}
\bibinfo{author}{\bibfnamefont{I.}~\bibnamefont{Loncaric}} \bibnamefont{and}
  \bibinfo{author}{\bibfnamefont{V.}~\bibnamefont{Despoja}},
  \bibinfo{journal}{Phys. Rev. B} \textbf{\bibinfo{volume}{90}},
  \bibinfo{pages}{075414} (\bibinfo{year}{2014}).

\bibitem[{\citenamefont{Zhu and L}(2014)}]{HAbsGrapheneOrigamiZhuACSNano2014}
\bibinfo{author}{\bibfnamefont{S.}~\bibnamefont{Zhu}} \bibnamefont{and}
  \bibinfo{author}{\bibfnamefont{T.}~\bibnamefont{L}}, \bibinfo{journal}{ACS
  Nano} \textbf{\bibinfo{volume}{8}}, \bibinfo{pages}{2864}
  (\bibinfo{year}{2014}).

\bibitem[{\citenamefont{Cunning et~al.}(2014)\citenamefont{Cunning, Ahmed,
  Mishra1, Kermany, Wood, and
  Iacopi}}]{GraphitizedSiCMicrobeamsCunningNanotech2014}
\bibinfo{author}{\bibfnamefont{B.~V.} \bibnamefont{Cunning}},
  \bibinfo{author}{\bibfnamefont{M.}~\bibnamefont{Ahmed}},
  \bibinfo{author}{\bibfnamefont{N.}~\bibnamefont{Mishra1}},
  \bibinfo{author}{\bibfnamefont{A.~R.} \bibnamefont{Kermany}},
  \bibinfo{author}{\bibfnamefont{B.}~\bibnamefont{Wood}}, \bibnamefont{and}
  \bibinfo{author}{\bibfnamefont{F.}~\bibnamefont{Iacopi}},
  \bibinfo{journal}{Nanotechnology} \textbf{\bibinfo{volume}{25}},
  \bibinfo{pages}{25301} (\bibinfo{year}{2014}).

\bibitem[{\citenamefont{Tao et~al.}(2015)\citenamefont{Tao, Cinquanta, Chiappe,
  Grazianetti, Fanciulli, and Dubey}}]{SiliceneTransistorsThTao}
\bibinfo{author}{\bibfnamefont{L.}~\bibnamefont{Tao}},
  \bibinfo{author}{\bibfnamefont{E.}~\bibnamefont{Cinquanta}},
  \bibinfo{author}{\bibfnamefont{D.}~\bibnamefont{Chiappe}},
  \bibinfo{author}{\bibfnamefont{C.}~\bibnamefont{Grazianetti}},
  \bibinfo{author}{\bibfnamefont{M.}~\bibnamefont{Fanciulli}},
  \bibnamefont{and} \bibinfo{author}{\bibfnamefont{M.}~\bibnamefont{Dubey}},
  \bibinfo{journal}{Nature Nanotechnology} \textbf{\bibinfo{volume}{10}},
  \bibinfo{pages}{227} (\bibinfo{year}{2015}).

\bibitem[{\citenamefont{Iacopi et~al.}(2015)\citenamefont{Iacopi, Mishra,
  Cunning, Goding, Dimitrijev, Brock, Dauskardt, wood, and
  Boeckl}}]{CatalGrapheneOnSiC_Si_Iacopi2015}
\bibinfo{author}{\bibfnamefont{F.}~\bibnamefont{Iacopi}},
  \bibinfo{author}{\bibfnamefont{N.}~\bibnamefont{Mishra}},
  \bibinfo{author}{\bibfnamefont{B.~V.} \bibnamefont{Cunning}},
  \bibinfo{author}{\bibfnamefont{D.}~\bibnamefont{Goding}},
  \bibinfo{author}{\bibfnamefont{S.}~\bibnamefont{Dimitrijev}},
  \bibinfo{author}{\bibfnamefont{R.}~\bibnamefont{Brock}},
  \bibinfo{author}{\bibfnamefont{R.~H.} \bibnamefont{Dauskardt}},
  \bibinfo{author}{\bibfnamefont{B.}~\bibnamefont{wood}}, \bibnamefont{and}
  \bibinfo{author}{\bibfnamefont{J.}~\bibnamefont{Boeckl}},
  \bibinfo{journal}{J. Mater. Res} \textbf{\bibinfo{volume}{30}},
  \bibinfo{pages}{609} (\bibinfo{year}{2015}).

\bibitem[{\citenamefont{Antony et~al.}(2010)\citenamefont{Antony, Grimme,
  Ehrlich, and Krieg}}]{Grimme+FT+DWithBondCountingJCP2010}
\bibinfo{author}{\bibfnamefont{J.}~\bibnamefont{Antony}},
  \bibinfo{author}{\bibfnamefont{S.}~\bibnamefont{Grimme}},
  \bibinfo{author}{\bibfnamefont{S.}~\bibnamefont{Ehrlich}}, \bibnamefont{and}
  \bibinfo{author}{\bibfnamefont{H.}~\bibnamefont{Krieg}}, \bibinfo{journal}{J.
  Chem. Phys.} \textbf{\bibinfo{volume}{132}}, \bibinfo{pages}{154104}
  (\bibinfo{year}{2010}).

\bibitem[{\citenamefont{Sure et~al.}(2014)\citenamefont{Sure, Antony, and
  Grimme}}]{AchievmentsShortcomingsDFT_D3_chargedSupramolecular}
\bibinfo{author}{\bibfnamefont{R.}~\bibnamefont{Sure}},
  \bibinfo{author}{\bibfnamefont{J.}~\bibnamefont{Antony}}, \bibnamefont{and}
  \bibinfo{author}{\bibfnamefont{S.}~\bibnamefont{Grimme}},
  \bibinfo{journal}{J. Phys. Chem. B} \textbf{\bibinfo{volume}{118}},
  \bibinfo{pages}{3431} (\bibinfo{year}{2014}).

\bibitem[{\citenamefont{Grimme}(2006)}]{Grimme:06}
\bibinfo{author}{\bibfnamefont{S.}~\bibnamefont{Grimme}},
  \bibinfo{journal}{Journal of Computational Chemistry}
  \textbf{\bibinfo{volume}{27}}, \bibinfo{pages}{1787} (\bibinfo{year}{2006}).

\bibitem[{\citenamefont{Tkatchenko and Scheffler}(2009)}]{Tkatchenko:09}
\bibinfo{author}{\bibfnamefont{A.}~\bibnamefont{Tkatchenko}} \bibnamefont{and}
  \bibinfo{author}{\bibfnamefont{M.}~\bibnamefont{Scheffler}},
  \bibinfo{journal}{Physical Review Letters} \textbf{\bibinfo{volume}{102}},
  \bibinfo{pages}{073005} (\bibinfo{year}{2009}).

\bibitem[{\citenamefont{Grimme et~al.}(2010)\citenamefont{Grimme, Antony,
  Ehrlich, and Krieg}}]{Grimme:10}
\bibinfo{author}{\bibfnamefont{S.}~\bibnamefont{Grimme}},
  \bibinfo{author}{\bibfnamefont{J.}~\bibnamefont{Antony}},
  \bibinfo{author}{\bibfnamefont{S.}~\bibnamefont{Ehrlich}}, \bibnamefont{and}
  \bibinfo{author}{\bibfnamefont{H.}~\bibnamefont{Krieg}}, \bibinfo{journal}{J.
  Chem. Phys.} \textbf{\bibinfo{volume}{132}}, \bibinfo{pages}{154104}
  (\bibinfo{year}{2010}).

\bibitem[{\citenamefont{Kim et~al.}(2012)\citenamefont{Kim, Choi, and
  Goddard}}]{kim_universal_2012}
\bibinfo{author}{\bibfnamefont{H.}~\bibnamefont{Kim}},
  \bibinfo{author}{\bibfnamefont{J.-M.} \bibnamefont{Choi}}, \bibnamefont{and}
  \bibinfo{author}{\bibfnamefont{W.~A.} \bibnamefont{Goddard}},
  \bibinfo{journal}{The Journal of Physical Chemistry Letters}
  \textbf{\bibinfo{volume}{3}}, \bibinfo{pages}{360} (\bibinfo{year}{2012}).

\bibitem[{\citenamefont{Dion et~al.}(2004)\citenamefont{Dion, Rydberg,
  Schr\"{o}der, Langreth, and Lundqvist}}]{vdWFnalGenGeomDionPRL04}
\bibinfo{author}{\bibfnamefont{M.}~\bibnamefont{Dion}},
  \bibinfo{author}{\bibfnamefont{H.}~\bibnamefont{Rydberg}},
  \bibinfo{author}{\bibfnamefont{E.}~\bibnamefont{Schr\"{o}der}},
  \bibinfo{author}{\bibfnamefont{D.~C.} \bibnamefont{Langreth}},
  \bibnamefont{and} \bibinfo{author}{\bibfnamefont{B.~I.}
  \bibnamefont{Lundqvist}}, \bibinfo{journal}{Phys. Rev. Lett.}
  \textbf{\bibinfo{volume}{92}}, \bibinfo{pages}{246401}
  (\bibinfo{year}{2004}).

\bibitem[{\citenamefont{Dobson et~al.}(2006)\citenamefont{Dobson, White, and
  Rubio}}]{AsyDispIntPRL2006JFDWhiteRubio}
\bibinfo{author}{\bibfnamefont{J.~F.} \bibnamefont{Dobson}},
  \bibinfo{author}{\bibfnamefont{A.}~\bibnamefont{White}}, \bibnamefont{and}
  \bibinfo{author}{\bibfnamefont{A.}~\bibnamefont{Rubio}},
  \bibinfo{journal}{Phys. Rev. Lett.} \textbf{\bibinfo{volume}{96}},
  \bibinfo{pages}{073201} (\bibinfo{year}{2006}).

\bibitem[{\citenamefont{Tkatchenko et~al.}(2012)\citenamefont{Tkatchenko,
  DiStasio, Jr., Car, and Scheffler}}]{TkatchenkoMB_vdW_PRL2012}
\bibinfo{author}{\bibfnamefont{A.}~\bibnamefont{Tkatchenko}},
  \bibinfo{author}{\bibfnamefont{R.~A.} \bibnamefont{DiStasio}},
  \bibinfo{author}{\bibnamefont{Jr.}},
  \bibinfo{author}{\bibfnamefont{R.}~\bibnamefont{Car}}, \bibnamefont{and}
  \bibinfo{author}{\bibfnamefont{M.}~\bibnamefont{Scheffler}},
  \bibinfo{journal}{Phys. Rev. Lett.} \textbf{\bibinfo{volume}{108}},
  \bibinfo{pages}{236402} (\bibinfo{year}{2012}).

\bibitem[{\citenamefont{Gobre and
  Tkatchenko}(2013)}]{GobreTkatvdWscalingNatComm2013}
\bibinfo{author}{\bibfnamefont{V.}~\bibnamefont{Gobre}} \bibnamefont{and}
  \bibinfo{author}{\bibfnamefont{A.}~\bibnamefont{Tkatchenko}},
  \bibinfo{journal}{Nature Communications} \textbf{\bibinfo{volume}{4}},
  \bibinfo{pages}{2341} (\bibinfo{year}{2013}).

\bibitem[{\citenamefont{Sernelius}(2011)}]{CasimirgrapheneSernelius2011}
\bibinfo{author}{\bibfnamefont{B.}~\bibnamefont{Sernelius}},
  \bibinfo{journal}{Europhys. Letts.} \textbf{\bibinfo{volume}{95}},
  \bibinfo{pages}{57003} (\bibinfo{year}{2011}).

\bibitem[{\citenamefont{Ruzsinszky et~al.}(2012)\citenamefont{Ruzsinszky,
  Perdew, Tao, Csonka, and
  Pitarke}}]{vdWFullerenesDefyConvWisdomRuzsEtalPRL2012}
\bibinfo{author}{\bibfnamefont{A.}~\bibnamefont{Ruzsinszky}},
  \bibinfo{author}{\bibfnamefont{J.}~\bibnamefont{Perdew}},
  \bibinfo{author}{\bibfnamefont{J.}~\bibnamefont{Tao}},
  \bibinfo{author}{\bibfnamefont{G.~I.} \bibnamefont{Csonka}},
  \bibnamefont{and} \bibinfo{author}{\bibfnamefont{J.~M.}
  \bibnamefont{Pitarke}}, \bibinfo{journal}{Phys. Rev. Lett.}
  \textbf{\bibinfo{volume}{109}}, \bibinfo{pages}{233203}
  (\bibinfo{year}{2012}).

\bibitem[{\citenamefont{Dobson and Wang}(1999)}]{DobsonWangPRL99}
\bibinfo{author}{\bibfnamefont{J.~F.} \bibnamefont{Dobson}} \bibnamefont{and}
  \bibinfo{author}{\bibfnamefont{J.}~\bibnamefont{Wang}},
  \bibinfo{journal}{Phys. Rev. Lett.} \textbf{\bibinfo{volume}{82}},
  \bibinfo{pages}{2123} (\bibinfo{year}{1999}).

\bibitem[{\citenamefont{Marini et~al.}(2006)\citenamefont{Marini,
  Garcia-Gonzalez, and Rubio}}]{RPAEnergyBN_MariniRubioPRL2006}
\bibinfo{author}{\bibfnamefont{A.}~\bibnamefont{Marini}},
  \bibinfo{author}{\bibfnamefont{P.}~\bibnamefont{Garcia-Gonzalez}},
  \bibnamefont{and} \bibinfo{author}{\bibfnamefont{A.}~\bibnamefont{Rubio}},
  \bibinfo{journal}{Phys. Rev. Lett.} \textbf{\bibinfo{volume}{96}},
  \bibinfo{pages}{136404} (\bibinfo{year}{2006}).

\bibitem[{\citenamefont{Harl and
  Kresse}(2009)}]{RPABulkPropertiesHarlKressePRL09}
\bibinfo{author}{\bibfnamefont{J.}~\bibnamefont{Harl}} \bibnamefont{and}
  \bibinfo{author}{\bibfnamefont{G.}~\bibnamefont{Kresse}},
  \bibinfo{journal}{Phys. Rev. Lett.} \textbf{\bibinfo{volume}{103}},
  \bibinfo{pages}{056401} (\bibinfo{year}{2009}).

\bibitem[{\citenamefont{Lebegue et~al.}(2010)\citenamefont{Lebegue, Harl,
  Gould, Angyan, Kresse, and Dobson}}]{GraphiteDispIntPRL2010LebegueEtal}
\bibinfo{author}{\bibfnamefont{S.}~\bibnamefont{Lebegue}},
  \bibinfo{author}{\bibfnamefont{J.}~\bibnamefont{Harl}},
  \bibinfo{author}{\bibfnamefont{T.}~\bibnamefont{Gould}},
  \bibinfo{author}{\bibfnamefont{J.~G.} \bibnamefont{Angyan}},
  \bibinfo{author}{\bibfnamefont{G.}~\bibnamefont{Kresse}}, \bibnamefont{and}
  \bibinfo{author}{\bibfnamefont{J.~F.} \bibnamefont{Dobson}},
  \bibinfo{journal}{Phys. Rev. Lett.} \textbf{\bibinfo{volume}{105}},
  \bibinfo{pages}{196401} (\bibinfo{year}{2010}).

\bibitem[{\citenamefont{Lu et~al.}(2009)\citenamefont{Lu, Li, Rocca, and
  Galli}}]{Lu:09}
\bibinfo{author}{\bibfnamefont{D.}~\bibnamefont{Lu}},
  \bibinfo{author}{\bibfnamefont{Y.}~\bibnamefont{Li}},
  \bibinfo{author}{\bibfnamefont{D.}~\bibnamefont{Rocca}}, \bibnamefont{and}
  \bibinfo{author}{\bibfnamefont{G.}~\bibnamefont{Galli}},
  \bibinfo{journal}{Physical Review Letters} \textbf{\bibinfo{volume}{102}},
  \bibinfo{pages}{206411} (\bibinfo{year}{2009}).

\bibitem[{\citenamefont{Gruneis et~al.}(2009)\citenamefont{Gruneis, Marsman,
  Harl, Schimka, and Kresse}}]{SOSEXtestsGruneis2009}
\bibinfo{author}{\bibfnamefont{A.}~\bibnamefont{Gruneis}},
  \bibinfo{author}{\bibfnamefont{M.}~\bibnamefont{Marsman}},
  \bibinfo{author}{\bibfnamefont{J.}~\bibnamefont{Harl}},
  \bibinfo{author}{\bibfnamefont{L.}~\bibnamefont{Schimka}}, \bibnamefont{and}
  \bibinfo{author}{\bibfnamefont{G.}~\bibnamefont{Kresse}},
  \bibinfo{journal}{J. Chem. Phys.} \textbf{\bibinfo{volume}{131}},
  \bibinfo{pages}{154115} (\bibinfo{year}{2009}).

\bibitem[{\citenamefont{Olsen and
  S.Thygesen}(2014)}]{OlsenThygesenEnergiesFrFxc_PRL2014}
\bibinfo{author}{\bibfnamefont{T.}~\bibnamefont{Olsen}} \bibnamefont{and}
  \bibinfo{author}{\bibfnamefont{K.}~\bibnamefont{S.Thygesen}},
  \bibinfo{journal}{Phys. rev. Lett.} \textbf{\bibinfo{volume}{112}},
  \bibinfo{pages}{203001} (\bibinfo{year}{2014}).

\bibitem[{\citenamefont{Dobson et~al.}(2014)\citenamefont{Dobson, Gould, and
  Vignale}}]{HowMBTAffectsGraphenevdW}
\bibinfo{author}{\bibfnamefont{J.~F.} \bibnamefont{Dobson}},
  \bibinfo{author}{\bibfnamefont{T.}~\bibnamefont{Gould}}, \bibnamefont{and}
  \bibinfo{author}{\bibfnamefont{G.}~\bibnamefont{Vignale}},
  \bibinfo{journal}{Phys. Rev. X} \textbf{\bibinfo{volume}{4}},
  \bibinfo{pages}{021040} (\bibinfo{year}{2014}).

\bibitem[{\citenamefont{Bj¨orkman et~al.}(2012)\citenamefont{Bj¨orkman, Gulans,
  Krasheninnikov, and Nieminen}}]{AreWevdWreadyBjorkmanJPCM2012}
\bibinfo{author}{\bibfnamefont{T.~.} \bibnamefont{Bj¨orkman}},
  \bibinfo{author}{\bibfnamefont{A.}~\bibnamefont{Gulans}},
  \bibinfo{author}{\bibfnamefont{A.~V.} \bibnamefont{Krasheninnikov}},
  \bibnamefont{and} \bibinfo{author}{\bibfnamefont{R.~M.}
  \bibnamefont{Nieminen}}, \bibinfo{journal}{J. Phys.: Condens. matter}
  \textbf{\bibinfo{volume}{24}}, \bibinfo{pages}{424218}
  (\bibinfo{year}{2012}).

\bibitem[{\citenamefont{Olsen and
  Thygesen}(2013)}]{RPASolidsvdWCoval_OlsenThygesenPRB2013}
\bibinfo{author}{\bibfnamefont{T.}~\bibnamefont{Olsen}} \bibnamefont{and}
  \bibinfo{author}{\bibfnamefont{K.~S.} \bibnamefont{Thygesen}},
  \bibinfo{journal}{Phys. Rev. B} \textbf{\bibinfo{volume}{87}},
  \bibinfo{pages}{075111} (\bibinfo{year}{2013}).

\bibitem[{\citenamefont{Liu et~al.}(2011)\citenamefont{Liu, Angyan, and
  Dobson}}]{HChainsLiuAngyanDobsonJCP2011}
\bibinfo{author}{\bibfnamefont{R.-F.} \bibnamefont{Liu}},
  \bibinfo{author}{\bibfnamefont{J.~G.} \bibnamefont{Angyan}},
  \bibnamefont{and} \bibinfo{author}{\bibfnamefont{J.~F.}
  \bibnamefont{Dobson}}, \bibinfo{journal}{J. Chem. Phys}
  \textbf{\bibinfo{volume}{134}}, \bibinfo{pages}{114106}
  (\bibinfo{year}{2011}).

\bibitem[{\citenamefont{Dzyaloshinskii
  et~al.}(1961)\citenamefont{Dzyaloshinskii, Lifshitz, and
  Pitaevskii}}]{vdWGenThDzyaloshinskiilLifshitzPitaevski1961}
\bibinfo{author}{\bibfnamefont{I.~E.} \bibnamefont{Dzyaloshinskii}},
  \bibinfo{author}{\bibfnamefont{E.~M.} \bibnamefont{Lifshitz}},
  \bibnamefont{and} \bibinfo{author}{\bibfnamefont{L.~P.}
  \bibnamefont{Pitaevskii}}, \bibinfo{journal}{Adv. Phys.}
  \textbf{\bibinfo{volume}{10}}, \bibinfo{pages}{165} (\bibinfo{year}{1961}).

\bibitem[{\citenamefont{Parsegian}(2006)}]{ParsegianvdW}
\bibinfo{author}{\bibfnamefont{V.~A.} \bibnamefont{Parsegian}},
  \emph{\bibinfo{title}{van der Waals Forces: a handbook for biologists,
  chemists, engineers, and physicists}} (\bibinfo{publisher}{Cambridge
  University Press}, \bibinfo{address}{Cambridge}, \bibinfo{year}{2006}), ISBN
  \bibinfo{isbn}{0521839068, 9780521839068}.

\bibitem[{\citenamefont{Israelachvili}(1974)}]{vdWBooks3}
\bibinfo{author}{\bibfnamefont{J.}~\bibnamefont{Israelachvili}},
  \bibinfo{journal}{\emph{Intermolecular and Surface Forces}, (Academic,
  London, 1992): J. Mahanty and B. Ninham, \emph{Dispersion Forces}, (Academic,
  London, 1976): D. Langbein, \emph{Theory of van der Waals Attraction}
  (Springer-Verlag, Berlin)}  (\bibinfo{year}{1974}).

\bibitem[{\citenamefont{Rahi et~al.}(2009)\citenamefont{Rahi, Emig, Graham,
  Jaffe, and Kardar}}]{ScattThCasimirRahiEmigEtalPRD09}
\bibinfo{author}{\bibfnamefont{S.~J.} \bibnamefont{Rahi}},
  \bibinfo{author}{\bibfnamefont{T.}~\bibnamefont{Emig}},
  \bibinfo{author}{\bibfnamefont{N.}~\bibnamefont{Graham}},
  \bibinfo{author}{\bibfnamefont{R.~L.} \bibnamefont{Jaffe}}, \bibnamefont{and}
  \bibinfo{author}{\bibfnamefont{M.}~\bibnamefont{Kardar}},
  \bibinfo{journal}{Phys. Rev. D} \textbf{\bibinfo{volume}{80}},
  \bibinfo{pages}{085021} (\bibinfo{year}{2009}).

\bibitem[{\citenamefont{Scheel and
  Buhmann}(2008)}]{MacroQEDLifshitzScheelBuhmannSlovakJ2008}
\bibinfo{author}{\bibfnamefont{S.}~\bibnamefont{Scheel}} \bibnamefont{and}
  \bibinfo{author}{\bibfnamefont{S.~Y.} \bibnamefont{Buhmann}},
  \bibinfo{journal}{Acta Physica Slovaca} \textbf{\bibinfo{volume}{58}},
  \bibinfo{pages}{675} (\bibinfo{year}{2008}).

\bibitem[{\citenamefont{Buhmann}(2012{\natexlab{a}})}]{BuhmannDispersionFrcBk1}
\bibinfo{author}{\bibfnamefont{S.~Y.} \bibnamefont{Buhmann}},
  \emph{\bibinfo{title}{Dispersion Forces I}}
  (\bibinfo{publisher}{Springer-Verlag}, \bibinfo{address}{Berlin},
  \bibinfo{year}{2012}{\natexlab{a}}), ISBN \bibinfo{isbn}{978-3-642-32483-3,
  978-3-642-32484-0}.

\bibitem[{\citenamefont{Buhmann}(2012{\natexlab{b}})}]{BuhmannDispersionFrcBk2}
\bibinfo{author}{\bibfnamefont{S.~Y.} \bibnamefont{Buhmann}},
  \emph{\bibinfo{title}{Dispersion Forces II}}
  (\bibinfo{publisher}{Springer-Verlag}, \bibinfo{address}{Berlin},
  \bibinfo{year}{2012}{\natexlab{b}}), ISBN \bibinfo{isbn}{978-3-642-32465-9,
  978-3-642-32466-6}.

\bibitem[{\citenamefont{Dobson and
  Gould}(2012)}]{CalcDispersionEnergiesDobsonJPCM2012}
\bibinfo{author}{\bibfnamefont{J.~F.} \bibnamefont{Dobson}} \bibnamefont{and}
  \bibinfo{author}{\bibfnamefont{T.}~\bibnamefont{Gould}}, \bibinfo{journal}{J.
  Phys. Condens. Matter} \textbf{\bibinfo{volume}{24}}, \bibinfo{pages}{073201}
  (\bibinfo{year}{2012}).

\bibitem[{\citenamefont{Kresse and Joubert}(1999)}]{VASP1}
\bibinfo{author}{\bibfnamefont{G.}~\bibnamefont{Kresse}} \bibnamefont{and}
  \bibinfo{author}{\bibfnamefont{D.}~\bibnamefont{Joubert}},
  \bibinfo{journal}{Phys. rev. B} \textbf{\bibinfo{volume}{59}},
  \bibinfo{pages}{1758} (\bibinfo{year}{1999}).

\bibitem[{\citenamefont{Kresse and Furthmuller}(1996)}]{VASP2}
\bibinfo{author}{\bibfnamefont{G.}~\bibnamefont{Kresse}} \bibnamefont{and}
  \bibinfo{author}{\bibfnamefont{J.}~\bibnamefont{Furthmuller}},
  \bibinfo{journal}{Phys. rev. B} \textbf{\bibinfo{volume}{54}},
  \bibinfo{pages}{11169} (\bibinfo{year}{1996}).

\bibitem[{\citenamefont{Nazarov}(2015)}]{LayerResponseQuantitiesUsefulExptNazarovNJP2015}
\bibinfo{author}{\bibfnamefont{V.~U.} \bibnamefont{Nazarov}},
  \bibinfo{journal}{New J. Phys.} \textbf{\bibinfo{volume}{17}},
  \bibinfo{pages}{073018} (\bibinfo{year}{2015}).

\bibitem[{\citenamefont{Andersen et~al.}(2015)\citenamefont{Andersen, Latini,
  and Thygesen}}]{DielectricGenomevdWHetAndersenThygesenNanolett2015}
\bibinfo{author}{\bibfnamefont{K.}~\bibnamefont{Andersen}},
  \bibinfo{author}{\bibfnamefont{S.}~\bibnamefont{Latini}}, \bibnamefont{and}
  \bibinfo{author}{\bibfnamefont{K.~S.} \bibnamefont{Thygesen}},
  \bibinfo{journal}{Nanoletters} \textbf{\bibinfo{volume}{15}},
  \bibinfo{pages}{4616} (\bibinfo{year}{2015}).

\bibitem[{\citenamefont{Dobson and Dinte}(1996)}]{JFDDintvdW96}
\bibinfo{author}{\bibfnamefont{J.~F.} \bibnamefont{Dobson}} \bibnamefont{and}
  \bibinfo{author}{\bibfnamefont{B.~P.} \bibnamefont{Dinte}},
  \bibinfo{journal}{Phys. Rev. Lett.} \textbf{\bibinfo{volume}{76}},
  \bibinfo{pages}{1780 } (\bibinfo{year}{1996}).

\bibitem[{\citenamefont{Dobson et~al.}(1998)\citenamefont{Dobson, Dinte, and
  Wang}}]{DobsonBrisvdWChap}
\bibinfo{author}{\bibfnamefont{J.~F.} \bibnamefont{Dobson}},
  \bibinfo{author}{\bibfnamefont{B.}~\bibnamefont{Dinte}}, \bibnamefont{and}
  \bibinfo{author}{\bibfnamefont{J.}~\bibnamefont{Wang}}, in
  \emph{\bibinfo{booktitle}{Electronic Density Functional Theory: Recent
  Progress and New Directions}}, edited by
  \bibinfo{editor}{\bibfnamefont{J.~F.} \bibnamefont{Dobson}},
  \bibinfo{editor}{\bibfnamefont{G.}~\bibnamefont{Vignale}}, \bibnamefont{and}
  \bibinfo{editor}{\bibfnamefont{M.~P.} \bibnamefont{Das}}
  (\bibinfo{publisher}{Plenum}, \bibinfo{address}{New York},
  \bibinfo{year}{1998}), chap.~\bibinfo{chapter}{18}, p. \bibinfo{pages}{261}.

\bibitem[{\citenamefont{Abramowitz and Stegun}(1965)}]{AbramowitzStegun}
\bibinfo{author}{\bibfnamefont{M.}~\bibnamefont{Abramowitz}} \bibnamefont{and}
  \bibinfo{author}{\bibfnamefont{I.~A.} \bibnamefont{Stegun}},
  \emph{\bibinfo{title}{Handbook of mathematical functions}}
  (\bibinfo{publisher}{Dover}, \bibinfo{address}{Toronto},
  \bibinfo{year}{1965}).

\bibitem[{\citenamefont{Gould et~al.}(2013{\natexlab{a}})\citenamefont{Gould,
  Dobson, and Lebegue}}]{GouldDobsongrapheneConesPRB2013}
\bibinfo{author}{\bibfnamefont{T.}~\bibnamefont{Gould}},
  \bibinfo{author}{\bibfnamefont{J.~F.} \bibnamefont{Dobson}},
  \bibnamefont{and} \bibinfo{author}{\bibfnamefont{S.}~\bibnamefont{Lebegue}},
  \bibinfo{journal}{Phys. Rev. B} \textbf{\bibinfo{volume}{87}},
  \bibinfo{pages}{165422} (\bibinfo{year}{2013}{\natexlab{a}}).

\bibitem[{\citenamefont{Dobson}(2011)}]{DispInductInteractGrapheneNanostrDobsonSurfSci2011}
\bibinfo{author}{\bibfnamefont{J.~F.} \bibnamefont{Dobson}},
  \bibinfo{journal}{Surf. Sci.} \textbf{\bibinfo{volume}{605}},
  \bibinfo{pages}{960} (\bibinfo{year}{2011}).

\bibitem[{\citenamefont{Dobson}(2009)}]{ValidityComparisonvdWDobsonJCompThNanosci2009}
\bibinfo{author}{\bibfnamefont{J.~F.} \bibnamefont{Dobson}},
  \bibinfo{journal}{J. Comp. Theoret. Nanosci.} \textbf{\bibinfo{volume}{6}},
  \bibinfo{pages}{960} (\bibinfo{year}{2009}).

\bibitem[{\citenamefont{Gould et~al.}(2013{\natexlab{b}})\citenamefont{Gould,
  Lebegue, and Dobson}}]{SimpleModelGrapheneBindingGould2013}
\bibinfo{author}{\bibfnamefont{T.}~\bibnamefont{Gould}},
  \bibinfo{author}{\bibfnamefont{S.}~\bibnamefont{Lebegue}}, \bibnamefont{and}
  \bibinfo{author}{\bibfnamefont{J.~F.} \bibnamefont{Dobson}},
  \bibinfo{journal}{J. Phys. Condens. Matter} \textbf{\bibinfo{volume}{25}},
  \bibinfo{pages}{445010} (\bibinfo{year}{2013}{\natexlab{b}}).

\bibitem[{\citenamefont{Olsen and
  S.Thygesen}(2013)}]{RenormalizedKernelFxcOlsenThygesenPRB2013}
\bibinfo{author}{\bibfnamefont{T.}~\bibnamefont{Olsen}} \bibnamefont{and}
  \bibinfo{author}{\bibfnamefont{K.}~\bibnamefont{S.Thygesen}},
  \bibinfo{journal}{Phys. Rev, B} \textbf{\bibinfo{volume}{88}},
  \bibinfo{pages}{115131} (\bibinfo{year}{2013}).

\bibitem[{\citenamefont{Gould}(2012)}]{BeyondRPAOnTHeCheapTGouldJCP2012}
\bibinfo{author}{\bibfnamefont{T.}~\bibnamefont{Gould}}, \bibinfo{journal}{J.
  Chem. Phys} \textbf{\bibinfo{volume}{137}}, \bibinfo{pages}{111101}
  (\bibinfo{year}{2012}).

\end{thebibliography}

%FIGURE LABELS AND CAPTIONS

\newpage

\newpage
\begin{figure}[tbp]
% currently Fig 2

%\includegraphics{bothx2.eps}
\includegraphics{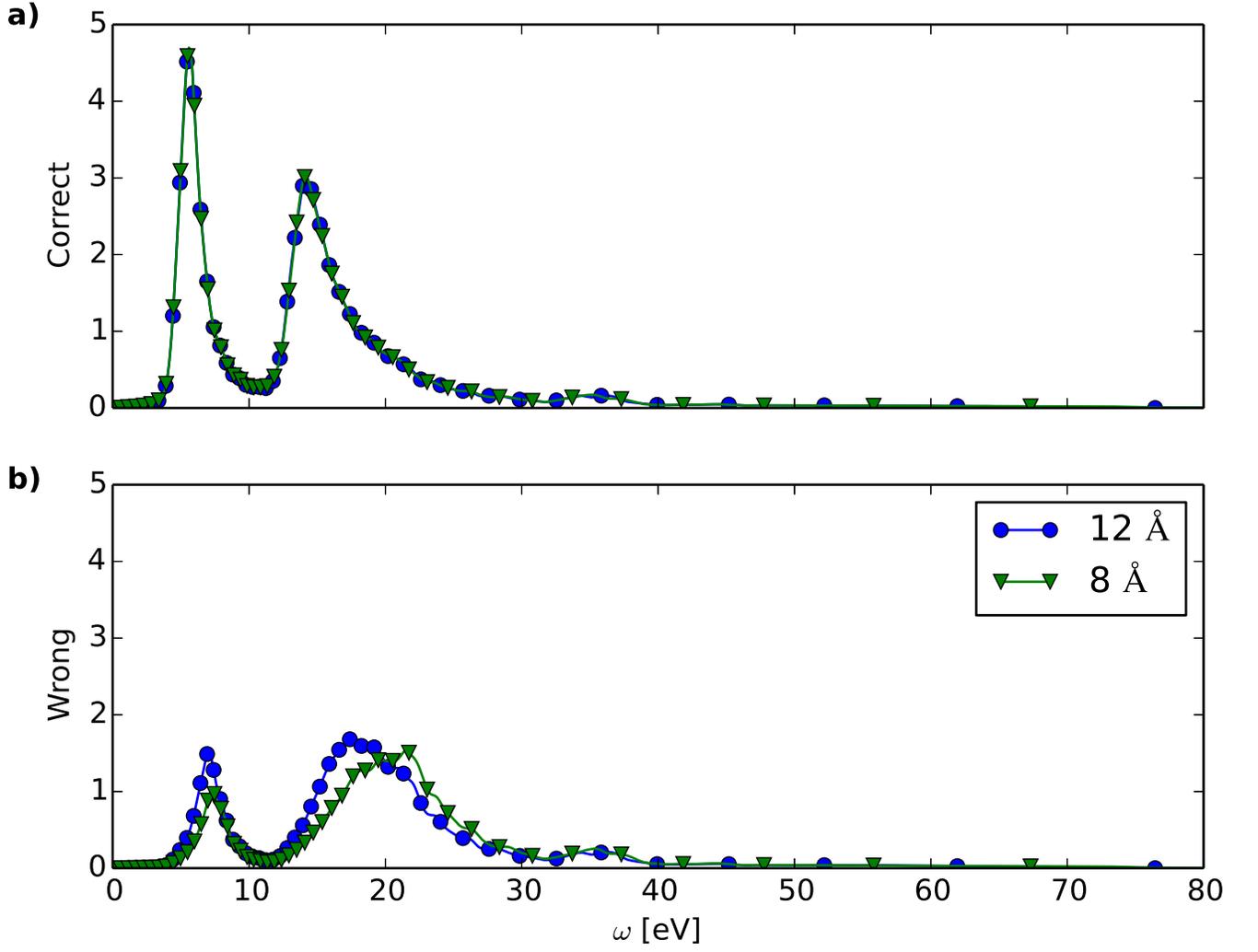}
\caption{
%(a)\ Squared parallel-field layer polarizabilites %
%$Im(\alpha _{xx}^{2Dscr}\left( \omega +i0\right)^2) $ 
(a)\ Parallel-field layer polarizabilites %
$Im(\alpha _{xx}^{2D,scr}\left( \omega +i0\right)) $ 
deduced, via the correct Eq (%
\ref{AlphaLayerScrxxFromEpsMacroxx}), from VASP $\varepsilon _{xx}^{macro}$
data for solid BN with stretched lattice spacings $D=0.8\ nm$ (triangles)
and $D=1.2\ nm$ (circles). The agreement for the two $D$ values is
excellent, verifying that (\ref{AlphaLayerScrxxFromEpsMacroxx}) correctly
allows for the interaction between the layers. \ 
(b) As in (a), except that the wrong formula (\ref{Aplha2DScrzzFromEpsMacro})
 (appropriate to $\epsilon _{zz}$ isntead of $\epsilon _{xx}$, and
reflecting lack of inter-layer interaction) was used. \ Now the cases %
$D=0.8\,\ nm$ and $D=1.2\ nm$ yield distinctly different results for 
$Im((\alpha_{xx}^{2D,scr})^2)$ , though one could still in principle obtain the correct 
$\alpha_{xx}^{2D,scr}\left( \omega +i0\right) $ by going to the $D\rightarrow \infty$ 
 case, where the layers truly do not interact.
 \label{fig:alphaxxFromepsMacro}
}
\end{figure}

\newpage
\begin{figure}[tbp]
% currently Fig 3
%\includegraphics{bothz2.eps}
\includegraphics{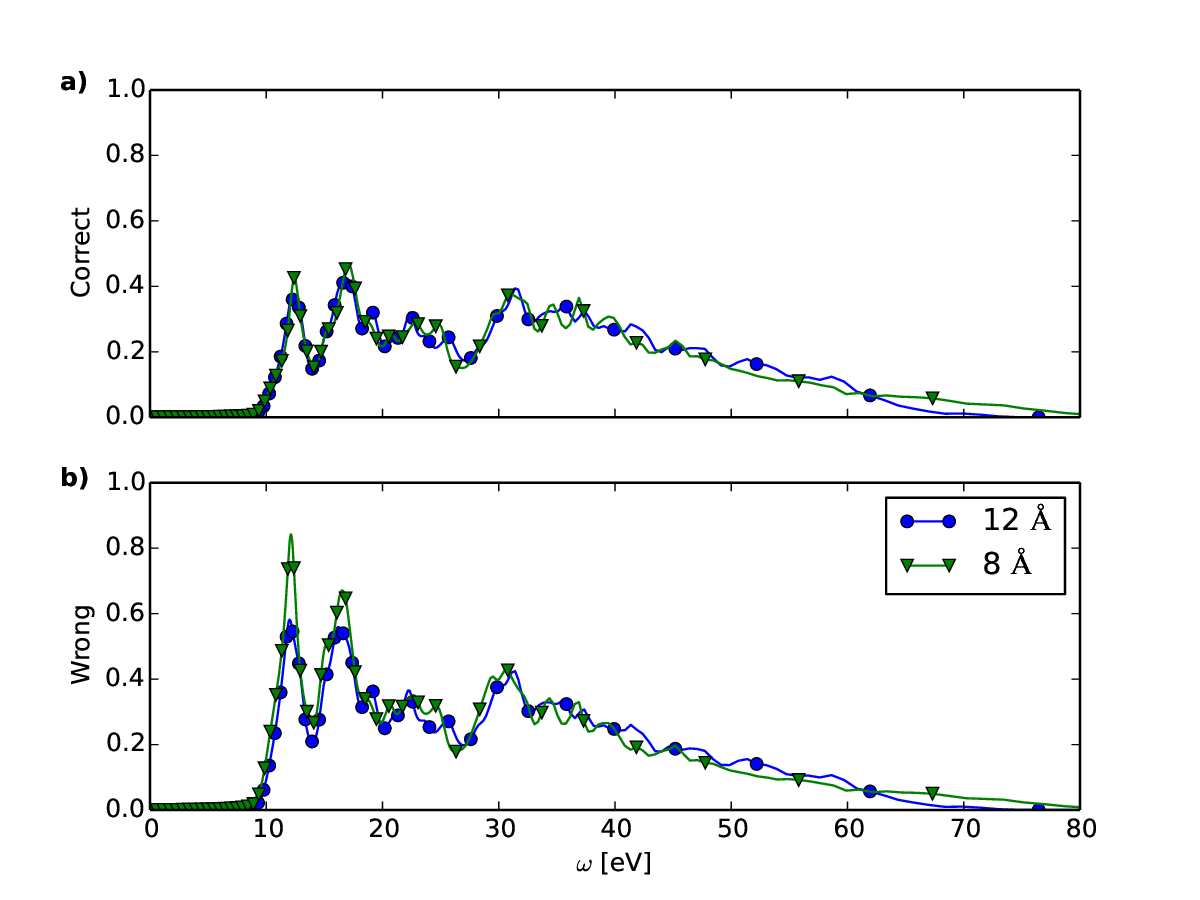}
\caption{
(a) Field-perpendicular layer polarizabilites %
$Im(\alpha _{zz}^{2D,scr}\left( \omega +i0\right)) $ 
deduced via 
Eq (\ref{Aplha2DScrzzFromEpsMacro}), from VASP $\varepsilon _{zz}^{macro}$ data for
solid BN with stretched lattice spacings $D=0.8\ nm$ (triangles) and $D=1.2\ nm$
(circles). The agreement for the two $D$ values is good. \\
(b) As in (a) but with unjustified use of the field-parallel formula 
(\ref{AlphaLayerScrxxFromEpsMacroxx}).  Agreement between $D = 0.8 nm$ and $D = 1.2 nm$ is now not so good.
\label{fig:alphazzFromEpsMacro}
}
\end{figure}

\newpage
\newpage
%Red dots: our asymptotic formula with 
%$\alpha_{zz}^{2D,scr}$ set to zero.

\newpage
%Red dots: with asymptotic correlation energy from 

%\newpage
%\begin{figure}
%% Not currently included
%\label{CompareLogApproxBNBilayer}
%\includegraphics{}
%\caption{
%Exploring the effect of expanding the logarithm in the expression %for the %interlayer correlation energy of 2-layer BN.  Solid %line:% $E_c(D)$ (meV/atom) %from the full logarithmic expression (\ref%%{EvdW2LayersFromAlphas}).  Line %with %symbols: $E_c(D)$ %from expansion %of the logarithm,Eq (\ref%{EcTwoLaersLogExpanded}).
%}
%\end{figure}
  
\newpage
%Red dots: our asymptotic formula using only the insulator-like contribution, which goes like $D^{-4}$%
%For the $D^{-3}$ prediction, arising from electronic transitions near the Dirac point and only significant compared with the %other terms for 
%$$D > 10 nm$, 

\newpage
%Red dots: our asymptotic formula using only the insulator-like $D^{-4}$ correlation energy contribution. 
\newpage
%Red dots: Our asymptotic formula.
\newpage

\end{document}